\documentclass[reprint,amsmath,amssymb,aps,prb,superscriptaddress]{revtex4-2}
\usepackage{array}
\usepackage{bm}
\usepackage{xcolor}
\usepackage{float}
\usepackage{graphicx}
\usepackage{hyperref}
\hypersetup{colorlinks=true,allcolors=blue}
\usepackage{natbib}
\usepackage{newtxtext}
\usepackage{newtxmath}
\usepackage{mathrsfs}
\usepackage[caption=false]{subfig}
\usepackage{tikz}
\begin{document}

\newcommand{\hexago}{\tikz[scale=0.1]{\draw (1,0) -- (0.5,0.866) -- (-0.5,0.866) -- (-1,0) -- (-0.5,-0.866) -- (0.5,-0.866) -- cycle;}}
\newcommand{\triang}{\tikz[scale=0.18]{\draw (0,0) -- (1,0) -- (0.5,0.866) -- cycle;}}
\newcommand{\squa}{\tikz[scale=0.18]{\draw (0,0) -- (1,0) -- (1,1) -- (0,1) -- cycle;}}
\newcommand{\rim}[1] {\textcolor{green}{#1}}

\title{Exact quantum spin liquids with topological order on maple-leaf and trellis lattices}

\author{Li Ern Chern}
\affiliation{Max Planck Institute for the Physics of Complex Systems, 01187 Dresden, Germany}

\author{Roderich Moessner}
\affiliation{Max Planck Institute for the Physics of Complex Systems, 01187 Dresden, Germany}

\author{Claudio Castelnovo}
\affiliation{T.C.M.~Group, Cavendish Laboratory, University of Cambridge, Cambridge CB3 0HE, United Kingdom}
%\email[]{}
%\thanks{}

\begin{abstract}
We construct spin models with bond-dependent anisotropic interactions on the penta-coordinated maple-leaf and trellis lattices, which yield exact $\mathbb{Z}_2$ quantum spin liquids akin to the Kitaev honeycomb model. We characterize the resulting ground states by their flux sectors, Chern numbers, and topological excitations. Using replica exchange quantum Monte Carlo simulations, we find that the gauge fluxes are ordered at sufficiently low temperatures such that every unit triangle has $\pm \pi/2$-flux, where the sign is uniform across the system, and every unit hexagon (square) has $0$-flux ($\pi$-flux) within the parameter space of interest. We map out the topological phase diagram for each model, which reveals parameter regimes hosting $\mathbb{Z}_2$ and Ising topological orders, and we derive an analytical expression for the mass term of the Majorana fermions, the vanishing of which indicates a transition between these phases. We further establish the correspondence between individual vortices (i.e., flux excitations) and two species of anyons in the dimer limit, overcoming the obstacle faced by degenerate perturbation theory in treating odd-length elementary plaquettes. Interestingly, we find that two dimer limits of the maple-leaf model with distinct assignments of anyon species can be smoothly connected to each other in the vortex-free sector, but they are separated by fermion-gap-closing transitions in certain two-vortex sectors.
\end{abstract}

\maketitle

% References should be done using the~\cite, \ref, and \label commands

\section{\label{section:introduce}Introduction}

Frustrated magnets provide fertile ground for the exploration of exciting collective phenomena, a case in point being emergent gauge theories, topological order, and fractionalization in quantum spin liquids~\cite{lacroixtextbook}. Frustration is oftentimes tied to specific lattice structures~\cite{annurev.ms.24.080194.002321}, with the iconic examples of the triangular, kagome, and pyrochlore lattices. Recent years have seen a rising interest in the magnetism of the geometrically frustrated maple-leaf lattice~\cite{PhysRevB.104.224415,PhysRevB.105.L180412,PhysRevB.108.L241116,PhysRevB.109.184422,Ghosh_2024,PhysRevB.110.014414,PhysRevB.110.085151,3zlw-hrbf,s43246-025-00904-1,jmgz-dsk9,1yv1-wtx8,zna-2025-0382,zna-2025-0376,zna-2025-0409,zna-2025-0389,2401.09422,2407.07145,2601.05308,2605.12592,2605.21587}, which is equivalent to a $1/7$-depleted triangular lattice~\cite{Proc.N.S.Inst.Sci.40.95,APhysPolA.97.971,PhysRevB.65.224405}, promoted in part by advances in numerical techniques and by the discovery of candidate materials with $S=1/2$ or $3/2$~\cite{anie.200502847,Fennell_2011,anie.201203775,PhysRevB.98.064412,PhysRevB.101.184429,PhysRevB.104.174439,PhysRevB.107.064419,PhysRevB.111.094439}.
Realistic models on the maple-leaf lattice consist of antiferromagnetic or mixed Heisenberg interactions that are isotropic in spin space, and their ground states cannot be analytically determined beyond a dimer regime~\cite{PhysRevB.105.L180412}, similar to the case in the Shastry-Sutherland model~\cite{SRIRAMSHASTRY19811069}. The hope is that the competing interactions arising from the triangular motifs may stabilize a quantum spin liquid in an extended parameter regime. Some works have indeed reported positive or promising results~\cite{PhysRevB.108.L241116,PhysRevB.110.085151,2401.09422,2407.07145,2601.05308}. However, different numerical methods do not always agree in their outcomes, even when they explore the same parameter space~\cite{zna-2025-0409}. Another structure of interest is the trellis lattice, which is made up of triangles and squares, and it offers yet another platform for geometric frustration~\cite{PhysRevB.49.8901,PhysRevB.56.R5736,PhysRevB.57.5005,JPSJ.67.3918,PhysRevLett.83.1387,PhysRevB.61.15185,PhysRevB.75.224414,PhysRevB.76.104429,JPSJ.86.124709,JPSJ.87.043701,xmtj-4kj9,0001147}. The Shastry-Sutherland, maple-leaf, and trellis lattices are the three penta-coordinated lattices among the eleven Archimedean lattices~\cite{PhysRevB.84.104406,PhysRevB.89.184407,PhysRevE.91.062121,PhysRevB.98.224402,SciPostPhys.19.1.025}.

The maple-leaf and trellis lattices are usually investigated under the lens of geometric frustration with antiferromagnetic Heisenberg or Ising interactions. Here, we take a different route and present them as venues for exact quantum-spin-liquid ground states that arise from bond-dependent anisotropic interactions similar to those in the Kitaev honeycomb model~\cite{KITAEV20062}. Each nearest-neighbor bond in the Kitaev honeycomb model is indexed by $\lambda \in \lbrace x,y,z \rbrace$ depending on its spatial orientation, such that (i) the pairwise interaction between local $S=1/2$ degrees of freedom connected by a $\lambda$ bond involves only the $\lambda$ component of spins, and (ii) the three bonds emanating from each site carry all distinct indices, leading to exact solvability upon a Majorana fermion representation. Similar models have been constructed on other networks with coordination number $z=3$, which coincides with the number of Pauli matrices, in both two and three spatial dimensions, such as the square-octagon~\cite{PhysRevB.76.180404}, star~\cite{PhysRevLett.99.247203}, hyperhoneycomb~\cite{PhysRevB.79.024426}, and hyperoctagon~\cite{PhysRevB.89.235102} lattices. Going beyond tri-coordination and $S=1/2$, Ref.~\onlinecite{PhysRevB.79.134427} introduced the idea of generalizing the $2 \times 2$ Pauli matrices to the $2^{n-1} \times 2^{n-1}$ $\Gamma$ matrices, which satisfy the Clifford algebra $\lbrace \Gamma^\mu, \Gamma^\nu \rbrace = 2 \delta^{\mu \nu}$, for $n>2$~\footnote{Pauli matrices are $\Gamma$ matrices with $n=2$.}. Such generalized Kitaev models are naturally defined on lattices with odd coordination number $z = 2n - 1$, which coincides with the number of $\Gamma$ matrices. Examples are the Shastry-Sutherland~\cite{PhysRevB.79.134427} and layered honeycomb~\cite{PhysRevB.81.125134} lattices, both of which have $z=5$ (i.e., $n=3$). Representing the $\Gamma$ matrices by Majorana fermions then yields exact $\mathbb{Z}_2$ quantum spin liquids. It is also possible to apply the $\Gamma$-matrix formalism to geometries with even coordination numbers by including slightly more sophisticated interactions, see, e.g., Refs.~\onlinecite{PhysRevB.79.075124,PhysRevLett.102.217202,PhysRevB.102.201111,PhysRevB.108.104208,km2j-3zy2}.

Taking advantage of their penta-coordination, we construct generalized Kitaev models on the maple-leaf and trellis lattices with bond-dependent interactions of $4 \times 4$ $\Gamma$ matrices, each of which can be interpreted as a local operator acting on two $S=1/2$ degrees of freedom. For each model, translational and rotational symmetries are imposed to reduce the number of independent couplings, which results in a two-dimensional parameter space. We then characterize the quantum-spin-liquid ground state by its flux sector, Chern number, and topological excitations. Although the model can be expressed as a quadratic Hamiltonian of Majorana fermions, the conserved $\mathbb{Z}_2$ gauge fluxes yield an exponentially large number of combinations, the most energetically favorable of which cannot be obtained by Lieb's theorem~\cite{PhysRevLett.73.2158,Macris1996}, because neither the maple-leaf nor trellis lattice is bipartite. We thus numerically determine the ground-state flux sector via replica exchange quantum Monte Carlo simulations~\cite{PhysRevLett.113.197205,PhysRevLett.115.087203,PhysRevB.92.115122,nphys3809,PhysRevB.96.125124,PhysRevB.98.054432,PhysRevResearch.1.032011,PhysRevB.101.045118,eschmannthesis,PhysRevB.102.075125,PhysRevResearch.2.043159}, which are free from the sign problem and sped up by the Green-function-based kernel polynomial method~\cite{RevModPhys.78.275,PhysRevLett.102.150604}. In the low-temperature regime of the Kitaev maple-leaf (trellis) model, we find that the fluxes of unit hexagons (squares) are all $0$ ($\pi$), while the fluxes of unit triangles are either all $+ \pi / 2$ or all $- \pi / 2$, in accordance with the flux phase conjecture~\cite{s41467-023-42105-9}. The two possibilities $\pm \pi / 2$ imply a twofold degeneracy and a spontaneous breaking of time-reversal symmetry in the ground state~\cite{KITAEV20062,PhysRevLett.99.247203}; the latter in particular manifests as a sharp peak in the heat capacity as the system is cooled down~\cite{PhysRevLett.115.087203,PhysRevResearch.2.043159}. 

We map out a topological phase diagram, which reveals the variation of the ground-state Chern number $\nu$ over the parameter space, for each model. We derive an effective two-band Hamiltonian for parameters near the gap-closing transition and identify the mass term of Majorana fermions that is responsible for the change in $\nu$~\cite{bernevigtextbook}. This allows for a precise demarcation of the $\nu = 0$ and $-1$ phases in the parameter space. According to Kitaev's sixteenfold way~\cite{KITAEV20062}, the former (latter) phase is described by $\mathbb{Z}_2$ (Ising) topological order at low energies and long distances, with flux excitations or vortices corresponding to two (one) different (single) species of anyons. As degenerate perturbation theory is unable to account for the anyon species of odd-length elementary plaquettes, we use the theory of Ref.~\onlinecite{2607.12027} (see also Ref.~\onlinecite{Wootton_2015}), which is based on the fusion rule of Abelian anyons and the fermion parity of physical states, to assign the vortices in the dimer limits of the Kitaev maple-leaf and trellis models to the electric and magnetic particles of the toric code model~\cite{KITAEV20032}. For the Kitaev maple-leaf model, we further show that a change of anyon species is effected by fermion-gap-closing transitions in some two-vortex sectors but not in the vortex-free sector~\cite{2607.12027}, making it the first time-reversal-symmetry-breaking example of topologically ordered states that exhibit such behavior. Our work thus combines aspects of quantum spin liquids, topological band theory, and topological order, in a way that had not been hitherto applied to the study of maple-leaf or trellis magnets.

The rest of this paper is organized as follows. In Sec.~\ref{section:model}, we define generalized Kitaev models on the maple-leaf and trellis lattices, and introduce the Majorana fermion representation of the $\Gamma$ matrices, which is the crucial ingredient for our exact solutions. In Sec.~\ref{section:numerics}, we present the results of numerical simulations and establish the ground-state flux sector of each model. The latter admits a translationally invariant gauge choice, which is useful for further analyses in the thermodynamic limit via Fourier transform. In Sec.~\ref{section:topology}, we study the ground-state band topology of the itinerant fermions and demarcate the parameter regimes with distinct Chern numbers. In Sec.~\ref{section:anyon}, we identify the topological order of each gapped parameter regime according to Kitaev's sixteenfold way, and investigate the correspondence between vortices and anyons. In Sec.~\ref{section:discuss}, we discuss the relevance of our present work and point out several directions that may be of interest for future studies. This paper is accompanied by a Supplemental Material~\cite{supply}, which contains details of various concepts, methods, and results presented in the main text.

\section{\label{section:model}Models}

Both the maple-leaf and trellis lattices are penta-coordinated. For $z=5$ or equivalently $n=3$, we have five $\Gamma$ matrices, each of which is four-dimensional~\cite{PhysRevB.79.134427}. We choose the five $\Gamma$ matrices to be~\cite{PhysRevB.98.054432}
\begin{equation} \label{gammadefine}
\begin{gathered}[b]
\Gamma^{x0} \equiv \tau^x \otimes \vmathbb{1} , \quad \Gamma^{y0} \equiv \tau^y \otimes \vmathbb{1} , \\
\Gamma^{zx} \equiv \tau^z \otimes \sigma^x , \quad \Gamma^{zy} \equiv \tau^z \otimes \sigma^y , \quad \Gamma^{zz} \equiv \tau^z \otimes \sigma^z 
\, , 
\end{gathered}
\end{equation}
where we have labeled each $\Gamma$ matrix according to the two Pauli matrices $\tau^\mu$ and $\sigma^\nu$ involved in its definition. It is easy to see from~\eqref{gammadefine} that the $\Gamma$ matrices anticommute with each other, and each of them squares to the identity, i.e., they obey the Clifford algebra as desired. The $\Gamma$ matrices can be interpreted as $S=3/2$ quadrupoles (cf.~the Pauli matrices as $S=1/2$ dipoles)~\cite{PhysRevB.79.134427,PhysRevLett.102.217202}, but we find no use of this interpretation in our current work. 

\begin{figure}
\subfloat[]{\label{figure:maplemodel}
\includegraphics[scale=0.24]{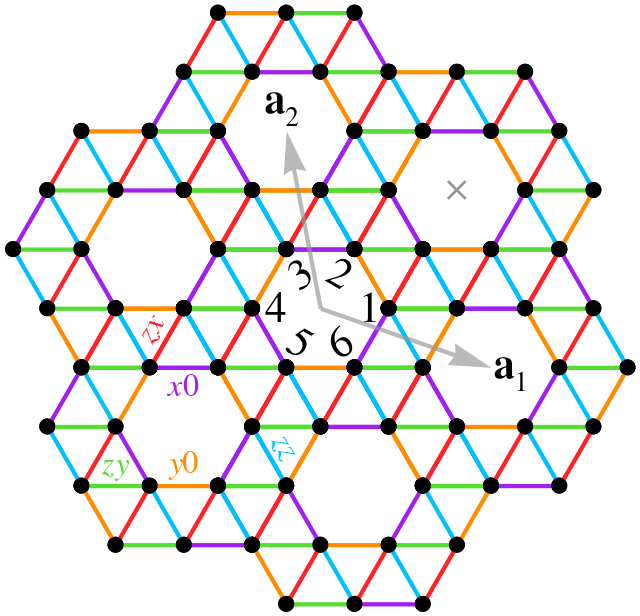}} \quad
\subfloat[]{\label{figure:trellismodel}
\includegraphics[scale=0.24]{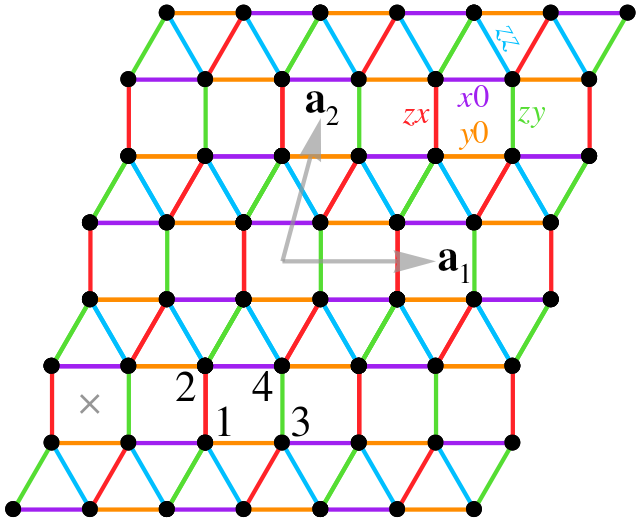}}
\caption{Generalized Kitaev models on the (a) maple-leaf and (b) trellis lattices. In each panel, the colors on the bonds indicate the types of pairwise interactions $J_\lambda$, see~\eqref{kitaevmodel}. The arrows represent the primitive lattice vectors $\mathbf{a}_1$ and $\mathbf{a}_2$. The numbers label the sites in a unit cell, i.e., the sublattices. The cross indicates a center of rotational symmetry.}
\end{figure}

To construct the generalized Kitaev model, we assign five different colors, which represent the indices of the $\Gamma$ matrices, to the bonds of the maple-leaf and trellis lattices. We require that the five bonds emanating from any site have all distinct colors for exact solvability. Furthermore, we would like the resulting models to have translational symmetries, such that there exist well-defined unit cells each containing a minimal number of sites. For the maple-leaf lattice, up to a permutation of colors, we find a unique assignment that preserve the translational symmetry of the lattice, as shown in Fig.~\ref{figure:maplemodel}. For the trellis lattice, the coloring of the bonds inevitably breaks the translational symmetry of the lattice, such that the unit cell is doubled along the $\mathbf{a}_1$ direction. We find eight such assignments that are not equivalent under a permutation of colors, and we choose the one shown in Fig.~\ref{figure:trellismodel}, where the implementation of a twofold symmetry is possible. Assigning the same interaction strength to bonds with the same color, we define the Kitaev maple-leaf and trellis models as
\begin{equation} \label{kitaevmodel}
H = - \sum_\lambda \sum_{\langle ij \rangle \in \lambda} J_\lambda \Gamma^\lambda_i \Gamma^\lambda_j 
\, , \quad J_\lambda > 0 \, .
\end{equation}

Either model contains five independent couplings, making it difficult to explore and visualize the full parameter space. Further simplifications can be made using point-group symmetries. For the Kitaev maple-leaf model, we impose a $C_3$ symmetry, which enforces $J_{zx} = J_{zy} = J_{zz}$, thus reducing the number of independent couplings by two. For the Kitaev trellis model, we impose a $C_2$ symmetry, which enforces $J_{x0} = J_{y0}$ and $J_{zx} = J_{zy}$, thus reducing the number of independent couplings by two also. Details of the symmetry analyses can be found in Sec.~\ref{section:symmetry} of the Supplemental Material~\cite{supply}. We only remark here that each local degree of freedom is treated as a composite of two $S=1/2$ moments, which do not have to transform in the same way under the point-group symmetry. We now have three independent couplings in each model. Since the overall energy scale is not meaningful, we further compactify the parameter space by $J_\mu + J_\nu + J_\lambda = 1$, where $(J_\mu , J_\nu , J_\lambda)=(J_{x0} , J_{y0} , J_{zx})$ and $(J_{x0} , J_{zx} , J_{zz})$ for the maple-leaf and trellis lattices, respectively. The plane equation $J_\mu + J_\nu + J_\lambda =1$ limited to the first octant defines a triangular parameter space. We also define $\mathbf{J} = (J_\mu , J_\nu , J_\lambda)$ for convenience. 

The model~\eqref{kitaevmodel} is exactly solvable via a Majorana fermion representation of the $\Gamma$ matrices~\cite{PhysRevB.79.134427,PhysRevB.79.075124,PhysRevLett.102.217202}. We need six species of Majorana fermions for the five local $\Gamma$ matrices~\footnote{For generic $n \geq 2$, the $2n-1$ local $\Gamma$ matrices can be represented in terms of $2n$ species of Majorana fermions \cite{PhysRevB.79.134427}.}. Substituting $\Gamma_i^\lambda = i b_i^\lambda c_i$ in \eqref{kitaevmodel} yields
\begin{equation} \label{kitaevmodelmajorana}
H = \sum_\lambda \sum_{\langle ij \rangle \in \lambda} i J_\lambda u_{ij}^\lambda c_i c_j 
\, , 
\quad u_{ij}^\lambda \equiv i b_i^\lambda b_j^\lambda 
\, . 
\end{equation}
The bond operators $u_{ij}^\lambda$ commute with each other as well as with $H$, so they can be replaced by their eigenvalues $\pm 1$. This reduces~\eqref{kitaevmodelmajorana} to a quadratic Hamiltonian
\begin{equation} \label{kitaevmodelquadratic}
H = \frac{i}{4} \sum_{ij} c_i A_{ij} c_j 
\, , \quad 
A_{ij} = \left \lbrace \begin{array}{ll} 2 J_\lambda u_{ij}^\lambda & \mathrm{if} \, \langle ij \rangle \in \lambda , \\ 0 & \mathrm{otherwise.} \end{array} \right.
\end{equation}
The quantities $u_{ij}^\lambda = \pm 1$ can be interpreted as a $\mathbb{Z}_2$ gauge field~\footnote{A more accurate name for $u_{ij}^\lambda$ would be the \textit{gauge vector potential}. Here, we follow the more common convention in the literature and call $u_{ij}^\lambda$ the \textit{gauge field}.}. The gauge field itself is not gauge invariant. Instead, the gauge flux
\begin{equation} \label{gaugefluxdefine}
W_p \equiv \prod_{\langle ij \rangle_\lambda \in \partial p} \Gamma_i^\lambda \Gamma_j^\lambda = \prod_{\langle ij \rangle_\lambda \in \partial p} (-i)^{\lvert \partial p \rvert} u_{ij}^\lambda
\end{equation}
defined around an elementary plaquette $p$ is gauge invariant~\footnote{It follows that the product of $\Gamma$ matrices involved in the pairwise interactions along any closed path, known as the Wilson loop, is also gauge invariant.}, where $\lvert \partial p \rvert$ denotes the number of bonds bounding $p$. Different plaquette operators $W_p$ commute with each other as well as with $H$. Therefore, the Hilbert space can be divided into sectors labeled by the flux eigenvalues of all elementary plaquettes. For future use, we define the phase $
\Phi_p = - i \ln W_p \in ( -\pi , \pi ]$ of the flux $W_p$, and we will refer to $\Phi_p$ also as the flux where no confusion arises.

\section{\label{section:numerics}Numerical Simulations}

We would like to obtain the ground-state flux sectors of the Kitaev maple-leaf and trellis models. Neither lattices is bipartite, so they do not fulfill the conditions under which Lieb's theorem~\cite{PhysRevLett.73.2158,Macris1996} is applicable to determine analytically the ground-state flux sector. For each model, we thus resort to replica exchange quantum Monte Carlo simulations~\cite{PhysRevLett.113.197205,PhysRevLett.115.087203,PhysRevB.92.115122,nphys3809,PhysRevB.96.125124,PhysRevB.98.054432,PhysRevResearch.1.032011,PhysRevB.101.045118,eschmannthesis,PhysRevB.102.075125,PhysRevResearch.2.043159} and solve for the ground-state flux sector numerically, on a regular grid of $28$ points in the triangular parameter space. 

The methodology is detailed in Sec.~\ref{section:montecarlo} of the Supplemental Material~\cite{supply}. We only mention here that, for the Monte Carlo part, the core step consists of a bond-flip proposal, where the sign of the gauge field on a randomly chosen bond is reversed, followed by the evaluation of the resulting change in free energy. The latter can be achieved by exactly diagonalizing the $N$-dimensional Hamiltonian~\eqref{kitaevmodelquadratic}, where $N$ is the total number of sites in the system. Since the exact diagonalization method has a time complexity of $O ( N^3 )$, computations become expensive when $N$ is large. To treat large systems while keeping numerical costs low, we employ the Green-function-based kernel polynomial method~\cite{RevModPhys.78.275,PhysRevLett.102.150604,PhysRevB.96.125124,PhysRevB.102.075125}, where the change in free energy is expressed in terms of the Green function, which in turn is approximated by a Chebyshev expansion up to some finite order $M$ (the larger $M$, the higher the accuracy) \cite{JPSJ.68.3853}. With a time complexity of $O ( M N )$, the kernel polynomial method enables a significant speedup in comparison to exact diagonalization if $M \lesssim N$. 

Our simulations are run on lattices with $L \times L$ unit cells and periodic boundary conditions. For $L \leq 5$, we use exact diagonalization to calculate the change in free energy. For $L > 5$, we use the kernel polynomial method to do so, with $M=256$ or $512$ for the order of the Chebyshev expansion. We measure the spatially averaged flux for each species of elementary plaquettes, e.g., $\Phi_{\hexago}= (1 / N_{\hexago}) \sum_{p \in \hexago} \Phi_p$ for the unit hexagons in the Kitaev maple-leaf model, where $N_{\hexago}$ is the total number of unit hexagons. However, we find that the fluxes of unit triangles are ordered at low temperatures such that the spatial average is $\Phi_{\triang} = \pm \pi / 2$, and the thermal average $\langle \Phi_{\triang} \rangle$ over different measurements in general does not reflect the saturated value due to cancellations of $+ \pi / 2$ and $- \pi / 2$. Therefore, for the unit triangles, we take the absolute value of the spatially averaged flux, $\lvert \Phi_{\triang} \rvert = (1 / N_{\triang}) \lvert \sum_{p \in \triang} \Phi_p \rvert$, at each measurement. We also measure the total heat capacity $C$, which includes contributions from both itinerant fermions and gauge fluxes~\cite{supply}. 

Our simulations suggest that the ground-state flux sector is such that $\Phi_p=0$ ($\pi$) for all unit hexagons (squares), while $\Phi_p= \pm \pi / 2$ for all unit triangles, independent of the choice of couplings~\footnote{The issue of domain walls is discussed in the Supplemental Material. The reader should also bear in mind that our conclusions are of course drawn from simulations performed only for a finite number of parameters.}. This is consistent with the extended flux phase conjecture~\cite{s41467-023-42105-9}, which states that $W_p = - ( \pm i )^{\lvert \partial p \rvert}$ in the ground state, where the $\pm$ sign inside the bracket is a global choice. In other words, the spatially averaged fluxes $\Phi_{\hexago}$, $\Phi_{\squa}$, and $\lvert \Phi_{\triang} \rvert$ reach the saturated values $0$, $\pi$, and $\pi / 2$, respectively, at sufficiently low temperatures. $\Phi_{\triang}=\pm \pi/2$ constitute a pair of time-reversal partners that are equal in energy. Time-reversal symmetry is spontaneously broken when the fluxes of unit triangles order in either way. In general, the presence of odd-length plaquettes in a generalized Kitaev model indicates that the energy eigenstates are at least twofold degenerate and that time-reversal symmetry is spontaneously broken in the ground state~\cite{KITAEV20062,PhysRevLett.99.247203}. 

\begin{figure}
\includegraphics[scale=0.24]{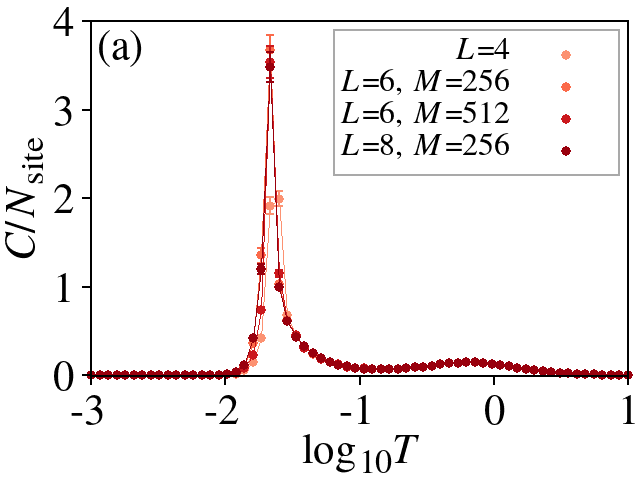}
\includegraphics[scale=0.24]{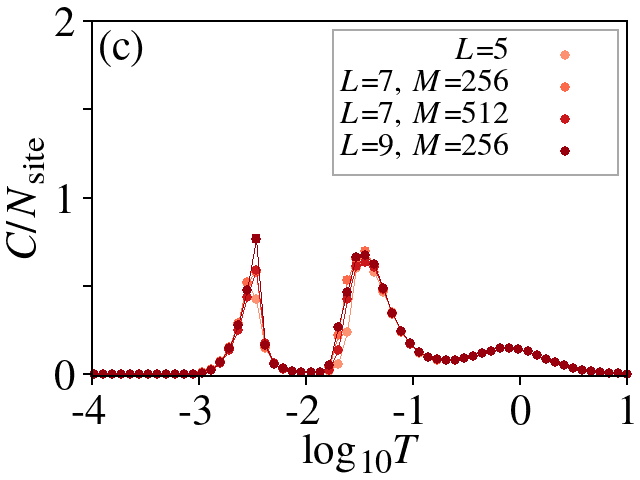} \\
\includegraphics[scale=0.24]{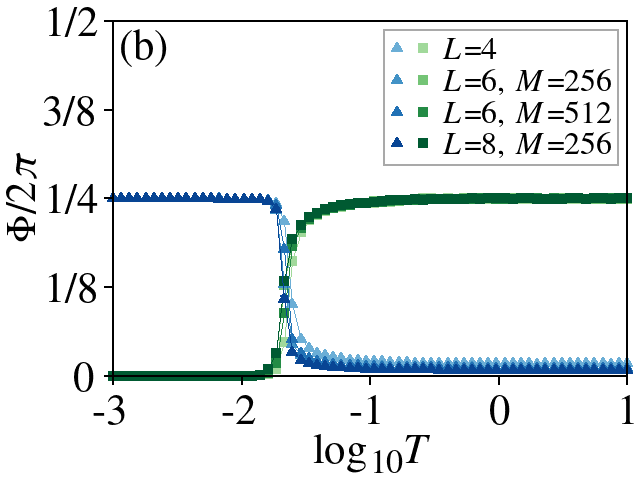}
\includegraphics[scale=0.24]{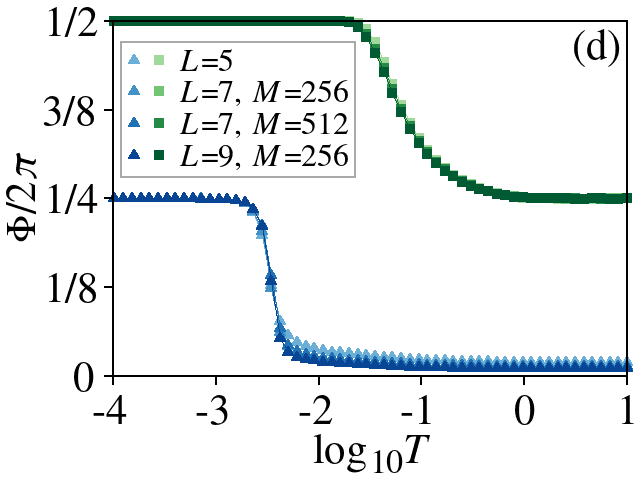}
\caption{\label{figure:numerics}(a) The heat capacity per site and (b) the spatially averaged fluxes of the unit hexagons and triangles, as functions of temperature, for the Kitaev maple-leaf model with $\mathbf{J}=(1/3,1/3,1/3)$. (c) The heat capacity per site and (d) the spatially averaged fluxes of the unit squares and triangles, as functions of temperature, for the Kitaev trellis model with $\mathbf{J}=(0.448,0.448,0.105)$. In (c,d), the green squares represent $\Phi_{\protect\hexago}$ or $\Phi_{\protect\squa}$, while the blue triangles represent $\lvert \Phi_{\protect\triang} \rvert$. $L$ and $M$ are the linear size of the system and the order of the Chebyshev expansion, respectively. Data with (without) $M$ indicated are obtained via the kernel polynomial (exact diagonalization) method.}
\end{figure}

We provide examples of how the heat capacity $C$ and the spatially averaged fluxes $\Phi$ evolve with the temperature $T$. Figs.~\ref{figure:numerics}a and~\ref{figure:numerics}b show data from simulations of the Kitaev maple-leaf model at the isotropic point in the parameter space, where $J_\lambda = 1/3$ for all $\lambda$. A small bump in the heat capacity at $T \sim 1$ indicates fractionalizations into Majorana fermions~\cite{PhysRevB.92.115122,PhysRevB.102.075125,PhysRevResearch.2.043159,JPSJ.89.012002}, while a sharp peak between $T=10^{-1}$ and $10^{-2}$ indicates a symmetry-breaking phase transition, see Fig.~\ref{figure:numerics}a. An inspection of Fig.~\ref{figure:numerics}b reveals that the sharp peak coincides with the simultaneous orderings of hexagon and triangle fluxes, i.e., $\Phi_{\hexago} = 0$ and $\lvert \Phi_{\triang} \rvert = \pi / 2$, the latter of which signals a spontaneous breaking of time-reversal symmetry. By tuning the anisotropy of the couplings, it is possible for different fluxes to order at different temperatures, as demonstrated by the next example. Fig.~\ref{figure:numerics}c and~\ref{figure:numerics}d show data from simulations of the Kitaev trellis model for the couplings $J_{x0}=J_{y0}=0.448$, $J_{zx}=J_{zy}=0.448$, and $J_{zz}=0.105$ (rounded to three decimal places). Similarly to the previous example, a small bump in $C ( T )$ at $T \sim 1$ indicates fractionalizations. In addition, $C ( T )$ exhibits a smooth peak and a sharp peak as $T$ is further lowered down to $10^{-4}$, see Fig.~\ref{figure:numerics}c. An inspection of Fig.~\ref{figure:numerics}d reveals that the smooth (sharp) peak at higher (lower) temperature coincides with the ordering of the square (triangle) fluxes, i.e., $\Phi_{\squa} = \pi$ ($\lvert \Phi_{\triang} \rvert = \pi / 2$). We note that this situation is similar to the partial flux ordering in the generalized Kitaev model on the Shastry-Sutherland lattice studied in Ref.~\onlinecite{PhysRevResearch.2.043159}. 

\begin{figure}
\subfloat[]{\label{figure:maplegauge}
\includegraphics[scale=0.18]{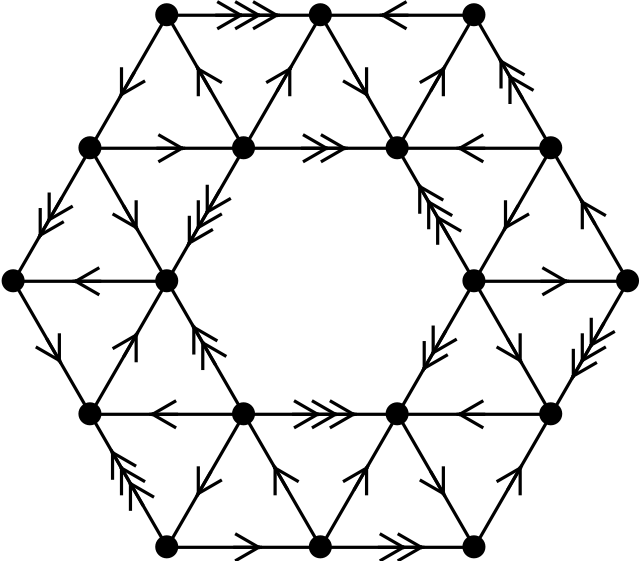}} \qquad
\subfloat[]{\label{figure:trellisgauge}
\includegraphics[scale=0.18]{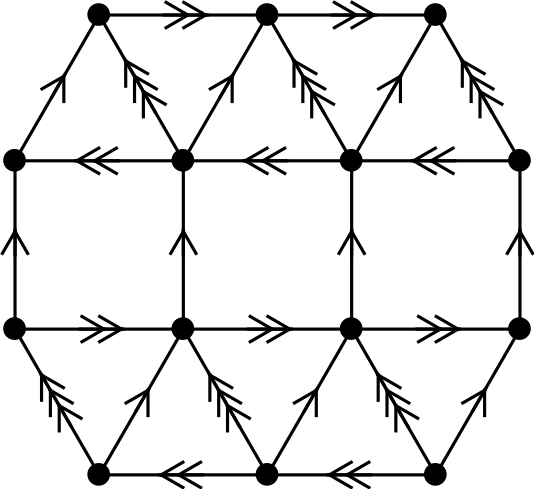}}
\caption{Translationally invariant gauge-field configurations that produce the ground-state flux sectors of the Kitaev (a) maple-leaf and (b) trellis models. Bonds that have the same number of arrows on them indicate that their respective couplings are equal in magnitude. On each bond $\langle ij \rangle_\lambda$, $u_{ij}^\lambda = +1$ ($-1$) along (against) the direction of the arrow(s). All triangle fluxes are set to be $+i$, which corresponds to one of the two degenerate sectors related by time reversal. Note in particular that the gauge-field configuration in (b) respects the translation symmetry not only of the Kitaev trellis model (Fig.~\ref{figure:trellismodel}), but also of the trellis lattice itself.}
\end{figure}

For each of the Kitaev maple-leaf and trellis models, we have thus established the ground-state flux sector at a finite number of points in the triangular parameter space \cite{supply}. It is uniform across the selected parameters and it is in accordance with the flux phase conjecture~\cite{s41467-023-42105-9}. We will henceforth assume that the same ground-state flux sector applies to the entire triangular parameter space. Note that the gauge-flux configuration is translationally invariant in the ground state. It turns out that one can also choose a translationally invariant gauge-field configuration that produces the respective ground-state flux sector of each model~\footnote{The gauge fields themselves are unphysical, but different gauge-field configurations yielding the same gauge-flux configuration must give the same set of physical quantities, such as the energy spectrum of the itinerant fermions and the Chern number indicating the net chiral edge mode(s).}, as shown in Figs.~\ref{figure:maplegauge} and~\ref{figure:trellisgauge}. For concreteness, we set all triangle fluxes to $W_p = +i$, which corresponds to one of the degenerate time-reversal partners, in the remainder of our work~\footnote{We also verified by explicit calculation that the fermion gap remains invariant, while the Chern number changes sign, upon reversing $W_p$ from $+i$ to $-i$ for all unit triangles.}. 

A translationally invariant gauge-field configuration is highly desirable as it enables further analyses in the thermodynamic limit via Fourier transform to momentum space. Of particular interest is the band topology of the Majorana fermions in the ground state, which is characterized by the Chern number. Apart from representing the net chiral edge mode(s), the Chern number also reveals the topological order---which in turn specifies the anyonic excitations---of the system via Kitaev's sixteenfold way~\cite{KITAEV20062}. Topological aspects of the Kitaev maple-leaf and trellis models are explored in the next two sections.

\section{\label{section:topology}Topological Phase Diagrams}

Given the role of the Chern number $\nu$ in characterizing the ground-state topology \cite{KITAEV20062}, we set off to determine how $\nu$ varies over the triangular parameter spaces of the Kitaev maple-leaf and trellis models. Since $\nu$ is a topological invariant that cannot change without a gap-closing transition, the first step is naturally to calculate the excitation gap $\Delta_\psi$ of the itinerant fermions across the parameter space. Before presenting the results specific to each model, we discuss some general features in the next two paragraphs. 

\begin{figure*}
\subfloat[]{\label{figure:maplegap}
\includegraphics[scale=0.23]{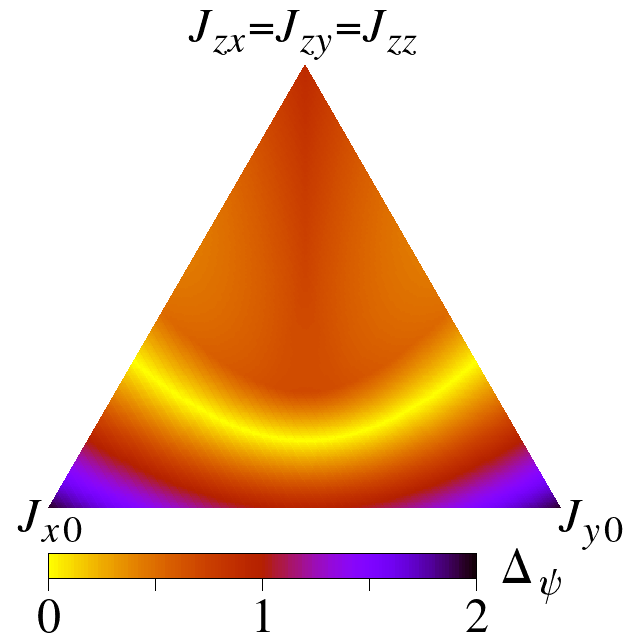}}
\subfloat[]{\label{figure:maplemass}
\includegraphics[scale=0.23]{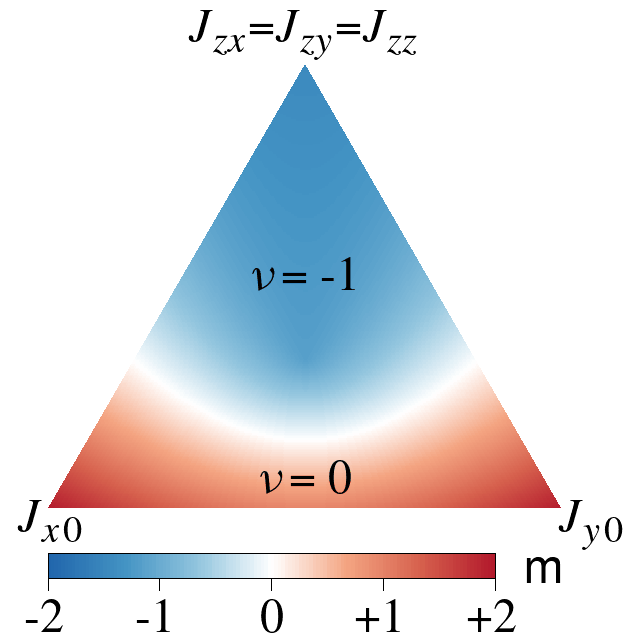}}
\subfloat[]{\label{figure:trellisgap}
\includegraphics[scale=0.23]{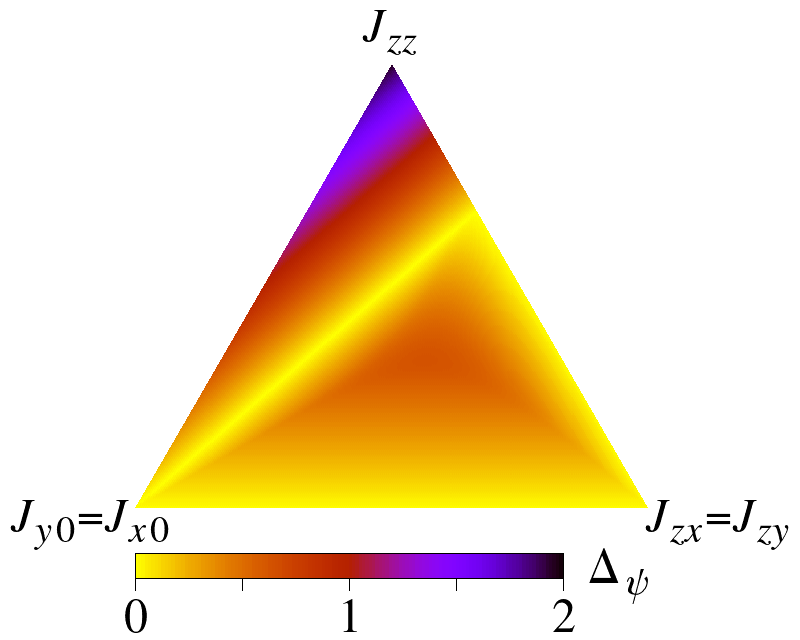}}
\subfloat[]{\label{figure:trellismass}
\includegraphics[scale=0.23]{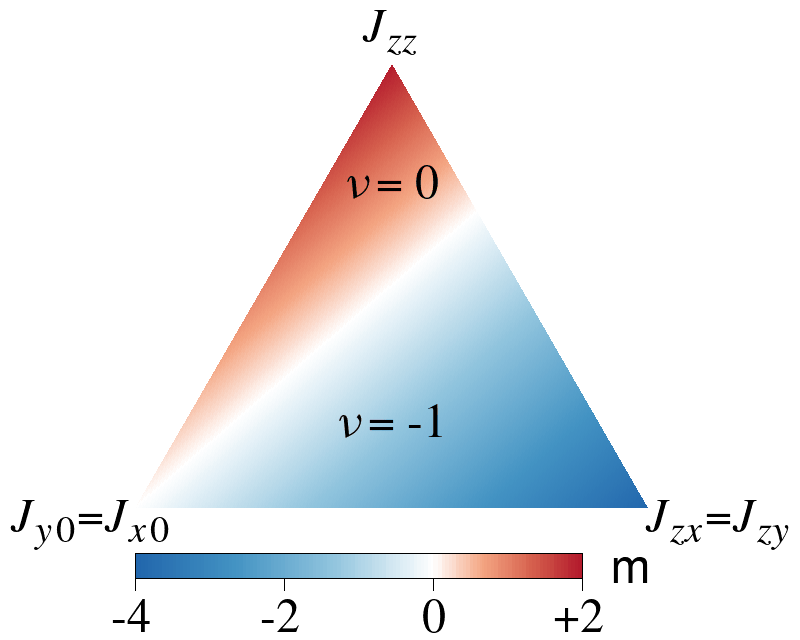}}
\caption{(a) The fermion gap $\Delta_\psi$ and (b) the mass term $\mathsf{m}$ of the Kitaev maple-leaf model, in the triangular parameter space $J_{x0} + J_{y0} +  J_{zx} = 1$. (c) The fermion gap $\Delta_\psi$ and (d) the mass term $\mathsf{m}$ of the Kitaev trellis model, in the triangular parameter space $J_{x0} + J_{zx} + J_{zz} = 1$. In (a-d), the vertex labeled by $J_\lambda$ corresponds to the limit $J_\lambda \longrightarrow 1$. In (b) and (d), the Chern number $\nu$ of each gapped regime is indicated. Note that only the values of $\mathsf{m}$ near $0$, i.e., in the vicinity of the gap-closing transition, are meaningful, while $\nu$, once determined, must be constant throughout each gapped parameter regime.}
\end{figure*}

Let $\mathcal{N}$ be the total number of unit cells, $\mathbf{R}$ be the coordinates of a site on the underlying Bravais lattice, and $\alpha$ be the sublattice index, such that an arbitrary site $i$ can be identified by $(\mathbf{R},\alpha)$, i.e., the unit cell and the sublattice to which it belongs. Upon Fourier transforming~\cite{KITAEV20062}
\begin{equation} \label{fourier}
c_{\mathbf{q}\alpha} = \frac{1}{\sqrt{2 \mathcal{N}}} \sum_{\mathbf{R}} c_{\mathbf{R}\alpha} e^{- i \mathbf{q} \cdot \mathbf{R}} 
\, , 
\end{equation}
the Hamiltonian~\eqref{kitaevmodelquadratic} takes the form
\begin{equation} \label{kitaevmodelfourier}
H = \frac{i}{2} \sum_{\mathbf{q}} \sum_{\alpha \beta} c_{- \mathbf{q} \alpha} A_{\alpha \beta} (\mathbf{q}) c_{\mathbf{q} \beta} 
\, ,
\end{equation}
where $A_{\alpha \beta} ( \mathbf{q} ) \equiv \sum_\mathbf{R} A_{(\mathbf{0},\alpha)(\mathbf{R},\beta)} e^{i \mathbf{q} \cdot \mathbf{R}}$. The excitation spectrum $\varepsilon ( \mathbf{q} ) \geq 0$ is given by the non-negative eigenvalues of the matrix $i A ( \mathbf{q} )$, the dimension of which is equal to the number of sublattices, i.e., the number of sites per unit cell. The fermion gap is defined as $\Delta_\psi = \min_\mathbf{q} \varepsilon ( \mathbf{q} )$. 

In the vicinity of a gap-closing transition, suppose that the effective Hamiltonian near zero energy can be expressed as
\begin{equation} \label{effectmodel}
\begin{aligned}[b]
H^\mathrm{eff} ( \mathbf{q} ) = & \, (a_{11} q_x + a_{21} q_y) \sigma^\mu + (a_{12} q_x + a_{22} q_y) \sigma^\nu \\
& + \mathsf{m} \, \sigma^\lambda \, , \\
\quad \mathsf{A} \equiv & \begin{pmatrix} a_{11} & a_{12} \\ a_{21} & a_{22} \end{pmatrix} \in \mathrm{GL} ( 2 , \mathbb{R} ) \, , \quad \mathsf{m} \in \mathbb{R} \, ,
\end{aligned}
\end{equation}
where $(\lambda , \mu , \nu)$ is a cyclic permutation of $(x,y,z)$. Then, the vanishing of the mass term $\mathsf{m}$ is responsible for the gap closing, and a sign change of $\mathsf{m}$ is related to a change in the Chern number as~\cite{bernevigtextbook}
\begin{equation} \label{chernchange}
\Delta \nu = - \frac{1}{2} [ \Delta \, \mathrm{sgn} (\mathsf{m}) ] \, \mathrm{sgn} (\det \mathsf{A}) 
\, ,
\end{equation}
where $\Delta \, \mathrm{sgn} ( \mathsf{m} ) = +2$ ($-2$) if $\mathsf{m}$ is changed from negative (positive) to positive (negative) values. A short review of the topology in a two-band model, which includes the derivation of~\eqref{chernchange}, can be found in Sec.~\ref{section:band} of the Supplemental Material~\cite{supply}.

For the Kitaev maple-leaf model, the fermion spectrum $\varepsilon ( \mathbf{q} )$ is obtained by numerically diagonalizing the $6 \times 6$ matrix $i A ( \mathbf{q} )$ in the Hamiltonian~\eqref{kitaevmodelfourier}, which is constructed according to the $C_3$ symmetric gauge in Fig.~\ref{figure:maplegauge}. The fermion gap $\Delta_\psi$ is plotted over the triangular parameter space in Fig.~\ref{figure:maplegap}. We find that the system is gapped for most of the parameters, except for a curve which separates the parameter space into an upper and lower parts. Within each gapped regime, we compute the total Chern number of the three negative-energy bands~\cite{JPSJ.74.1674}, allowing them to cross each other as long as $\Delta_\psi$ remains finite. The Chern numbers are $\nu=-1$ and $0$ in the upper and lower parts, respectively. Moreover, we find that the fermion gap always closes at the $\Gamma$ point $\mathbf{q}=0$ with our gauge choice. This allows us to derive an effective two-band Hamiltonian of the form~\eqref{effectmodel} at low energies, with the mass term given by
\begin{equation} \label{maplemass}
\begin{aligned}[b]
\frac{\mathsf{m}}{\sqrt{2}} = & \left[ (J_{zx} - J_{x0})^2 + (J_{x0} - J_{y0})^2 + (J_{y0} - J_{zx})^2 \right]^{1/2} \\
& - \sqrt{6} J_{zx} 
\, .
\end{aligned}
\end{equation}
The condition $\mathsf{m}=0$ thus provides an analytical expression for the topological phase boundary, which separates the $\nu=-1$ and $\nu=0$ regimes. We plot $\mathsf{m}$ over the triangular parameter space in Fig.~\ref{figure:maplemass}. Details of the computation of $\nu$ and the derivation leading to~\eqref{maplemass} are presented in Sec.~\ref{section:band} of the Supplemental Material~\cite{supply}. 

The Kitaev trellis model, on the other hand, offers a simple yet illuminating demonstration of topological band theory. At the level of the Majorana fermion representation, our gauge choice in Fig.~\ref{figure:trellisgauge} restores the translational symmetry of the trellis lattice with two sites per unit cell, even though the original model~\eqref{kitaevmodel} defined via the $\Gamma$ matrices has a reduced translational symmetry with four sites per unit cell. We construct the Hamiltonian~\eqref{kitaevmodelfourier} accordingly and obtain the $2 \times 2$ matrix $i A (\mathbf{q}) = \sum_\lambda d_\lambda ( \mathbf{q} ) \sigma^\lambda$, with
\begin{subequations}
\begin{align}
d_x ( \mathbf{q} ) &= - 2 J_{zx} \sin q_2 + 2 J_{zz} \sin (q_1 - q_2) , \\
d_y ( \mathbf{q} ) &= - 2 J_{zx} ( 1 - \cos q_2 ) + 2 J_{zz} \cos (q_1 - q_2) , \\
d_z ( \mathbf{q} ) &= - 4 J_{x0} \sin q_1 
\, ,
\end{align}
\end{subequations}
where
\begin{equation*}
q_i \in [0 , 2 \pi) , \quad \begin{pmatrix} q_1 \\ q_2 \end{pmatrix} = \mathsf{B} \begin{pmatrix} q_x \\ q_y \end{pmatrix} , \quad \mathsf{B} \equiv \begin{pmatrix} 1 & 0 \\ 1/2 & 1 + \sqrt{3}/2 \end{pmatrix} 
\, .
\end{equation*}
The eigenvalues are given by $\pm d ( \mathbf{q} ) = \pm [\sum_\lambda d_\lambda^2 ( \mathbf{q} ) ]^{1/2}$. We plot the fermion gap $\Delta_\psi = \min_\mathbf{q} d ( \mathbf{q} )$ over the triangular parameter space in Fig.~\ref{figure:trellisgap}. The set of critical parameters, at which the system becomes gapless, form a line separating the parameter space into an upper and lower parts~\footnote{Fig.~\ref{figure:trellisgap} suggests that the system is also gapless on segments of the boundary of the triangular parameter space where $J_\lambda = 0$ for some $\lambda$, which are excluded from our considerations as the lattice geometry would be altered by missing bonds.}. The gap-closing condition is $- 4 J_{zx} + 2 J_{zz} = 0$, where $d ( \mathbf{q} ) = 0$ at the $\mathrm{M}$ point $( q_1 , q_2 ) = ( \pi , \pi )$~\cite{supply}. We perform a small $\mathbf{q}$ expansion around the $\mathrm{M}$ point up to linear order and we find
\begin{subequations}
\begin{align}
d_x (\mathrm{M} + \mathbf{q}) & \approx 2 J_{zx} q_2 + 2 J_{zz} ( q_1 - q_2 ) \, , \\
d_y (\mathrm{M} + \mathbf{q}) & \approx - 4 J_{zx} + 2 J_{zz} \equiv \mathsf{m} \, , \label{trellismass} \\
d_z (\mathrm{M} + \mathbf{q}) & \approx 4 J_{x0} q_1 \, ,
\end{align}
\end{subequations}
which leads to an effective model of the form~\eqref{effectmodel} with $(\lambda , \mu , \nu) = (y,z,x)$ and
\begin{equation*}
\mathsf{A} = 2 \mathsf{B}^\mathrm{T} \begin{pmatrix} 2 J_{x0} & J_{zz} \\ 0 & J_{zx} - J_{zz} \end{pmatrix} \, .
\end{equation*}
Note that $J_{zx} - J_{zz} < 0$, which implies $\det \mathsf{A} < 0$, in the vicinity of $\mathsf{m}=0$. We plot the mass term $\mathsf{m}$ defined in~\eqref{trellismass} over the triangular parameter space in Fig.~\ref{figure:trellismass}. With this information at hand, we are now ready to determine the Chern number $\nu$ of each gapped regime. First, notice that the upper part is continuously connected to the $J_{zz} \longrightarrow 1$ limit without closing the gap. There the Majorana fermions are strongly localized on the $zz$ bonds, akin to electrons bound to individual atoms in an atomic insulator. Therefore, the upper part should have a trivial Chern number $\nu = 0$. Crossing the gap-closing transition, $\mathsf{m}$ changes from positive to negative, which, by virtue of~\eqref{chernchange}, decreases $\nu$ by 1. We thus deduce that the lower part has $\nu = -1$. We further verified this analytical argument by numerically calculating $\nu$~\cite{supply}. 

\section{\label{section:anyon}Vortices as anyons}

According to Kitaev's sixteenfold way~\cite{KITAEV20062}, the $\nu = 0$ phase is equivalent to a toric code phase, which has four superselection sectors $1$, $e$, $m$, and $\psi$, where $1$ is the vacuum and the others are anyons. $e$ and $m$ are known as the electric and magnetic particles~\footnote{They are originally known as the electric charge and the magnetic vortex. To avoid confusion, we reserve the terminology ``vortex'' for a generic flux excitation, which can be either $e$ or $m$, rather than specifically referring to $m$ with it.}, respectively, while $\psi$ is a composite of $e$ and $m$. Although $e$ ($m$) is bosonic with respect to its own species, it has nontrivial statistics with $m$ ($e$). As a result, $\psi$ is fermionic with respect to its own species, and it has nontrivial statistics with $e$ and $m$. In the context of generalized Kitaev models, a complex fermion made up of two Majorana fermions is equivalent to $\psi$, while a flux excitation, or \textit{vortex}, on an elementary plaquette corresponds to either $e$ or $m$~\cite{KITAEV20062,PhysRevB.90.134404}. Note that a vortex is not necessarily a $\pi$-flux, and the ground-state flux sector is always the vortex-free sector.

In this section, we study the mapping of vortices to $e$ and $m$ anyons in the anisotropic limit $J_\lambda \gg J_{\lambda' \neq \lambda}$, where the strong bonds indexed by $\lambda$ form a dimer covering of the lattice in question. More specifically, we consider the strong $J_{x0}$ and $J_{y0}$ limits of the Kitaev maple-leaf model, and the strong $J_{zz}$ limit of the Kitaev trellis model. Note that the dimers are disconnected from each other, so the Majorana fermions are strongly localized. The topological phase diagrams Figs.~\ref{figure:maplemass} and~\ref{figure:trellismass} reveal that these dimer limits indeed lie in the $\nu=0$ regime, as expected. Before proceeding, we remark that the toric code phase possesses an anyon permutation symmetry, such that a global exchange of the $e$ and $m$ labels leaves the physics invariant~\cite{simontextbook}. However, once the anyon species of a reference plaquette is fixed, the anyon species of all other plaquettes are no longer arbitrary. 

\begin{figure}
\subfloat[]{\label{figure:mapleanyonx0}
\includegraphics[scale=0.2]{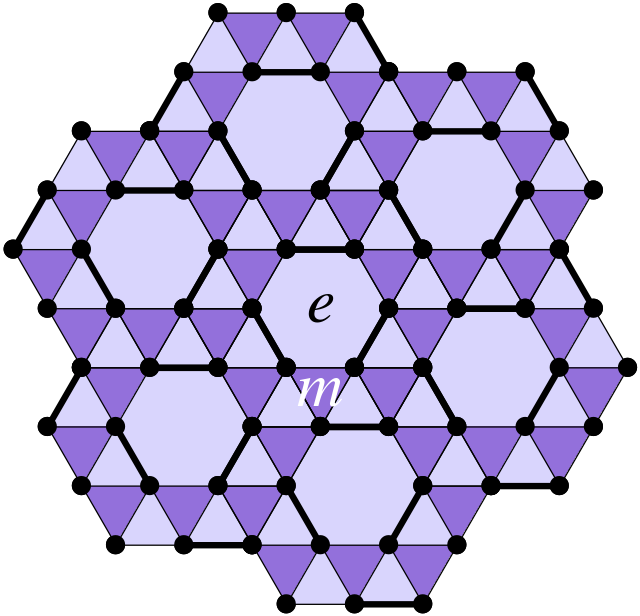}} \quad
\subfloat[]{\label{figure:mapleanyony0}
\includegraphics[scale=0.2]{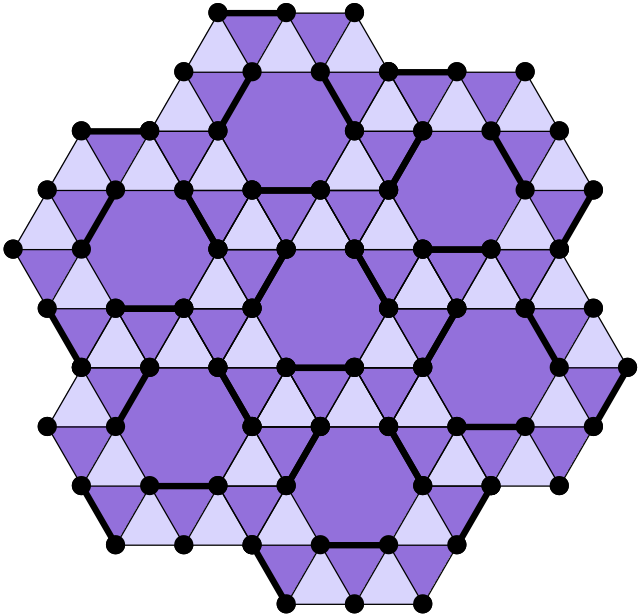}} \\
\subfloat[]{\label{figure:maplegapht}
\includegraphics[scale=0.23]{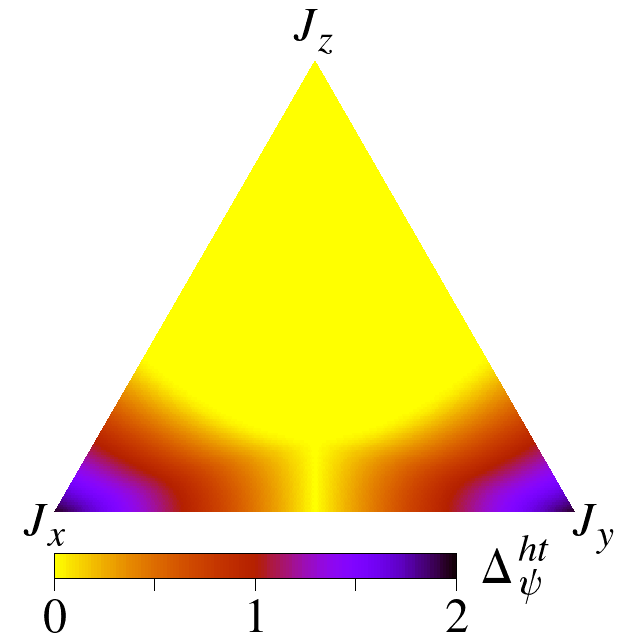}}
\subfloat[]{\label{figure:maplegaptt}
\includegraphics[scale=0.23]{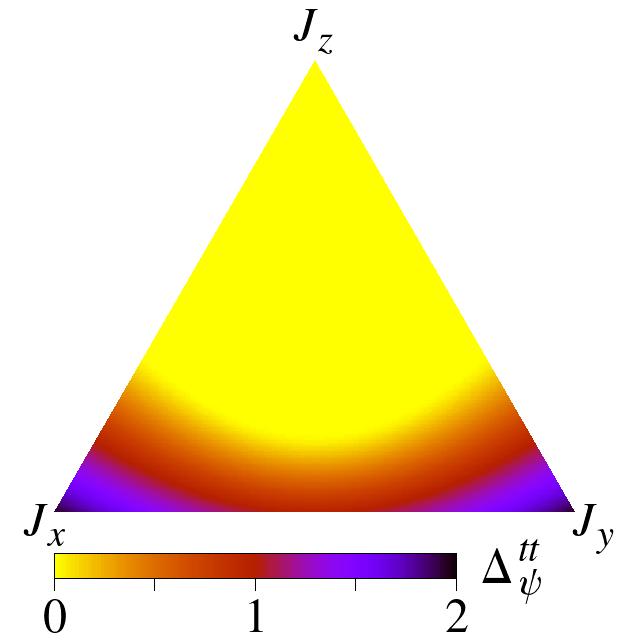}}
\caption{The assignments of anyon species in the strong (a) $J_{x0}$ and (b) $J_{y0}$ limits of the Kitaev maple-leaf model. Thick lines indicate the $\lambda$ bonds such that $J_\lambda \gg J_{\lambda'}$ for every $\lambda' \neq \lambda$. Elementary plaquettes filled with light (dark) purple colors indicate that the respective vortices are $e$ ($m$) particles. The fermion gaps $\Delta_\psi^{ht}$ and $\Delta_\psi^{tt}$ of two-vortex sectors (c) with one vortex at a unit hexagon and the other at a unit triangle and (d) with both vortices at unit triangles, respectively, for the Kitaev maple-leaf model.}
\end{figure}

The assignment of the two anyon species to the elementary plaquettes is conventionally obtained with degenerate perturbation theory, by which one explicitly derives an effective Hamiltonian at low energies and demonstrates its equivalence to a toric code model~\cite{KITAEV20062}. However, a limitation of degenerate perturbation theory is that, in the dimer limit, it can never yield an operator that acts nontrivially only on an odd-length elementary plaquette~\cite{PhysRevB.90.134404,2607.12027}. This leads to difficulties in interpreting the effective Hamiltonian as a toric code model and determining the anyon species of some vortices when the lattice contains odd-length elementary plaquettes, such as the unit triangles in the maple-leaf and trellis lattices. The theory of Ref.~\onlinecite{2607.12027} (see also Ref.~\onlinecite{Wootton_2015}), which utilizes the fusion rules of Abelian anyons~\cite{PhysRevResearch.2.023334} and the fermion parity of physical states~\cite{PhysRevB.84.165414,PhysRevB.92.014403}, enables one to circumvent degenerate perturbation theory together with its limitations. We will use the two following corollaries of Ref.~\onlinecite{2607.12027}, which are applicable to an arbitrary planar graph with a constant coordination number $z=2n-1$. Corollary 1 applies to the dimer limit and states that if two neighboring elementary plaquettes are separated by a strong (weak) bond, then they belong to the same (different) anyon species. Corollary 2 states that the assignment of anyon species to the elementary plaquettes is invariant under a continuous change of model parameters that does not close the fermion gaps of both the vortex-free and two-vortex sectors.

\begin{figure}
\includegraphics[scale=0.2]{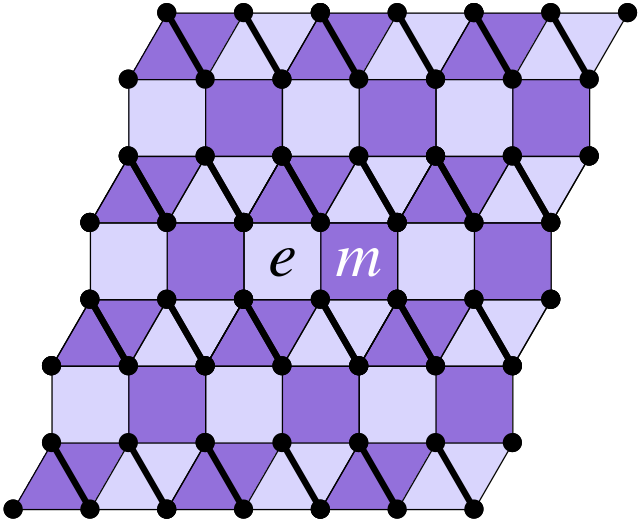}
\caption{\label{figure:trellisanyonzz}The assignment of anyon species in the strong $J_{zz}$ limit, where $J_{zz} \gg J_{\lambda}$ for every $\lambda \neq zz$, of the Kitaev trellis model. Thick lines indicate the $zz$ bonds. Elementary plaquettes filled with light (dark) purple colors indicate that the respective vortices are $e$ ($m$) particles.}
\end{figure}

Using Corollary 1, we can easily map out the anyon species of each elementary plaquette in the aforementioned dimer limits. The results are shown in Figs.~\ref{figure:mapleanyonx0}, \ref{figure:mapleanyony0}, and \ref{figure:trellisanyonzz}. For the Kitaev maple-leaf model, the strong $J_{x0}$ and $J_{y0}$ limits have different assignments of anyon species, even though they can be smoothly connected to each other in the vortex-free sector, see Fig.~\ref{figure:maplegap}. By Corollary 2, these two limits must be separated by fermion-gap-closing transitions in some two-vortex sectors. We verify this numerically by computing the fermion gaps of distinct two-vortex sectors on a torus with $16 \times 12$ unit cells. Starting from the vortex-free sector, we create two well-separated vortices \footnote{Ideally, the distance between the vortices should be much greater than the correlation length, such that their braiding statistics is well-defined.} by flipping a string of bonds that connects two elementary plaquettes of types $\alpha$ and $\beta$. We denote the fermion gap of such a two-vortex sector as $\Delta_\psi^{\alpha \beta}$. Since each unit cell contains $9$ elementary plaquettes, namely $1$ unit hexagon and $8$ unit triangles, we have $9 + {_9 C_2} = 45$ distinct combinations of $\alpha$ and $\beta$. We find that $\Delta_\psi^{\alpha \beta}$ vanishes in the $\nu=-1$ regime as well as on the topological phase boundary $\mathsf{m}=0$ for all $\alpha \beta$. In addition, when at least one of the vortices is hosted by a unit hexagon, we observe $\Delta_\psi^{\alpha \beta}=0$ at parameters satisfying $J_{x0} = J_{y0} \geq (1 + \sqrt{3}) J_{zx}$, which is a line joining the lowest point of the topological phase boundary and the midpoint of the bottom edge in the triangular parameter space, see for example Fig.~\ref{figure:maplegapht}. The two-vortex sectors can thus be organized into two groups according to whether both vortices are hosted by unit triangles or at least one vortex is hosted by a unit hexagon. The fermion gaps of the two-vortex sectors within each group are found to be nearly (if not exactly) equal throughout the triangular parameter space.

We show the fermion gap $\Delta_\psi^{ht}$ ($\Delta_\psi^{tt}$) of a two-vortex sector with one vortex at a unit hexagon and the other at a unit triangle (with both vortices at unit triangles) in Fig.~\ref{figure:maplegapht} (Fig.~\ref{figure:maplegaptt}). The set of critical parameters where $\Delta_\psi^{\alpha \beta}=0$ for some $\alpha \beta$ thus splits the $\nu = 0$ regime (cf.~Fig.~\ref{figure:maplegap}) into two parts, which contain the strong $J_{x0}$ and $J_{y0}$ limits, respectively. The Kitaev maple-leaf model thus provides an interesting example where a change of anyon species can indeed occur without a fermion-gap-closing transition in the vortex-free sector, beside the Kitaev square-octagon and Kekul\'{e} models analyzed in Ref.~\onlinecite{2607.12027}. Using Proposition 2 of Ref.~\onlinecite{2607.12027}, one can further deduce that the change of anyon species only takes place on all unit hexagons but not on any unit triangle, cf.~Figs.~\ref{figure:mapleanyonx0} and \ref{figure:mapleanyony0} \footnote{We refer interested readers to Ref.~\onlinecite{2607.12027} for a detailed discussion of Proposition 2.}. In Sec.~\ref{section:toriccode} of the Supplemental Material~\cite{supply}, we perform a convergence analysis of the fermion gaps of two-vortex sectors, each of which involves at least one unit hexagon, with respect to the system size along a path in the triangular parameter space.

On the other hand, the $\nu = - 1$ phase is identified with the Ising topological order, which has three superselection sectors $1$, $\psi$, and $\sigma$, where $1$ is the vacuum and the rest are anyons. In the context of generalized Kitaev models, a complex fermion is equivalent to $\psi$, while each vortex carries an unpaired Majorana mode and corresponds to $\sigma$~\cite{KITAEV20062}. In other words, the elementary plaquettes all map to a single anyon species~\cite{PhysRevResearch.2.023334}. Two vortices contribute to a zero-energy complex fermion mode (see Figs.~\ref{figure:maplegapht} and \ref{figure:maplegaptt}, cf.~Fig.~\ref{figure:maplemass}) that can be either occupied or unoccupied, which corresponds to the outcome $\psi$ or $1$ of fusing two $\sigma$. Since the outcome is not unique, $\sigma$ is known as a non-Abelian anyon~\cite{simontextbook}. The $\nu = \pm 1$ phase is thus often known as a non-Abelian Kitaev spin liquid, which has been the subject of intense investigations in recent years~\cite{s41586-018-0274-0,science.aay5551,s41567-021-01501-y,s41567-021-01488-6,s41563-022-01397-w} due to its possible realization in the honeycomb magnet $\alpha$-RuCl$_3$~\cite{JPSJ.89.012002,3m4m-3v59}.

\section{\label{section:discuss}Discussion}

In summary, we have studied generalized Kitaev models on the penta-coordinated maple-leaf and trellis lattices, which are characterized by bond-dependent interactions of $\Gamma$ matrices~\cite{PhysRevB.79.134427}. A representation in terms of Majorana fermions leads to exact quantum spin liquids with fractionalized excitations coupled to $\mathbb{Z}_2$ gauge fields. For each model, we exploit spatial symmetries to reduce the number of independent couplings and obtain a two-dimensional parameter space. Due to the inapplicability of Lieb's theorem~\cite{PhysRevLett.73.2158,Macris1996}, we perform replica exchange quantum Monte Carlo simulations~\cite{PhysRevLett.113.197205,PhysRevLett.115.087203,PhysRevB.92.115122,nphys3809,PhysRevB.96.125124,PhysRevB.98.054432,PhysRevResearch.1.032011,PhysRevB.101.045118,eschmannthesis,PhysRevB.102.075125,PhysRevResearch.2.043159} to solve for the ground-state flux sector. We then map out a topological phase diagram, which reveals parameter regimes with Chern numbers $0$ and $- 1$. We also assign the vortices in the dimer limit, which has trivial Chern number, to the two species of anyons in the toric code model. In the following, we elaborate on the significance of our results and suggest several potentially interesting directions for future investigations.

From our simulations of the Kitaev maple-leaf (trellis) model at a discrete set of parameters, we find that the $\mathbb{Z}_2$ gauge fluxes are ordered at low temperatures according to the flux phase conjecture $W_p = - (\pm i)^{\lvert \partial p \rvert}$~\cite{s41467-023-42105-9}, i.e., $W_p = + 1$ ($-1$) for all unit hexagons (squares) and $W_p= \pm i$ for all unit triangles. This provides further support to the conjecture, following several previous studies~\cite{PhysRevB.79.024426,PhysRevB.89.235102,PhysRevLett.115.087203,PhysRevB.98.054432,PhysRevB.102.075125,PhysRevB.108.104208} of generalized Kitaev models defined on lattices where Lieb's theorem does not apply. Despite strong numerical evidence, the flux phase conjecture remains unproven for generic graphs. A thorough understanding of why and when it applies to graphs that are non-bipartite and/or non-reflection-symmetric would greatly simplify future studies of the ground-state properties of generalized Kitaev models.

We discuss the relation between time-reversal symmetry breaking and Chern number, as well as a potential route to obtain topological orders beyond those found in this work. Let $\nu$ be the (total) Chern number of the negative energy band(s) when the fermion spectrum is gapped. While a state with $\nu \neq 0$ necessarily breaks time-reversal symmetry, a state that breaks time-reversal symmetry does not necessarily have $\nu \neq 0$. Indeed, our topological phase diagrams contain parameter regimes with $\nu = 0$ and $-1$, both of which break time-reversal symmetry due to the presence of unit triangles in the maple-leaf and trellis lattices. Similar to the generalized Kitaev models on the star \cite{PhysRevLett.99.247203} and Shastry-Sutherland \cite{PhysRevB.79.134427} lattices, the $\nu = -1$ phase can be obtained by merely tuning the nearest-neighbor coupling strengths, without introducing additional terms in the Hamiltonian~\eqref{kitaevmodel}. This is in contrast to the Kitaev honeycomb model, where one has to apply a magnetic field to explicitly break time-reversal symmetry~\cite{KITAEV20062}. It would be interesting to investigate whether the inclusion of farther-neighbor hoppings generated by three- or four-body interactions~\cite{PhysRevResearch.2.023334} can stabilize different flux sectors and realize higher Chern numbers, and therefore different topological orders, on these penta-coordinated geometries. 

According to Kitaev's sixteenfold way~\cite{KITAEV20062}, the $\nu = 0$ phase of a generalized Kitaev model is equivalent to a toric code model with the Abelian anyons $e$, $m$, and $\psi$. A vortex is either $e$ or $m$ but not both. Due to the limitation of degenerate perturbation theory on the maple-leaf and trellis lattices, we use the theory of Ref.~\onlinecite{2607.12027}, which is derived from the fusion rules of anyons and the fermion parity of physical states, to map vortices to anyons in the dimer limits of the respective Kitaev models. Although the strong $J_{x0}$ and $J_{y0}$ limits of the Kitaev maple-leaf model have different mappings, they can be connected to each other without a fermion-gap-closing transition in the ground-state vortex-free sector. Instead, we demonstrate that it is the vanishing of fermion gaps in some two-vortex sectors that allows for a change in the assignment of anyon species. The Kitaev maple-leaf model thus offers an excellent illustration of the theory in Ref.~\onlinecite{2607.12027}. 

Finally, we have constructed Kitaev-like models on the maple-leaf and trellis lattices that yield exact quantum spin liquids. The existing literature mostly focuses on isotropic spin models on these frustrated geometries, especially the maple-leaf lattice~\cite{APhysPolA.97.971,PhysRevB.65.224405,PhysRevB.104.224415,PhysRevB.105.L180412,PhysRevB.108.L241116,PhysRevB.109.184422,Ghosh_2024,PhysRevB.110.014414,PhysRevB.110.085151,3zlw-hrbf,s43246-025-00904-1,jmgz-dsk9,zna-2025-0382,zna-2025-0376,zna-2025-0409,zna-2025-0389,2401.09422,2407.07145,2601.05308,PhysRevB.84.104406,PhysRevB.89.184407,PhysRevB.98.224402,2605.12592,2605.21587}, which are motivated by a number of minerals and synthetic compounds~\cite{anie.200502847,Fennell_2011,anie.201203775,PhysRevB.98.064412,PhysRevB.101.184429,PhysRevB.104.174439,PhysRevB.107.064419,PhysRevB.111.094439}. These models consist of antiferromagnetic or mixed Heisenberg exchanges, and they are not exactly solvable in general. Various numerical studies using exact diagonalization, tensor network, pseudofermion functional renormalization group etc.~have been conducted to identify parameter ranges that may stabilize quantum spin liquids, but the outcomes of different methods do not always agree. Investigations of anisotropic spin models on these lattices~\cite{1yv1-wtx8,PhysRevResearch.6.033080,7psd-1zvr}, on the other hand, are rare, very likely because of the lack of candidate materials. Although the materialization of even a proximate Kitaev honeycomb model remains a challenging issue, we hope that our work will inspire future searches for mechanisms of realizing bond-dependent $\Gamma$-matrix interactions on penta-coordinated geometries such as the maple-leaf and trellis lattices, which might involve spin-orbital or quadrupolar interactions~\cite{PU1982v025n04ABEH004537,s41535-025-00744-9,PhysRevB.79.134427,PhysRevLett.102.217202}.

\begin{acknowledgements}
This work was supported in part by the Deutsche Forschungsgemeinschaft via the cluster of excellence ctd.qmat (EXC 2147, project-id 390858490) and SFB 1143 (Project-ID No.~247310070), and by the Engineering and Physical Sciences Research Council (EPSRC) grant No.~EP/V062654/1.
\end{acknowledgements}

\bibliography{reference260825}

\begin{thebibliography}{135}%
\makeatletter
\providecommand \@ifxundefined [1]{%
 \@ifx{#1\undefined}
}%
\providecommand \@ifnum [1]{%
 \ifnum #1\expandafter \@firstoftwo
 \else \expandafter \@secondoftwo
 \fi
}%
\providecommand \@ifx [1]{%
 \ifx #1\expandafter \@firstoftwo
 \else \expandafter \@secondoftwo
 \fi
}%
\providecommand \natexlab [1]{#1}%
\providecommand \enquote  [1]{``#1''}%
\providecommand \bibnamefont  [1]{#1}%
\providecommand \bibfnamefont [1]{#1}%
\providecommand \citenamefont [1]{#1}%
\providecommand \href@noop [0]{\@secondoftwo}%
\providecommand \href [0]{\begingroup \@sanitize@url \@href}%
\providecommand \@href[1]{\@@startlink{#1}\@@href}%
\providecommand \@@href[1]{\endgroup#1\@@endlink}%
\providecommand \@sanitize@url [0]{\catcode `\\12\catcode `\$12\catcode
  `\&12\catcode `\#12\catcode `\^12\catcode `\_12\catcode `\%12\relax}%
\providecommand \@@startlink[1]{}%
\providecommand \@@endlink[0]{}%
\providecommand \url  [0]{\begingroup\@sanitize@url \@url }%
\providecommand \@url [1]{\endgroup\@href {#1}{\urlprefix }}%
\providecommand \urlprefix  [0]{URL }%
\providecommand \Eprint [0]{\href }%
\providecommand \doibase [0]{https://doi.org/}%
\providecommand \selectlanguage [0]{\@gobble}%
\providecommand \bibinfo  [0]{\@secondoftwo}%
\providecommand \bibfield  [0]{\@secondoftwo}%
\providecommand \translation [1]{[#1]}%
\providecommand \BibitemOpen [0]{}%
\providecommand \bibitemStop [0]{}%
\providecommand \bibitemNoStop [0]{.\EOS\space}%
\providecommand \EOS [0]{\spacefactor3000\relax}%
\providecommand \BibitemShut  [1]{\csname bibitem#1\endcsname}%
\let\auto@bib@innerbib\@empty
%</preamble>
\bibitem [{\citenamefont {Lacroix}\ \emph {et~al.}(2011)\citenamefont
  {Lacroix}, \citenamefont {Mendels},\ and\ \citenamefont
  {Mila}}]{lacroixtextbook}%
  \BibitemOpen
  \bibinfo {editor} {\bibfnamefont {C.}~\bibnamefont {Lacroix}}, \bibinfo
  {editor} {\bibfnamefont {P.}~\bibnamefont {Mendels}},\ and\ \bibinfo {editor}
  {\bibfnamefont {F.}~\bibnamefont {Mila}},\ eds.,\ \href@noop {} {\emph
  {\bibinfo {title} {Introduction to Frustrated Magnetism}}}\ (\bibinfo
  {publisher} {Springer},\ \bibinfo {year} {2011})\BibitemShut {NoStop}%
\bibitem [{\citenamefont {Ramirez}(1994)}]{annurev.ms.24.080194.002321}%
  \BibitemOpen
  \bibfield  {author} {\bibinfo {author} {\bibfnamefont {A.~P.}\ \bibnamefont
  {Ramirez}},\ }\bibfield  {title} {\bibinfo {title} {Strongly geometrically
  frustrated magnets},\ }\href
  {https://doi.org/https://doi.org/10.1146/annurev.ms.24.080194.002321}
  {\bibfield  {journal} {\bibinfo  {journal} {Annual Review of Materials
  Research}\ }\textbf {\bibinfo {volume} {24}},\ \bibinfo {pages} {453}
  (\bibinfo {year} {1994})}\BibitemShut {NoStop}%
\bibitem [{\citenamefont {Makuta}\ and\ \citenamefont
  {Hotta}(2021)}]{PhysRevB.104.224415}%
  \BibitemOpen
  \bibfield  {author} {\bibinfo {author} {\bibfnamefont {R.}~\bibnamefont
  {Makuta}}\ and\ \bibinfo {author} {\bibfnamefont {C.}~\bibnamefont {Hotta}},\
  }\bibfield  {title} {\bibinfo {title} {Dimensional reduction in quantum
  spin-$\frac{1}{2}$ system on a $\frac{1}{7}$-depleted triangular lattice},\
  }\href {https://doi.org/10.1103/PhysRevB.104.224415} {\bibfield  {journal}
  {\bibinfo  {journal} {Phys. Rev. B}\ }\textbf {\bibinfo {volume} {104}},\
  \bibinfo {pages} {224415} (\bibinfo {year} {2021})}\BibitemShut {NoStop}%
\bibitem [{\citenamefont {Ghosh}\ \emph {et~al.}(2022)\citenamefont {Ghosh},
  \citenamefont {M\"uller},\ and\ \citenamefont
  {Thomale}}]{PhysRevB.105.L180412}%
  \BibitemOpen
  \bibfield  {author} {\bibinfo {author} {\bibfnamefont {P.}~\bibnamefont
  {Ghosh}}, \bibinfo {author} {\bibfnamefont {T.}~\bibnamefont {M\"uller}},\
  and\ \bibinfo {author} {\bibfnamefont {R.}~\bibnamefont {Thomale}},\
  }\bibfield  {title} {\bibinfo {title} {Another exact ground state of a
  two-dimensional quantum antiferromagnet},\ }\href
  {https://doi.org/10.1103/PhysRevB.105.L180412} {\bibfield  {journal}
  {\bibinfo  {journal} {Phys. Rev. B}\ }\textbf {\bibinfo {volume} {105}},\
  \bibinfo {pages} {L180412} (\bibinfo {year} {2022})}\BibitemShut {NoStop}%
\bibitem [{\citenamefont {Gresista}\ \emph {et~al.}(2023)\citenamefont
  {Gresista}, \citenamefont {Hickey}, \citenamefont {Trebst},\ and\
  \citenamefont {Iqbal}}]{PhysRevB.108.L241116}%
  \BibitemOpen
  \bibfield  {author} {\bibinfo {author} {\bibfnamefont {L.}~\bibnamefont
  {Gresista}}, \bibinfo {author} {\bibfnamefont {C.}~\bibnamefont {Hickey}},
  \bibinfo {author} {\bibfnamefont {S.}~\bibnamefont {Trebst}},\ and\ \bibinfo
  {author} {\bibfnamefont {Y.}~\bibnamefont {Iqbal}},\ }\bibfield  {title}
  {\bibinfo {title} {Candidate quantum disordered intermediate phase in the
  {H}eisenberg antiferromagnet on the maple-leaf lattice},\ }\href
  {https://doi.org/10.1103/PhysRevB.108.L241116} {\bibfield  {journal}
  {\bibinfo  {journal} {Phys. Rev. B}\ }\textbf {\bibinfo {volume} {108}},\
  \bibinfo {pages} {L241116} (\bibinfo {year} {2023})}\BibitemShut {NoStop}%
\bibitem [{\citenamefont {Beck}\ \emph {et~al.}(2024)\citenamefont {Beck},
  \citenamefont {Bodky}, \citenamefont {Motruk}, \citenamefont {M\"uller},
  \citenamefont {Thomale},\ and\ \citenamefont {Ghosh}}]{PhysRevB.109.184422}%
  \BibitemOpen
  \bibfield  {author} {\bibinfo {author} {\bibfnamefont {J.}~\bibnamefont
  {Beck}}, \bibinfo {author} {\bibfnamefont {J.}~\bibnamefont {Bodky}},
  \bibinfo {author} {\bibfnamefont {J.}~\bibnamefont {Motruk}}, \bibinfo
  {author} {\bibfnamefont {T.}~\bibnamefont {M\"uller}}, \bibinfo {author}
  {\bibfnamefont {R.}~\bibnamefont {Thomale}},\ and\ \bibinfo {author}
  {\bibfnamefont {P.}~\bibnamefont {Ghosh}},\ }\bibfield  {title} {\bibinfo
  {title} {Phase diagram of the {$J\text{\ensuremath{-}}{J}_{d}$} {H}eisenberg
  model on the maple leaf lattice: Neural networks and density matrix
  renormalization group},\ }\href {https://doi.org/10.1103/PhysRevB.109.184422}
  {\bibfield  {journal} {\bibinfo  {journal} {Phys. Rev. B}\ }\textbf {\bibinfo
  {volume} {109}},\ \bibinfo {pages} {184422} (\bibinfo {year}
  {2024})}\BibitemShut {NoStop}%
\bibitem [{\citenamefont {Ghosh}(2024)}]{Ghosh_2024}%
  \BibitemOpen
  \bibfield  {author} {\bibinfo {author} {\bibfnamefont {P.}~\bibnamefont
  {Ghosh}},\ }\bibfield  {title} {\bibinfo {title} {Triplon analysis of
  magnetic disorder and order in maple-leaf {H}eisenberg magnet},\ }\href
  {https://doi.org/10.1088/1361-648X/ad69f4} {\bibfield  {journal} {\bibinfo
  {journal} {Journal of Physics: Condensed Matter}\ }\textbf {\bibinfo {volume}
  {36}},\ \bibinfo {pages} {455803} (\bibinfo {year} {2024})}\BibitemShut
  {NoStop}%
\bibitem [{\citenamefont {Sonnenschein}\ \emph {et~al.}(2024)\citenamefont
  {Sonnenschein}, \citenamefont {Maity}, \citenamefont {Liu}, \citenamefont
  {Thomale}, \citenamefont {Ferrari},\ and\ \citenamefont
  {Iqbal}}]{PhysRevB.110.014414}%
  \BibitemOpen
  \bibfield  {author} {\bibinfo {author} {\bibfnamefont {J.}~\bibnamefont
  {Sonnenschein}}, \bibinfo {author} {\bibfnamefont {A.}~\bibnamefont {Maity}},
  \bibinfo {author} {\bibfnamefont {C.}~\bibnamefont {Liu}}, \bibinfo {author}
  {\bibfnamefont {R.}~\bibnamefont {Thomale}}, \bibinfo {author} {\bibfnamefont
  {F.}~\bibnamefont {Ferrari}},\ and\ \bibinfo {author} {\bibfnamefont
  {Y.}~\bibnamefont {Iqbal}},\ }\bibfield  {title} {\bibinfo {title} {Candidate
  quantum spin liquids on the maple-leaf lattice},\ }\href
  {https://doi.org/10.1103/PhysRevB.110.014414} {\bibfield  {journal} {\bibinfo
   {journal} {Phys. Rev. B}\ }\textbf {\bibinfo {volume} {110}},\ \bibinfo
  {pages} {014414} (\bibinfo {year} {2024})}\BibitemShut {NoStop}%
\bibitem [{\citenamefont {Gemb\'e}\ \emph {et~al.}(2024)\citenamefont
  {Gemb\'e}, \citenamefont {Gresista}, \citenamefont {Schmidt}, \citenamefont
  {Hickey}, \citenamefont {Iqbal},\ and\ \citenamefont
  {Trebst}}]{PhysRevB.110.085151}%
  \BibitemOpen
  \bibfield  {author} {\bibinfo {author} {\bibfnamefont {M.}~\bibnamefont
  {Gemb\'e}}, \bibinfo {author} {\bibfnamefont {L.}~\bibnamefont {Gresista}},
  \bibinfo {author} {\bibfnamefont {H.-J.}\ \bibnamefont {Schmidt}}, \bibinfo
  {author} {\bibfnamefont {C.}~\bibnamefont {Hickey}}, \bibinfo {author}
  {\bibfnamefont {Y.}~\bibnamefont {Iqbal}},\ and\ \bibinfo {author}
  {\bibfnamefont {S.}~\bibnamefont {Trebst}},\ }\bibfield  {title} {\bibinfo
  {title} {Noncoplanar orders and quantum disordered states in maple-leaf
  antiferromagnets},\ }\href {https://doi.org/10.1103/PhysRevB.110.085151}
  {\bibfield  {journal} {\bibinfo  {journal} {Phys. Rev. B}\ }\textbf {\bibinfo
  {volume} {110}},\ \bibinfo {pages} {085151} (\bibinfo {year}
  {2024})}\BibitemShut {NoStop}%
\bibitem [{\citenamefont {Ghosh}(2025)}]{3zlw-hrbf}%
  \BibitemOpen
  \bibfield  {author} {\bibinfo {author} {\bibfnamefont {P.}~\bibnamefont
  {Ghosh}},\ }\bibfield  {title} {\bibinfo {title} {Chiral crossroads in
  {${\mathrm{Ho}}_{3}{\mathrm{ScO}}_{6}$}: Competing interactions on the
  maple-leaf lattice},\ }\href {https://doi.org/10.1103/3zlw-hrbf} {\bibfield
  {journal} {\bibinfo  {journal} {Phys. Rev. B}\ }\textbf {\bibinfo {volume}
  {111}},\ \bibinfo {pages} {224431} (\bibinfo {year} {2025})}\BibitemShut
  {NoStop}%
\bibitem [{\citenamefont {Schmoll}\ \emph {et~al.}(2025)\citenamefont
  {Schmoll}, \citenamefont {Jeschke},\ and\ \citenamefont
  {Iqbal}}]{s43246-025-00904-1}%
  \BibitemOpen
  \bibfield  {author} {\bibinfo {author} {\bibfnamefont {P.}~\bibnamefont
  {Schmoll}}, \bibinfo {author} {\bibfnamefont {H.~O.}\ \bibnamefont
  {Jeschke}},\ and\ \bibinfo {author} {\bibfnamefont {Y.}~\bibnamefont
  {Iqbal}},\ }\bibfield  {title} {\bibinfo {title} {Tensor network analysis of
  the maple-leaf antiferromagnet spangolite},\ }\href
  {https://doi.org/10.1038/s43246-025-00904-1} {\bibfield  {journal} {\bibinfo
  {journal} {Communications Materials}\ }\textbf {\bibinfo {volume} {6}},\
  \bibinfo {pages} {178} (\bibinfo {year} {2025})}\BibitemShut {NoStop}%
\bibitem [{\citenamefont {Hutak}(2025)}]{jmgz-dsk9}%
  \BibitemOpen
  \bibfield  {author} {\bibinfo {author} {\bibfnamefont {T.}~\bibnamefont
  {Hutak}},\ }\bibfield  {title} {\bibinfo {title} {Thermodynamics of the
  {$S=\frac{1}{2}$} maple-leaf {H}eisenberg antiferromagnet},\ }\href
  {https://doi.org/10.1103/jmgz-dsk9} {\bibfield  {journal} {\bibinfo
  {journal} {Phys. Rev. B}\ }\textbf {\bibinfo {volume} {112}},\ \bibinfo
  {pages} {104405} (\bibinfo {year} {2025})}\BibitemShut {NoStop}%
\bibitem [{\citenamefont {Esaki}\ \emph {et~al.}(2025)\citenamefont {Esaki},
  \citenamefont {Akagi}, \citenamefont {Penc},\ and\ \citenamefont
  {Katsura}}]{1yv1-wtx8}%
  \BibitemOpen
  \bibfield  {author} {\bibinfo {author} {\bibfnamefont {N.}~\bibnamefont
  {Esaki}}, \bibinfo {author} {\bibfnamefont {Y.}~\bibnamefont {Akagi}},
  \bibinfo {author} {\bibfnamefont {K.}~\bibnamefont {Penc}},\ and\ \bibinfo
  {author} {\bibfnamefont {H.}~\bibnamefont {Katsura}},\ }\bibfield  {title}
  {\bibinfo {title} {Spin {N}ernst and thermal {H}all effects of topological
  triplons in quantum dimer magnets on the maple-leaf and star lattices},\
  }\href {https://doi.org/10.1103/1yv1-wtx8} {\bibfield  {journal} {\bibinfo
  {journal} {Phys. Rev. B}\ }\textbf {\bibinfo {volume} {112}},\ \bibinfo
  {pages} {134435} (\bibinfo {year} {2025})}\BibitemShut {NoStop}%
\bibitem [{\citenamefont {Sch\"{a}fer}\ \emph {et~al.}(2026)\citenamefont
  {Sch\"{a}fer}, \citenamefont {Ebert}, \citenamefont {Hassan}, \citenamefont
  {Reuther}, \citenamefont {Luitz},\ and\ \citenamefont
  {Wietek}}]{zna-2025-0382}%
  \BibitemOpen
  \bibfield  {author} {\bibinfo {author} {\bibfnamefont {R.}~\bibnamefont
  {Sch\"{a}fer}}, \bibinfo {author} {\bibfnamefont {P.~L.}\ \bibnamefont
  {Ebert}}, \bibinfo {author} {\bibfnamefont {N.}~\bibnamefont {Hassan}},
  \bibinfo {author} {\bibfnamefont {J.}~\bibnamefont {Reuther}}, \bibinfo
  {author} {\bibfnamefont {D.~J.}\ \bibnamefont {Luitz}},\ and\ \bibinfo
  {author} {\bibfnamefont {A.}~\bibnamefont {Wietek}},\ }\bibfield  {title}
  {\bibinfo {title} {Thermodynamics of the {H}eisenberg antiferromagnet on the
  maple-leaf lattice},\ }\href {https://doi.org/doi:10.1515/zna-2025-0382}
  {\bibfield  {journal} {\bibinfo  {journal} {Zeitschrift f\"{u}r
  Naturforschung A}\ }\textbf {\bibinfo {volume} {81}},\ \bibinfo {pages} {471}
  (\bibinfo {year} {2026})}\BibitemShut {NoStop}%
\bibitem [{\citenamefont {Gresista}\ \emph {et~al.}(2026)\citenamefont
  {Gresista}, \citenamefont {Kiese}, \citenamefont {Trebst},\ and\
  \citenamefont {Iqbal}}]{zna-2025-0376}%
  \BibitemOpen
  \bibfield  {author} {\bibinfo {author} {\bibfnamefont {L.}~\bibnamefont
  {Gresista}}, \bibinfo {author} {\bibfnamefont {D.}~\bibnamefont {Kiese}},
  \bibinfo {author} {\bibfnamefont {S.}~\bibnamefont {Trebst}},\ and\ \bibinfo
  {author} {\bibfnamefont {Y.}~\bibnamefont {Iqbal}},\ }\bibfield  {title}
  {\bibinfo {title} {Unconventional orders in the maple-leaf
  ferro-antiferromagnetic {H}eisenberg model},\ }\href
  {https://doi.org/doi:10.1515/zna-2025-0376} {\bibfield  {journal} {\bibinfo
  {journal} {Zeitschrift f\"{u}r Naturforschung A}\ }\textbf {\bibinfo {volume}
  {81}},\ \bibinfo {pages} {433} (\bibinfo {year} {2026})}\BibitemShut
  {NoStop}%
\bibitem [{\citenamefont {Nyckees}\ \emph {et~al.}(2026)\citenamefont
  {Nyckees}, \citenamefont {Ghosh},\ and\ \citenamefont
  {Mila}}]{zna-2025-0409}%
  \BibitemOpen
  \bibfield  {author} {\bibinfo {author} {\bibfnamefont {S.}~\bibnamefont
  {Nyckees}}, \bibinfo {author} {\bibfnamefont {P.}~\bibnamefont {Ghosh}},\
  and\ \bibinfo {author} {\bibfnamefont {F.}~\bibnamefont {Mila}},\ }\bibfield
  {title} {\bibinfo {title} {Tensor-network study of the ground state of the
  maple-leaf {H}eisenberg antiferromagnet},\ }\href
  {https://doi.org/doi:10.1515/zna-2025-0409} {\bibfield  {journal} {\bibinfo
  {journal} {Zeitschrift f\"{u}r Naturforschung A}\ }\textbf {\bibinfo {volume}
  {81}},\ \bibinfo {pages} {453} (\bibinfo {year} {2026})}\BibitemShut
  {NoStop}%
\bibitem [{\citenamefont {Nakano}\ and\ \citenamefont
  {Sakai}(2026)}]{zna-2025-0389}%
  \BibitemOpen
  \bibfield  {author} {\bibinfo {author} {\bibfnamefont {H.}~\bibnamefont
  {Nakano}}\ and\ \bibinfo {author} {\bibfnamefont {T.}~\bibnamefont {Sakai}},\
  }\bibfield  {title} {\bibinfo {title} {Spin excitation of the {H}eisenberg
  antiferromagnet with frustration: from the bounce-lattice antiferromagnet
  through the maple-leaf-lattice antiferromagnet to the exact-dimer system},\
  }\href {https://doi.org/10.1515/zna-2025-0389} {\bibfield  {journal}
  {\bibinfo  {journal} {Zeitschrift f\"{u}r Naturforschung A}\ }\textbf
  {\bibinfo {volume} {81}},\ \bibinfo {pages} {461} (\bibinfo {year}
  {2026})}\BibitemShut {NoStop}%
\bibitem [{\citenamefont {Ghosh}()}]{2401.09422}%
  \BibitemOpen
  \bibfield  {author} {\bibinfo {author} {\bibfnamefont {P.}~\bibnamefont
  {Ghosh}},\ }\bibfield  {title} {\bibinfo {title} {Where is the spin liquid in
  maple-leaf quantum magnet?},\ }\href {https://arxiv.org/abs/2401.09422} {\
  }\Eprint {https://arxiv.org/abs/2401.09422} {arXiv:2401.09422} \BibitemShut
  {NoStop}%
\bibitem [{\citenamefont {Schmoll}\ \emph {et~al.}()\citenamefont {Schmoll},
  \citenamefont {Naumann}, \citenamefont {Weerda}, \citenamefont {Eisert},\
  and\ \citenamefont {Iqbal}}]{2407.07145}%
  \BibitemOpen
  \bibfield  {author} {\bibinfo {author} {\bibfnamefont {P.}~\bibnamefont
  {Schmoll}}, \bibinfo {author} {\bibfnamefont {J.}~\bibnamefont {Naumann}},
  \bibinfo {author} {\bibfnamefont {E.~L.}\ \bibnamefont {Weerda}}, \bibinfo
  {author} {\bibfnamefont {J.}~\bibnamefont {Eisert}},\ and\ \bibinfo {author}
  {\bibfnamefont {Y.}~\bibnamefont {Iqbal}},\ }\bibfield  {title} {\bibinfo
  {title} {An extended non-magnetic phase in the spin-$1/2$ {H}eisenberg
  antiferromagnet from the ruby to the maple-leaf lattice},\ }\href
  {https://arxiv.org/abs/2407.07145} {\ }\Eprint
  {https://arxiv.org/abs/2407.07145} {arXiv:2407.07145} \BibitemShut {NoStop}%
\bibitem [{\citenamefont {Ebert}\ \emph {et~al.}({\natexlab{a}})\citenamefont
  {Ebert}, \citenamefont {Iqbal},\ and\ \citenamefont {Wietek}}]{2601.05308}%
  \BibitemOpen
  \bibfield  {author} {\bibinfo {author} {\bibfnamefont {P.~L.}\ \bibnamefont
  {Ebert}}, \bibinfo {author} {\bibfnamefont {Y.}~\bibnamefont {Iqbal}},\ and\
  \bibinfo {author} {\bibfnamefont {A.}~\bibnamefont {Wietek}},\ }\bibfield
  {title} {\bibinfo {title} {Competing paramagnetic phases in the maple-leaf
  {H}eisenberg antiferromagnet},\ }\href {https://arxiv.org/abs/2601.05308} {\
  ({\natexlab{a}})},\ \Eprint {https://arxiv.org/abs/2601.05308}
  {arXiv:2601.05308} \BibitemShut {NoStop}%
\bibitem [{\citenamefont {Ebert}\ \emph {et~al.}({\natexlab{b}})\citenamefont
  {Ebert}, \citenamefont {Iqbal},\ and\ \citenamefont {Wietek}}]{2605.12592}%
  \BibitemOpen
  \bibfield  {author} {\bibinfo {author} {\bibfnamefont {P.~L.}\ \bibnamefont
  {Ebert}}, \bibinfo {author} {\bibfnamefont {Y.}~\bibnamefont {Iqbal}},\ and\
  \bibinfo {author} {\bibfnamefont {A.}~\bibnamefont {Wietek}},\ }\bibfield
  {title} {\bibinfo {title} {Incommensurate spin-density waves in a frustrated
  maple-leaf lattice ferromagnet},\ }\href {https://arxiv.org/abs/2605.12592}
  {\  ({\natexlab{b}})},\ \Eprint {https://arxiv.org/abs/2605.12592}
  {arXiv:2605.12592} \BibitemShut {NoStop}%
\bibitem [{\citenamefont {Zhang}\ \emph {et~al.}()\citenamefont {Zhang},
  \citenamefont {Feuerpfeil}, \citenamefont {Sachdev}, \citenamefont
  {Thomale},\ and\ \citenamefont {Iqbal}}]{2605.21587}%
  \BibitemOpen
  \bibfield  {author} {\bibinfo {author} {\bibfnamefont {Y.}~\bibnamefont
  {Zhang}}, \bibinfo {author} {\bibfnamefont {A.}~\bibnamefont {Feuerpfeil}},
  \bibinfo {author} {\bibfnamefont {S.}~\bibnamefont {Sachdev}}, \bibinfo
  {author} {\bibfnamefont {R.}~\bibnamefont {Thomale}},\ and\ \bibinfo {author}
  {\bibfnamefont {Y.}~\bibnamefont {Iqbal}},\ }\bibfield  {title} {\bibinfo
  {title} {Large-flavor route to a stable {$U(1)$} {D}irac spin liquid on the
  maple-leaf lattice},\ }\href {https://arxiv.org/abs/2605.21587} {\ }\Eprint
  {https://arxiv.org/abs/2605.21587} {arXiv:2605.21587} \BibitemShut {NoStop}%
\bibitem [{\citenamefont {Betts}(1995)}]{Proc.N.S.Inst.Sci.40.95}%
  \BibitemOpen
  \bibfield  {author} {\bibinfo {author} {\bibfnamefont {D.~D.}\ \bibnamefont
  {Betts}},\ }\bibfield  {title} {\bibinfo {title} {A new two-dimensional
  lattice of coordination number five},\ }\href
  {https://hdl.handle.net/10222/35332} {\bibfield  {journal} {\bibinfo
  {journal} {Proceedings of the Nova Scotian Institute of Science}\ }\textbf
  {\bibinfo {volume} {40}},\ \bibinfo {pages} {95} (\bibinfo {year}
  {1995})}\BibitemShut {NoStop}%
\bibitem [{\citenamefont {Schulenburg}\ \emph {et~al.}(2000)\citenamefont
  {Schulenburg}, \citenamefont {Richter},\ and\ \citenamefont
  {Betts}}]{APhysPolA.97.971}%
  \BibitemOpen
  \bibfield  {author} {\bibinfo {author} {\bibfnamefont {J.}~\bibnamefont
  {Schulenburg}}, \bibinfo {author} {\bibfnamefont {J.}~\bibnamefont
  {Richter}},\ and\ \bibinfo {author} {\bibfnamefont {D.~D.}\ \bibnamefont
  {Betts}},\ }\bibfield  {title} {\bibinfo {title} {Heisenberg antiferromagnet
  on a $1/7$-depleted triangular lattice},\ }\href
  {https://doi.org/10.12693/APhysPolA.97.971} {\bibfield  {journal} {\bibinfo
  {journal} {Acta Physica Polonica A}\ }\textbf {\bibinfo {volume} {97}},\
  \bibinfo {pages} {971} (\bibinfo {year} {2000})}\BibitemShut {NoStop}%
\bibitem [{\citenamefont {Schmalfu\ss{}}\ \emph {et~al.}(2002)\citenamefont
  {Schmalfu\ss{}}, \citenamefont {Tomczak}, \citenamefont {Schulenburg},\ and\
  \citenamefont {Richter}}]{PhysRevB.65.224405}%
  \BibitemOpen
  \bibfield  {author} {\bibinfo {author} {\bibfnamefont {D.}~\bibnamefont
  {Schmalfu\ss{}}}, \bibinfo {author} {\bibfnamefont {P.}~\bibnamefont
  {Tomczak}}, \bibinfo {author} {\bibfnamefont {J.}~\bibnamefont
  {Schulenburg}},\ and\ \bibinfo {author} {\bibfnamefont {J.}~\bibnamefont
  {Richter}},\ }\bibfield  {title} {\bibinfo {title} {The spin-$\frac{1}{2}$
  {H}eisenberg antiferromagnet on a $\frac{1}{7}$-depleted triangular lattice:
  Ground-state properties},\ }\href
  {https://doi.org/10.1103/PhysRevB.65.224405} {\bibfield  {journal} {\bibinfo
  {journal} {Phys. Rev. B}\ }\textbf {\bibinfo {volume} {65}},\ \bibinfo
  {pages} {224405} (\bibinfo {year} {2002})}\BibitemShut {NoStop}%
\bibitem [{\citenamefont {Cave}\ \emph {et~al.}(2006)\citenamefont {Cave},
  \citenamefont {Coomer}, \citenamefont {Molinos}, \citenamefont {Klauss},\
  and\ \citenamefont {Wood}}]{anie.200502847}%
  \BibitemOpen
  \bibfield  {author} {\bibinfo {author} {\bibfnamefont {D.}~\bibnamefont
  {Cave}}, \bibinfo {author} {\bibfnamefont {F.~C.}\ \bibnamefont {Coomer}},
  \bibinfo {author} {\bibfnamefont {E.}~\bibnamefont {Molinos}}, \bibinfo
  {author} {\bibfnamefont {H.-H.}\ \bibnamefont {Klauss}},\ and\ \bibinfo
  {author} {\bibfnamefont {P.~T.}\ \bibnamefont {Wood}},\ }\bibfield  {title}
  {\bibinfo {title} {Compounds with the ``maple leaf'' lattice: Synthesis,
  structure, and magnetism of
  {M$_x$[Fe(O$_2$CCH$_2$)$_2$NCH$_2$PO$_3$]$_6$$\cdot n$H$_2$O}},\ }\href
  {https://doi.org/https://doi.org/10.1002/anie.200502847} {\bibfield
  {journal} {\bibinfo  {journal} {Angewandte Chemie International Edition}\
  }\textbf {\bibinfo {volume} {45}},\ \bibinfo {pages} {803} (\bibinfo {year}
  {2006})}\BibitemShut {NoStop}%
\bibitem [{\citenamefont {Fennell}\ \emph {et~al.}(2011)\citenamefont
  {Fennell}, \citenamefont {Piatek}, \citenamefont {Stephenson}, \citenamefont
  {Nilsen},\ and\ \citenamefont {R\o{}nnow}}]{Fennell_2011}%
  \BibitemOpen
  \bibfield  {author} {\bibinfo {author} {\bibfnamefont {T.}~\bibnamefont
  {Fennell}}, \bibinfo {author} {\bibfnamefont {J.~O.}\ \bibnamefont {Piatek}},
  \bibinfo {author} {\bibfnamefont {R.~A.}\ \bibnamefont {Stephenson}},
  \bibinfo {author} {\bibfnamefont {G.~J.}\ \bibnamefont {Nilsen}},\ and\
  \bibinfo {author} {\bibfnamefont {H.~M.}\ \bibnamefont {R\o{}nnow}},\
  }\bibfield  {title} {\bibinfo {title} {Spangolite: an s = 1/2 maple leaf
  lattice antiferromagnet?},\ }\href
  {https://doi.org/10.1088/0953-8984/23/16/164201} {\bibfield  {journal}
  {\bibinfo  {journal} {Journal of Physics: Condensed Matter}\ }\textbf
  {\bibinfo {volume} {23}},\ \bibinfo {pages} {164201} (\bibinfo {year}
  {2011})}\BibitemShut {NoStop}%
\bibitem [{\citenamefont {Aliev}\ \emph {et~al.}(2012)\citenamefont {Aliev},
  \citenamefont {Huv\'{e}}, \citenamefont {Colis}, \citenamefont {Colmont},
  \citenamefont {Dinia},\ and\ \citenamefont {Mentr\'{e}}}]{anie.201203775}%
  \BibitemOpen
  \bibfield  {author} {\bibinfo {author} {\bibfnamefont {A.}~\bibnamefont
  {Aliev}}, \bibinfo {author} {\bibfnamefont {M.}~\bibnamefont {Huv\'{e}}},
  \bibinfo {author} {\bibfnamefont {S.}~\bibnamefont {Colis}}, \bibinfo
  {author} {\bibfnamefont {M.}~\bibnamefont {Colmont}}, \bibinfo {author}
  {\bibfnamefont {A.}~\bibnamefont {Dinia}},\ and\ \bibinfo {author}
  {\bibfnamefont {O.}~\bibnamefont {Mentr\'{e}}},\ }\bibfield  {title}
  {\bibinfo {title} {Two-dimensional antiferromagnetism in the
  {[Mn$_{3+x}$O$_7$][Bi$_4$O$_{4.5-y}$]} compound with a maple-leaf lattice},\
  }\href {https://doi.org/https://doi.org/10.1002/anie.201203775} {\bibfield
  {journal} {\bibinfo  {journal} {Angewandte Chemie International Edition}\
  }\textbf {\bibinfo {volume} {51}},\ \bibinfo {pages} {9393} (\bibinfo {year}
  {2012})}\BibitemShut {NoStop}%
\bibitem [{\citenamefont {Haraguchi}\ \emph {et~al.}(2018)\citenamefont
  {Haraguchi}, \citenamefont {Matsuo}, \citenamefont {Kindo},\ and\
  \citenamefont {Hiroi}}]{PhysRevB.98.064412}%
  \BibitemOpen
  \bibfield  {author} {\bibinfo {author} {\bibfnamefont {Y.}~\bibnamefont
  {Haraguchi}}, \bibinfo {author} {\bibfnamefont {A.}~\bibnamefont {Matsuo}},
  \bibinfo {author} {\bibfnamefont {K.}~\bibnamefont {Kindo}},\ and\ \bibinfo
  {author} {\bibfnamefont {Z.}~\bibnamefont {Hiroi}},\ }\bibfield  {title}
  {\bibinfo {title} {Frustrated magnetism of the maple-leaf-lattice
  antiferromagnet
  {${\mathrm{MgMn}}_{3}{\mathrm{O}}_{7}\ifmmode\cdot\else\textperiodcentered\fi{}{3\mathrm{H}}_{2}\mathrm{O}$}},\
  }\href {https://doi.org/10.1103/PhysRevB.98.064412} {\bibfield  {journal}
  {\bibinfo  {journal} {Phys. Rev. B}\ }\textbf {\bibinfo {volume} {98}},\
  \bibinfo {pages} {064412} (\bibinfo {year} {2018})}\BibitemShut {NoStop}%
\bibitem [{\citenamefont {Venkatesh}\ \emph {et~al.}(2020)\citenamefont
  {Venkatesh}, \citenamefont {Bandyopadhyay}, \citenamefont {Midya},
  \citenamefont {Mahalingam}, \citenamefont {Ganesan},\ and\ \citenamefont
  {Mandal}}]{PhysRevB.101.184429}%
  \BibitemOpen
  \bibfield  {author} {\bibinfo {author} {\bibfnamefont {C.}~\bibnamefont
  {Venkatesh}}, \bibinfo {author} {\bibfnamefont {B.}~\bibnamefont
  {Bandyopadhyay}}, \bibinfo {author} {\bibfnamefont {A.}~\bibnamefont
  {Midya}}, \bibinfo {author} {\bibfnamefont {K.}~\bibnamefont {Mahalingam}},
  \bibinfo {author} {\bibfnamefont {V.}~\bibnamefont {Ganesan}},\ and\ \bibinfo
  {author} {\bibfnamefont {P.}~\bibnamefont {Mandal}},\ }\bibfield  {title}
  {\bibinfo {title} {Magnetic properties of the one-dimensional
  {$S=\frac{3}{2}$} {H}eisenberg antiferromagnetic spin-chain compound
  {${\mathrm{Na}}_{2}{\mathrm{Mn}}_{3}{\mathrm{O}}_{7}$}},\ }\href
  {https://doi.org/10.1103/PhysRevB.101.184429} {\bibfield  {journal} {\bibinfo
   {journal} {Phys. Rev. B}\ }\textbf {\bibinfo {volume} {101}},\ \bibinfo
  {pages} {184429} (\bibinfo {year} {2020})}\BibitemShut {NoStop}%
\bibitem [{\citenamefont {Haraguchi}\ \emph {et~al.}(2021)\citenamefont
  {Haraguchi}, \citenamefont {Matsuo}, \citenamefont {Kindo},\ and\
  \citenamefont {Hiroi}}]{PhysRevB.104.174439}%
  \BibitemOpen
  \bibfield  {author} {\bibinfo {author} {\bibfnamefont {Y.}~\bibnamefont
  {Haraguchi}}, \bibinfo {author} {\bibfnamefont {A.}~\bibnamefont {Matsuo}},
  \bibinfo {author} {\bibfnamefont {K.}~\bibnamefont {Kindo}},\ and\ \bibinfo
  {author} {\bibfnamefont {Z.}~\bibnamefont {Hiroi}},\ }\bibfield  {title}
  {\bibinfo {title} {Quantum antiferromagnet bluebellite comprising a
  maple-leaf lattice made of
  spin-{$1/2\phantom{\rule{0.16em}{0ex}}{\mathrm{Cu}}^{2+}$} ions},\ }\href
  {https://doi.org/10.1103/PhysRevB.104.174439} {\bibfield  {journal} {\bibinfo
   {journal} {Phys. Rev. B}\ }\textbf {\bibinfo {volume} {104}},\ \bibinfo
  {pages} {174439} (\bibinfo {year} {2021})}\BibitemShut {NoStop}%
\bibitem [{\citenamefont {Saha}\ \emph {et~al.}(2023)\citenamefont {Saha},
  \citenamefont {Bera}, \citenamefont {Yusuf},\ and\ \citenamefont
  {Hoser}}]{PhysRevB.107.064419}%
  \BibitemOpen
  \bibfield  {author} {\bibinfo {author} {\bibfnamefont {B.}~\bibnamefont
  {Saha}}, \bibinfo {author} {\bibfnamefont {A.~K.}\ \bibnamefont {Bera}},
  \bibinfo {author} {\bibfnamefont {S.~M.}\ \bibnamefont {Yusuf}},\ and\
  \bibinfo {author} {\bibfnamefont {A.}~\bibnamefont {Hoser}},\ }\bibfield
  {title} {\bibinfo {title} {Two-dimensional short-range spin-spin correlations
  in the layered spin-$\frac{3}{2}$ maple leaf lattice antiferromagnet
  {${\mathrm{Na}}_{2}{\mathrm{Mn}}_{3}{\mathrm{O}}_{7}$} with crystal stacking
  disorder},\ }\href {https://doi.org/10.1103/PhysRevB.107.064419} {\bibfield
  {journal} {\bibinfo  {journal} {Phys. Rev. B}\ }\textbf {\bibinfo {volume}
  {107}},\ \bibinfo {pages} {064419} (\bibinfo {year} {2023})}\BibitemShut
  {NoStop}%
\bibitem [{\citenamefont {Aguilar-Maldonado}\ \emph {et~al.}(2025)\citenamefont
  {Aguilar-Maldonado}, \citenamefont {Feyerherm}, \citenamefont
  {Proke\ifmmode~\check{s}\else \v{s}\fi{}}, \citenamefont {Keller},\ and\
  \citenamefont {Lake}}]{PhysRevB.111.094439}%
  \BibitemOpen
  \bibfield  {author} {\bibinfo {author} {\bibfnamefont {C.}~\bibnamefont
  {Aguilar-Maldonado}}, \bibinfo {author} {\bibfnamefont {R.}~\bibnamefont
  {Feyerherm}}, \bibinfo {author} {\bibfnamefont {K.}~\bibnamefont
  {Proke\ifmmode~\check{s}\else \v{s}\fi{}}}, \bibinfo {author} {\bibfnamefont
  {L.}~\bibnamefont {Keller}},\ and\ \bibinfo {author} {\bibfnamefont
  {B.}~\bibnamefont {Lake}},\ }\bibfield  {title} {\bibinfo {title} {Structure
  and magnetic properties of the maple leaf antiferromagnet
  {${\mathrm{Ho}}_{3}{\mathrm{ScO}}_{6}$}},\ }\href
  {https://doi.org/10.1103/PhysRevB.111.094439} {\bibfield  {journal} {\bibinfo
   {journal} {Phys. Rev. B}\ }\textbf {\bibinfo {volume} {111}},\ \bibinfo
  {pages} {094439} (\bibinfo {year} {2025})}\BibitemShut {NoStop}%
\bibitem [{\citenamefont {{Sriram Shastry}}\ and\ \citenamefont
  {Sutherland}(1981)}]{SRIRAMSHASTRY19811069}%
  \BibitemOpen
  \bibfield  {author} {\bibinfo {author} {\bibfnamefont {B.}~\bibnamefont
  {{Sriram Shastry}}}\ and\ \bibinfo {author} {\bibfnamefont {B.}~\bibnamefont
  {Sutherland}},\ }\bibfield  {title} {\bibinfo {title} {Exact ground state of
  a quantum mechanical antiferromagnet},\ }\href
  {https://doi.org/https://doi.org/10.1016/0378-4363(81)90838-X} {\bibfield
  {journal} {\bibinfo  {journal} {Physica B+C}\ }\textbf {\bibinfo {volume}
  {108}},\ \bibinfo {pages} {1069} (\bibinfo {year} {1981})}\BibitemShut
  {NoStop}%
\bibitem [{\citenamefont {Gopalan}\ \emph {et~al.}(1994)\citenamefont
  {Gopalan}, \citenamefont {Rice},\ and\ \citenamefont
  {Sigrist}}]{PhysRevB.49.8901}%
  \BibitemOpen
  \bibfield  {author} {\bibinfo {author} {\bibfnamefont {S.}~\bibnamefont
  {Gopalan}}, \bibinfo {author} {\bibfnamefont {T.~M.}\ \bibnamefont {Rice}},\
  and\ \bibinfo {author} {\bibfnamefont {M.}~\bibnamefont {Sigrist}},\
  }\bibfield  {title} {\bibinfo {title} {Spin ladders with spin gaps: A
  description of a class of cuprates},\ }\href
  {https://doi.org/10.1103/PhysRevB.49.8901} {\bibfield  {journal} {\bibinfo
  {journal} {Phys. Rev. B}\ }\textbf {\bibinfo {volume} {49}},\ \bibinfo
  {pages} {8901} (\bibinfo {year} {1994})}\BibitemShut {NoStop}%
\bibitem [{\citenamefont {Normand}\ \emph {et~al.}(1997)\citenamefont
  {Normand}, \citenamefont {Penc}, \citenamefont {Albrecht},\ and\
  \citenamefont {Mila}}]{PhysRevB.56.R5736}%
  \BibitemOpen
  \bibfield  {author} {\bibinfo {author} {\bibfnamefont {B.}~\bibnamefont
  {Normand}}, \bibinfo {author} {\bibfnamefont {K.}~\bibnamefont {Penc}},
  \bibinfo {author} {\bibfnamefont {M.}~\bibnamefont {Albrecht}},\ and\
  \bibinfo {author} {\bibfnamefont {F.}~\bibnamefont {Mila}},\ }\bibfield
  {title} {\bibinfo {title} {Phase diagram of the {$S$=$\frac{1}{2}$}
  frustrated coupled ladder system},\ }\href
  {https://doi.org/10.1103/PhysRevB.56.R5736} {\bibfield  {journal} {\bibinfo
  {journal} {Phys. Rev. B}\ }\textbf {\bibinfo {volume} {56}},\ \bibinfo
  {pages} {R5736} (\bibinfo {year} {1997})}\BibitemShut {NoStop}%
\bibitem [{\citenamefont {Millet}\ \emph {et~al.}(1998)\citenamefont {Millet},
  \citenamefont {Satto}, \citenamefont {Bonvoisin}, \citenamefont {Normand},
  \citenamefont {Penc}, \citenamefont {Albrecht},\ and\ \citenamefont
  {Mila}}]{PhysRevB.57.5005}%
  \BibitemOpen
  \bibfield  {author} {\bibinfo {author} {\bibfnamefont {P.}~\bibnamefont
  {Millet}}, \bibinfo {author} {\bibfnamefont {C.}~\bibnamefont {Satto}},
  \bibinfo {author} {\bibfnamefont {J.}~\bibnamefont {Bonvoisin}}, \bibinfo
  {author} {\bibfnamefont {B.}~\bibnamefont {Normand}}, \bibinfo {author}
  {\bibfnamefont {K.}~\bibnamefont {Penc}}, \bibinfo {author} {\bibfnamefont
  {M.}~\bibnamefont {Albrecht}},\ and\ \bibinfo {author} {\bibfnamefont
  {F.}~\bibnamefont {Mila}},\ }\bibfield  {title} {\bibinfo {title} {Magnetic
  properties of the coupled ladder system
  {${\mathrm{MgV}}_{2}{\mathrm{O}}_{5}$}},\ }\href
  {https://doi.org/10.1103/PhysRevB.57.5005} {\bibfield  {journal} {\bibinfo
  {journal} {Phys. Rev. B}\ }\textbf {\bibinfo {volume} {57}},\ \bibinfo
  {pages} {5005} (\bibinfo {year} {1998})}\BibitemShut {NoStop}%
\bibitem [{\citenamefont {Miyahara}\ \emph {et~al.}(1998)\citenamefont
  {Miyahara}, \citenamefont {Troyer}, \citenamefont {Johnston},\ and\
  \citenamefont {Ueda}}]{JPSJ.67.3918}%
  \BibitemOpen
  \bibfield  {author} {\bibinfo {author} {\bibfnamefont {S.}~\bibnamefont
  {Miyahara}}, \bibinfo {author} {\bibfnamefont {M.}~\bibnamefont {Troyer}},
  \bibinfo {author} {\bibfnamefont {D.~C.}\ \bibnamefont {Johnston}},\ and\
  \bibinfo {author} {\bibfnamefont {K.}~\bibnamefont {Ueda}},\ }\bibfield
  {title} {\bibinfo {title} {Quantum {M}onte {C}arlo simulation of the trellis
  lattice {H}eisenberg model for {SrCu$_2$O$_3$} and {CaV$_2$O$_5$}},\ }\href
  {https://doi.org/10.1143/JPSJ.67.3918} {\bibfield  {journal} {\bibinfo
  {journal} {Journal of the Physical Society of Japan}\ }\textbf {\bibinfo
  {volume} {67}},\ \bibinfo {pages} {3918} (\bibinfo {year}
  {1998})}\BibitemShut {NoStop}%
\bibitem [{\citenamefont {Korotin}\ \emph {et~al.}(1999)\citenamefont
  {Korotin}, \citenamefont {Elfimov}, \citenamefont {Anisimov}, \citenamefont
  {Troyer},\ and\ \citenamefont {Khomskii}}]{PhysRevLett.83.1387}%
  \BibitemOpen
  \bibfield  {author} {\bibinfo {author} {\bibfnamefont {M.~A.}\ \bibnamefont
  {Korotin}}, \bibinfo {author} {\bibfnamefont {I.~S.}\ \bibnamefont
  {Elfimov}}, \bibinfo {author} {\bibfnamefont {V.~I.}\ \bibnamefont
  {Anisimov}}, \bibinfo {author} {\bibfnamefont {M.}~\bibnamefont {Troyer}},\
  and\ \bibinfo {author} {\bibfnamefont {D.~I.}\ \bibnamefont {Khomskii}},\
  }\bibfield  {title} {\bibinfo {title} {Exchange interactions and magnetic
  properties of the layered vanadates {${\mathrm{CaV}}_{2}{\mathrm{O}}_{5}$},
  {${\mathrm{MgV}}_{2}{\mathrm{O}}_{5}$},
  {${\mathrm{CaV}}_{3}{\mathrm{O}}_{7}$}, and
  {${\mathrm{CaV}}_{4}{\mathrm{O}}_{9}$}},\ }\href
  {https://doi.org/10.1103/PhysRevLett.83.1387} {\bibfield  {journal} {\bibinfo
   {journal} {Phys. Rev. Lett.}\ }\textbf {\bibinfo {volume} {83}},\ \bibinfo
  {pages} {1387} (\bibinfo {year} {1999})}\BibitemShut {NoStop}%
\bibitem [{\citenamefont {Konstantinovi\ifmmode~\acute{c}\else \'{c}\fi{}}\
  \emph {et~al.}(2000)\citenamefont {Konstantinovi\ifmmode~\acute{c}\else
  \'{c}\fi{}}, \citenamefont {Popovi\ifmmode~\acute{c}\else \'{c}\fi{}},
  \citenamefont {Isobe},\ and\ \citenamefont {Ueda}}]{PhysRevB.61.15185}%
  \BibitemOpen
  \bibfield  {author} {\bibinfo {author} {\bibfnamefont {M.~J.}\ \bibnamefont
  {Konstantinovi\ifmmode~\acute{c}\else \'{c}\fi{}}}, \bibinfo {author}
  {\bibfnamefont {Z.~V.}\ \bibnamefont {Popovi\ifmmode~\acute{c}\else
  \'{c}\fi{}}}, \bibinfo {author} {\bibfnamefont {M.}~\bibnamefont {Isobe}},\
  and\ \bibinfo {author} {\bibfnamefont {Y.}~\bibnamefont {Ueda}},\ }\bibfield
  {title} {\bibinfo {title} {Raman scattering from magnetic excitations in the
  spin-ladder compounds {${\mathrm{CaV}}_{2}{\mathrm{O}}_{5}$} and
  {${\mathrm{MgV}}_{2}{\mathrm{O}}_{5}$}},\ }\href
  {https://doi.org/10.1103/PhysRevB.61.15185} {\bibfield  {journal} {\bibinfo
  {journal} {Phys. Rev. B}\ }\textbf {\bibinfo {volume} {61}},\ \bibinfo
  {pages} {15185} (\bibinfo {year} {2000})}\BibitemShut {NoStop}%
\bibitem [{\citenamefont {Schmidt}\ and\ \citenamefont
  {Uhrig}(2007)}]{PhysRevB.75.224414}%
  \BibitemOpen
  \bibfield  {author} {\bibinfo {author} {\bibfnamefont {K.~P.}\ \bibnamefont
  {Schmidt}}\ and\ \bibinfo {author} {\bibfnamefont {G.~S.}\ \bibnamefont
  {Uhrig}},\ }\bibfield  {title} {\bibinfo {title} {Two dimensionality of
  magnetic excitations on the trellis lattice:
  {${(\mathrm{La},\mathrm{Sr},\mathrm{Ca})}_{14}{\mathrm{Cu}}_{24}{\mathrm{O}}_{41}$}
  and {$\mathrm{Sr}{\mathrm{Cu}}_{2}{\mathrm{O}}_{3}$}},\ }\href
  {https://doi.org/10.1103/PhysRevB.75.224414} {\bibfield  {journal} {\bibinfo
  {journal} {Phys. Rev. B}\ }\textbf {\bibinfo {volume} {75}},\ \bibinfo
  {pages} {224414} (\bibinfo {year} {2007})}\BibitemShut {NoStop}%
\bibitem [{\citenamefont {Tsirlin}\ \emph {et~al.}(2007)\citenamefont
  {Tsirlin}, \citenamefont {Shpanchenko}, \citenamefont {Antipov},
  \citenamefont {Bougerol}, \citenamefont {Hadermann}, \citenamefont
  {Van~Tendeloo}, \citenamefont {Schnelle},\ and\ \citenamefont
  {Rosner}}]{PhysRevB.76.104429}%
  \BibitemOpen
  \bibfield  {author} {\bibinfo {author} {\bibfnamefont {A.~A.}\ \bibnamefont
  {Tsirlin}}, \bibinfo {author} {\bibfnamefont {R.~V.}\ \bibnamefont
  {Shpanchenko}}, \bibinfo {author} {\bibfnamefont {E.~V.}\ \bibnamefont
  {Antipov}}, \bibinfo {author} {\bibfnamefont {C.}~\bibnamefont {Bougerol}},
  \bibinfo {author} {\bibfnamefont {J.}~\bibnamefont {Hadermann}}, \bibinfo
  {author} {\bibfnamefont {G.}~\bibnamefont {Van~Tendeloo}}, \bibinfo {author}
  {\bibfnamefont {W.}~\bibnamefont {Schnelle}},\ and\ \bibinfo {author}
  {\bibfnamefont {H.}~\bibnamefont {Rosner}},\ }\bibfield  {title} {\bibinfo
  {title} {Spin ladder compound
  {${\mathrm{Pb}}_{0.55}{\mathrm{Cd}}_{0.45}{\mathrm{V}}_{2}{\mathrm{O}}_{5}$}:
  Synthesis and investigation},\ }\href
  {https://doi.org/10.1103/PhysRevB.76.104429} {\bibfield  {journal} {\bibinfo
  {journal} {Phys. Rev. B}\ }\textbf {\bibinfo {volume} {76}},\ \bibinfo
  {pages} {104429} (\bibinfo {year} {2007})}\BibitemShut {NoStop}%
\bibitem [{\citenamefont {Sakakida}\ and\ \citenamefont
  {Shimahara}(2017)}]{JPSJ.86.124709}%
  \BibitemOpen
  \bibfield  {author} {\bibinfo {author} {\bibfnamefont {K.}~\bibnamefont
  {Sakakida}}\ and\ \bibinfo {author} {\bibfnamefont {H.}~\bibnamefont
  {Shimahara}},\ }\bibfield  {title} {\bibinfo {title} {Spin structures and
  phase diagrams of extended spatially completely anisotropic triangular
  lattice antiferromagnets},\ }\href {https://doi.org/10.7566/JPSJ.86.124709}
  {\bibfield  {journal} {\bibinfo  {journal} {Journal of the Physical Society
  of Japan}\ }\textbf {\bibinfo {volume} {86}},\ \bibinfo {pages} {124709}
  (\bibinfo {year} {2017})}\BibitemShut {NoStop}%
\bibitem [{\citenamefont {Yamaguchi}\ \emph {et~al.}(2018)\citenamefont
  {Yamaguchi}, \citenamefont {Yoshizawa}, \citenamefont {Kida}, \citenamefont
  {Hagiwara}, \citenamefont {Matsuo}, \citenamefont {Kono}, \citenamefont
  {Sakakibara}, \citenamefont {Tamekuni}, \citenamefont {Miyagai},\ and\
  \citenamefont {Hosokoshi}}]{JPSJ.87.043701}%
  \BibitemOpen
  \bibfield  {author} {\bibinfo {author} {\bibfnamefont {H.}~\bibnamefont
  {Yamaguchi}}, \bibinfo {author} {\bibfnamefont {D.}~\bibnamefont
  {Yoshizawa}}, \bibinfo {author} {\bibfnamefont {T.}~\bibnamefont {Kida}},
  \bibinfo {author} {\bibfnamefont {M.}~\bibnamefont {Hagiwara}}, \bibinfo
  {author} {\bibfnamefont {A.}~\bibnamefont {Matsuo}}, \bibinfo {author}
  {\bibfnamefont {Y.}~\bibnamefont {Kono}}, \bibinfo {author} {\bibfnamefont
  {T.}~\bibnamefont {Sakakibara}}, \bibinfo {author} {\bibfnamefont
  {Y.}~\bibnamefont {Tamekuni}}, \bibinfo {author} {\bibfnamefont
  {H.}~\bibnamefont {Miyagai}},\ and\ \bibinfo {author} {\bibfnamefont
  {Y.}~\bibnamefont {Hosokoshi}},\ }\bibfield  {title} {\bibinfo {title}
  {Magnetic-field-induced quantum phase in {$S=1/2$} frustrated trellis
  lattice},\ }\href {https://doi.org/10.7566/JPSJ.87.043701} {\bibfield
  {journal} {\bibinfo  {journal} {Journal of the Physical Society of Japan}\
  }\textbf {\bibinfo {volume} {87}},\ \bibinfo {pages} {043701} (\bibinfo
  {year} {2018})}\BibitemShut {NoStop}%
\bibitem [{\citenamefont {Chatterjee}\ \emph {et~al.}(2026)\citenamefont
  {Chatterjee}, \citenamefont {Maity}, \citenamefont {Potten}, \citenamefont
  {M\"uller}, \citenamefont {Feuerpfeil}, \citenamefont {Thomale},
  \citenamefont {Penc}, \citenamefont {Jeschke}, \citenamefont {Samajdar},\
  and\ \citenamefont {Iqbal}}]{xmtj-4kj9}%
  \BibitemOpen
  \bibfield  {author} {\bibinfo {author} {\bibfnamefont {S.}~\bibnamefont
  {Chatterjee}}, \bibinfo {author} {\bibfnamefont {A.}~\bibnamefont {Maity}},
  \bibinfo {author} {\bibfnamefont {J.}~\bibnamefont {Potten}}, \bibinfo
  {author} {\bibfnamefont {T.}~\bibnamefont {M\"uller}}, \bibinfo {author}
  {\bibfnamefont {A.}~\bibnamefont {Feuerpfeil}}, \bibinfo {author}
  {\bibfnamefont {R.}~\bibnamefont {Thomale}}, \bibinfo {author} {\bibfnamefont
  {K.}~\bibnamefont {Penc}}, \bibinfo {author} {\bibfnamefont {H.~O.}\
  \bibnamefont {Jeschke}}, \bibinfo {author} {\bibfnamefont {R.}~\bibnamefont
  {Samajdar}},\ and\ \bibinfo {author} {\bibfnamefont {Y.}~\bibnamefont
  {Iqbal}},\ }\bibfield  {title} {\bibinfo {title} {Semi-{D}irac spin liquids
  and frustrated quantum magnetism on the trellis lattice},\ }\href
  {https://doi.org/10.1103/xmtj-4kj9} {\bibfield  {journal} {\bibinfo
  {journal} {Phys. Rev. Res.}\ }\textbf {\bibinfo {volume} {8}},\ \bibinfo
  {pages} {013191} (\bibinfo {year} {2026})}\BibitemShut {NoStop}%
\bibitem [{\citenamefont {Johnston}\ \emph {et~al.}()\citenamefont {Johnston},
  \citenamefont {Troyer}, \citenamefont {Miyahara}, \citenamefont {Lidsky},
  \citenamefont {Ueda}, \citenamefont {Azuma}, \citenamefont {Hiroi},
  \citenamefont {Takano}, \citenamefont {Isobe}, \citenamefont {Ueda},
  \citenamefont {Korotin}, \citenamefont {Anisimov}, \citenamefont {Mahajan},\
  and\ \citenamefont {Miller}}]{0001147}%
  \BibitemOpen
  \bibfield  {author} {\bibinfo {author} {\bibfnamefont {D.}~\bibnamefont
  {Johnston}}, \bibinfo {author} {\bibfnamefont {M.}~\bibnamefont {Troyer}},
  \bibinfo {author} {\bibfnamefont {S.}~\bibnamefont {Miyahara}}, \bibinfo
  {author} {\bibfnamefont {D.}~\bibnamefont {Lidsky}}, \bibinfo {author}
  {\bibfnamefont {K.}~\bibnamefont {Ueda}}, \bibinfo {author} {\bibfnamefont
  {M.}~\bibnamefont {Azuma}}, \bibinfo {author} {\bibfnamefont
  {Z.}~\bibnamefont {Hiroi}}, \bibinfo {author} {\bibfnamefont
  {M.}~\bibnamefont {Takano}}, \bibinfo {author} {\bibfnamefont
  {M.}~\bibnamefont {Isobe}}, \bibinfo {author} {\bibfnamefont
  {Y.}~\bibnamefont {Ueda}}, \bibinfo {author} {\bibfnamefont {M.}~\bibnamefont
  {Korotin}}, \bibinfo {author} {\bibfnamefont {V.}~\bibnamefont {Anisimov}},
  \bibinfo {author} {\bibfnamefont {A.}~\bibnamefont {Mahajan}},\ and\ \bibinfo
  {author} {\bibfnamefont {L.}~\bibnamefont {Miller}},\ }\bibfield  {title}
  {\bibinfo {title} {Magnetic susceptibilities of spin-$1/2$ antiferromagnetic
  {H}eisenberg ladders and applications to ladder oxide compounds},\ }\href
  {https://arxiv.org/abs/cond-mat/0001147} {\ }\Eprint
  {https://arxiv.org/abs/cond-mat/0001147} {arXiv:cond-mat/0001147}
  \BibitemShut {NoStop}%
\bibitem [{\citenamefont {Farnell}\ \emph {et~al.}(2011)\citenamefont
  {Farnell}, \citenamefont {Darradi}, \citenamefont {Schmidt},\ and\
  \citenamefont {Richter}}]{PhysRevB.84.104406}%
  \BibitemOpen
  \bibfield  {author} {\bibinfo {author} {\bibfnamefont {D.~J.~J.}\
  \bibnamefont {Farnell}}, \bibinfo {author} {\bibfnamefont {R.}~\bibnamefont
  {Darradi}}, \bibinfo {author} {\bibfnamefont {R.}~\bibnamefont {Schmidt}},\
  and\ \bibinfo {author} {\bibfnamefont {J.}~\bibnamefont {Richter}},\
  }\bibfield  {title} {\bibinfo {title} {Spin-half {H}eisenberg antiferromagnet
  on two archimedian lattices: From the bounce lattice to the maple-leaf
  lattice and beyond},\ }\href {https://doi.org/10.1103/PhysRevB.84.104406}
  {\bibfield  {journal} {\bibinfo  {journal} {Phys. Rev. B}\ }\textbf {\bibinfo
  {volume} {84}},\ \bibinfo {pages} {104406} (\bibinfo {year}
  {2011})}\BibitemShut {NoStop}%
\bibitem [{\citenamefont {Farnell}\ \emph {et~al.}(2014)\citenamefont
  {Farnell}, \citenamefont {G\"otze}, \citenamefont {Richter}, \citenamefont
  {Bishop},\ and\ \citenamefont {Li}}]{PhysRevB.89.184407}%
  \BibitemOpen
  \bibfield  {author} {\bibinfo {author} {\bibfnamefont {D.~J.~J.}\
  \bibnamefont {Farnell}}, \bibinfo {author} {\bibfnamefont {O.}~\bibnamefont
  {G\"otze}}, \bibinfo {author} {\bibfnamefont {J.}~\bibnamefont {Richter}},
  \bibinfo {author} {\bibfnamefont {R.~F.}\ \bibnamefont {Bishop}},\ and\
  \bibinfo {author} {\bibfnamefont {P.~H.~Y.}\ \bibnamefont {Li}},\ }\bibfield
  {title} {\bibinfo {title} {Quantum $s=\frac{1}{2}$ antiferromagnets on
  {A}rchimedean lattices: The route from semiclassical magnetic order to
  nonmagnetic quantum states},\ }\href
  {https://doi.org/10.1103/PhysRevB.89.184407} {\bibfield  {journal} {\bibinfo
  {journal} {Phys. Rev. B}\ }\textbf {\bibinfo {volume} {89}},\ \bibinfo
  {pages} {184407} (\bibinfo {year} {2014})}\BibitemShut {NoStop}%
\bibitem [{\citenamefont {Yu}(2015)}]{PhysRevE.91.062121}%
  \BibitemOpen
  \bibfield  {author} {\bibinfo {author} {\bibfnamefont {U.}~\bibnamefont
  {Yu}},\ }\bibfield  {title} {\bibinfo {title} {Ising antiferromagnet on the
  {A}rchimedean lattices},\ }\href {https://doi.org/10.1103/PhysRevE.91.062121}
  {\bibfield  {journal} {\bibinfo  {journal} {Phys. Rev. E}\ }\textbf {\bibinfo
  {volume} {91}},\ \bibinfo {pages} {062121} (\bibinfo {year}
  {2015})}\BibitemShut {NoStop}%
\bibitem [{\citenamefont {Farnell}\ \emph {et~al.}(2018)\citenamefont
  {Farnell}, \citenamefont {G\"otze}, \citenamefont {Schulenburg},
  \citenamefont {Zinke}, \citenamefont {Bishop},\ and\ \citenamefont
  {Li}}]{PhysRevB.98.224402}%
  \BibitemOpen
  \bibfield  {author} {\bibinfo {author} {\bibfnamefont {D.~J.~J.}\
  \bibnamefont {Farnell}}, \bibinfo {author} {\bibfnamefont {O.}~\bibnamefont
  {G\"otze}}, \bibinfo {author} {\bibfnamefont {J.}~\bibnamefont
  {Schulenburg}}, \bibinfo {author} {\bibfnamefont {R.}~\bibnamefont {Zinke}},
  \bibinfo {author} {\bibfnamefont {R.~F.}\ \bibnamefont {Bishop}},\ and\
  \bibinfo {author} {\bibfnamefont {P.~H.~Y.}\ \bibnamefont {Li}},\ }\bibfield
  {title} {\bibinfo {title} {Interplay between lattice topology, frustration,
  and spin quantum number in quantum antiferromagnets on {A}rchimedean
  lattices},\ }\href {https://doi.org/10.1103/PhysRevB.98.224402} {\bibfield
  {journal} {\bibinfo  {journal} {Phys. Rev. B}\ }\textbf {\bibinfo {volume}
  {98}},\ \bibinfo {pages} {224402} (\bibinfo {year} {2018})}\BibitemShut
  {NoStop}%
\bibitem [{\citenamefont {Pierre}\ \emph {et~al.}(2025)\citenamefont {Pierre},
  \citenamefont {Bernu},\ and\ \citenamefont {Messio}}]{SciPostPhys.19.1.025}%
  \BibitemOpen
  \bibfield  {author} {\bibinfo {author} {\bibfnamefont {L.}~\bibnamefont
  {Pierre}}, \bibinfo {author} {\bibfnamefont {B.}~\bibnamefont {Bernu}},\ and\
  \bibinfo {author} {\bibfnamefont {L.}~\bibnamefont {Messio}},\ }\bibfield
  {title} {\bibinfo {title} {{Derivation of free energy, entropy and specific
  heat for planar Ising models: Application to Archimedean lattices and their
  duals}},\ }\href {https://doi.org/10.21468/SciPostPhys.19.1.025} {\bibfield
  {journal} {\bibinfo  {journal} {SciPost Phys.}\ }\textbf {\bibinfo {volume}
  {19}},\ \bibinfo {pages} {025} (\bibinfo {year} {2025})}\BibitemShut
  {NoStop}%
\bibitem [{\citenamefont {Kitaev}(2006)}]{KITAEV20062}%
  \BibitemOpen
  \bibfield  {author} {\bibinfo {author} {\bibfnamefont {A.}~\bibnamefont
  {Kitaev}},\ }\bibfield  {title} {\bibinfo {title} {Anyons in an exactly
  solved model and beyond},\ }\href {https://doi.org/10.1016/j.aop.2005.10.005}
  {\bibfield  {journal} {\bibinfo  {journal} {Annals of Physics}\ }\textbf
  {\bibinfo {volume} {321}},\ \bibinfo {pages} {2} (\bibinfo {year}
  {2006})}\BibitemShut {NoStop}%
\bibitem [{\citenamefont {Yang}\ \emph {et~al.}(2007)\citenamefont {Yang},
  \citenamefont {Zhou},\ and\ \citenamefont {Sun}}]{PhysRevB.76.180404}%
  \BibitemOpen
  \bibfield  {author} {\bibinfo {author} {\bibfnamefont {S.}~\bibnamefont
  {Yang}}, \bibinfo {author} {\bibfnamefont {D.~L.}\ \bibnamefont {Zhou}},\
  and\ \bibinfo {author} {\bibfnamefont {C.~P.}\ \bibnamefont {Sun}},\
  }\bibfield  {title} {\bibinfo {title} {Mosaic spin models with topological
  order},\ }\href {https://doi.org/10.1103/PhysRevB.76.180404} {\bibfield
  {journal} {\bibinfo  {journal} {Phys. Rev. B}\ }\textbf {\bibinfo {volume}
  {76}},\ \bibinfo {pages} {180404} (\bibinfo {year} {2007})}\BibitemShut
  {NoStop}%
\bibitem [{\citenamefont {Yao}\ and\ \citenamefont
  {Kivelson}(2007)}]{PhysRevLett.99.247203}%
  \BibitemOpen
  \bibfield  {author} {\bibinfo {author} {\bibfnamefont {H.}~\bibnamefont
  {Yao}}\ and\ \bibinfo {author} {\bibfnamefont {S.~A.}\ \bibnamefont
  {Kivelson}},\ }\bibfield  {title} {\bibinfo {title} {Exact chiral spin liquid
  with non-{A}belian anyons},\ }\href
  {https://doi.org/10.1103/PhysRevLett.99.247203} {\bibfield  {journal}
  {\bibinfo  {journal} {Phys. Rev. Lett.}\ }\textbf {\bibinfo {volume} {99}},\
  \bibinfo {pages} {247203} (\bibinfo {year} {2007})}\BibitemShut {NoStop}%
\bibitem [{\citenamefont {Mandal}\ and\ \citenamefont
  {Surendran}(2009)}]{PhysRevB.79.024426}%
  \BibitemOpen
  \bibfield  {author} {\bibinfo {author} {\bibfnamefont {S.}~\bibnamefont
  {Mandal}}\ and\ \bibinfo {author} {\bibfnamefont {N.}~\bibnamefont
  {Surendran}},\ }\bibfield  {title} {\bibinfo {title} {Exactly solvable
  {K}itaev model in three dimensions},\ }\href
  {https://doi.org/10.1103/PhysRevB.79.024426} {\bibfield  {journal} {\bibinfo
  {journal} {Phys. Rev. B}\ }\textbf {\bibinfo {volume} {79}},\ \bibinfo
  {pages} {024426} (\bibinfo {year} {2009})}\BibitemShut {NoStop}%
\bibitem [{\citenamefont {Hermanns}\ and\ \citenamefont
  {Trebst}(2014)}]{PhysRevB.89.235102}%
  \BibitemOpen
  \bibfield  {author} {\bibinfo {author} {\bibfnamefont {M.}~\bibnamefont
  {Hermanns}}\ and\ \bibinfo {author} {\bibfnamefont {S.}~\bibnamefont
  {Trebst}},\ }\bibfield  {title} {\bibinfo {title} {Quantum spin liquid with a
  {M}ajorana {F}ermi surface on the three-dimensional hyperoctagon lattice},\
  }\href {https://doi.org/10.1103/PhysRevB.89.235102} {\bibfield  {journal}
  {\bibinfo  {journal} {Phys. Rev. B}\ }\textbf {\bibinfo {volume} {89}},\
  \bibinfo {pages} {235102} (\bibinfo {year} {2014})}\BibitemShut {NoStop}%
\bibitem [{\citenamefont {Wu}\ \emph {et~al.}(2009)\citenamefont {Wu},
  \citenamefont {Arovas},\ and\ \citenamefont {Hung}}]{PhysRevB.79.134427}%
  \BibitemOpen
  \bibfield  {author} {\bibinfo {author} {\bibfnamefont {C.}~\bibnamefont
  {Wu}}, \bibinfo {author} {\bibfnamefont {D.}~\bibnamefont {Arovas}},\ and\
  \bibinfo {author} {\bibfnamefont {H.-H.}\ \bibnamefont {Hung}},\ }\bibfield
  {title} {\bibinfo {title} {{$\Gamma$}-matrix generalization of the {K}itaev
  model},\ }\href {https://doi.org/10.1103/PhysRevB.79.134427} {\bibfield
  {journal} {\bibinfo  {journal} {Phys. Rev. B}\ }\textbf {\bibinfo {volume}
  {79}},\ \bibinfo {pages} {134427} (\bibinfo {year} {2009})}\BibitemShut
  {NoStop}%
\bibitem [{Note1()}]{Note1}%
  \BibitemOpen
  \bibinfo {note} {Pauli matrices are $\Gamma $ matrices with
  $n=2$.}\BibitemShut {Stop}%
\bibitem [{\citenamefont {Chern}(2010)}]{PhysRevB.81.125134}%
  \BibitemOpen
  \bibfield  {author} {\bibinfo {author} {\bibfnamefont {G.-W.}\ \bibnamefont
  {Chern}},\ }\bibfield  {title} {\bibinfo {title} {Three-dimensional
  topological phases in a layered honeycomb spin-orbital model},\ }\href
  {https://doi.org/10.1103/PhysRevB.81.125134} {\bibfield  {journal} {\bibinfo
  {journal} {Phys. Rev. B}\ }\textbf {\bibinfo {volume} {81}},\ \bibinfo
  {pages} {125134} (\bibinfo {year} {2010})}\BibitemShut {NoStop}%
\bibitem [{\citenamefont {Ryu}(2009)}]{PhysRevB.79.075124}%
  \BibitemOpen
  \bibfield  {author} {\bibinfo {author} {\bibfnamefont {S.}~\bibnamefont
  {Ryu}},\ }\bibfield  {title} {\bibinfo {title} {Three-dimensional topological
  phase on the diamond lattice},\ }\href
  {https://doi.org/10.1103/PhysRevB.79.075124} {\bibfield  {journal} {\bibinfo
  {journal} {Phys. Rev. B}\ }\textbf {\bibinfo {volume} {79}},\ \bibinfo
  {pages} {075124} (\bibinfo {year} {2009})}\BibitemShut {NoStop}%
\bibitem [{\citenamefont {Yao}\ \emph {et~al.}(2009)\citenamefont {Yao},
  \citenamefont {Zhang},\ and\ \citenamefont
  {Kivelson}}]{PhysRevLett.102.217202}%
  \BibitemOpen
  \bibfield  {author} {\bibinfo {author} {\bibfnamefont {H.}~\bibnamefont
  {Yao}}, \bibinfo {author} {\bibfnamefont {S.-C.}\ \bibnamefont {Zhang}},\
  and\ \bibinfo {author} {\bibfnamefont {S.~A.}\ \bibnamefont {Kivelson}},\
  }\bibfield  {title} {\bibinfo {title} {Algebraic spin liquid in an exactly
  solvable spin model},\ }\href
  {https://doi.org/10.1103/PhysRevLett.102.217202} {\bibfield  {journal}
  {\bibinfo  {journal} {Phys. Rev. Lett.}\ }\textbf {\bibinfo {volume} {102}},\
  \bibinfo {pages} {217202} (\bibinfo {year} {2009})}\BibitemShut {NoStop}%
\bibitem [{\citenamefont {Chulliparambil}\ \emph {et~al.}(2020)\citenamefont
  {Chulliparambil}, \citenamefont {Seifert}, \citenamefont {Vojta},
  \citenamefont {Janssen},\ and\ \citenamefont {Tu}}]{PhysRevB.102.201111}%
  \BibitemOpen
  \bibfield  {author} {\bibinfo {author} {\bibfnamefont {S.}~\bibnamefont
  {Chulliparambil}}, \bibinfo {author} {\bibfnamefont {U.~F.~P.}\ \bibnamefont
  {Seifert}}, \bibinfo {author} {\bibfnamefont {M.}~\bibnamefont {Vojta}},
  \bibinfo {author} {\bibfnamefont {L.}~\bibnamefont {Janssen}},\ and\ \bibinfo
  {author} {\bibfnamefont {H.-H.}\ \bibnamefont {Tu}},\ }\bibfield  {title}
  {\bibinfo {title} {Microscopic models for {K}itaev's sixteenfold way of anyon
  theories},\ }\href {https://doi.org/10.1103/PhysRevB.102.201111} {\bibfield
  {journal} {\bibinfo  {journal} {Phys. Rev. B}\ }\textbf {\bibinfo {volume}
  {102}},\ \bibinfo {pages} {201111} (\bibinfo {year} {2020})}\BibitemShut
  {NoStop}%
\bibitem [{\citenamefont {Keskiner}\ \emph {et~al.}(2023)\citenamefont
  {Keskiner}, \citenamefont {Erten},\ and\ \citenamefont
  {Oktel}}]{PhysRevB.108.104208}%
  \BibitemOpen
  \bibfield  {author} {\bibinfo {author} {\bibfnamefont {M.~A.}\ \bibnamefont
  {Keskiner}}, \bibinfo {author} {\bibfnamefont {O.}~\bibnamefont {Erten}},\
  and\ \bibinfo {author} {\bibfnamefont {M.~O.}\ \bibnamefont {Oktel}},\
  }\bibfield  {title} {\bibinfo {title} {Kitaev-type spin liquid on a
  quasicrystal},\ }\href {https://doi.org/10.1103/PhysRevB.108.104208}
  {\bibfield  {journal} {\bibinfo  {journal} {Phys. Rev. B}\ }\textbf {\bibinfo
  {volume} {108}},\ \bibinfo {pages} {104208} (\bibinfo {year}
  {2023})}\BibitemShut {NoStop}%
\bibitem [{\citenamefont {Vijayvargia}\ \emph {et~al.}(2025)\citenamefont
  {Vijayvargia}, \citenamefont {Day-Roberts}, \citenamefont {Botana},\ and\
  \citenamefont {Erten}}]{km2j-3zy2}%
  \BibitemOpen
  \bibfield  {author} {\bibinfo {author} {\bibfnamefont {A.}~\bibnamefont
  {Vijayvargia}}, \bibinfo {author} {\bibfnamefont {E.}~\bibnamefont
  {Day-Roberts}}, \bibinfo {author} {\bibfnamefont {A.~S.}\ \bibnamefont
  {Botana}},\ and\ \bibinfo {author} {\bibfnamefont {O.}~\bibnamefont
  {Erten}},\ }\bibfield  {title} {\bibinfo {title} {Altermagnets with
  topological order in {K}itaev bilayers},\ }\href
  {https://doi.org/10.1103/km2j-3zy2} {\bibfield  {journal} {\bibinfo
  {journal} {Phys. Rev. Lett.}\ }\textbf {\bibinfo {volume} {135}},\ \bibinfo
  {pages} {166701} (\bibinfo {year} {2025})}\BibitemShut {NoStop}%
\bibitem [{\citenamefont {Lieb}(1994)}]{PhysRevLett.73.2158}%
  \BibitemOpen
  \bibfield  {author} {\bibinfo {author} {\bibfnamefont {E.~H.}\ \bibnamefont
  {Lieb}},\ }\bibfield  {title} {\bibinfo {title} {Flux phase of the
  half-filled band},\ }\href {https://doi.org/10.1103/PhysRevLett.73.2158}
  {\bibfield  {journal} {\bibinfo  {journal} {Phys. Rev. Lett.}\ }\textbf
  {\bibinfo {volume} {73}},\ \bibinfo {pages} {2158} (\bibinfo {year}
  {1994})}\BibitemShut {NoStop}%
\bibitem [{\citenamefont {Macris}\ and\ \citenamefont
  {Nachtergaele}(1996)}]{Macris1996}%
  \BibitemOpen
  \bibfield  {author} {\bibinfo {author} {\bibfnamefont {N.}~\bibnamefont
  {Macris}}\ and\ \bibinfo {author} {\bibfnamefont {B.}~\bibnamefont
  {Nachtergaele}},\ }\bibfield  {title} {\bibinfo {title} {On the flux phase
  conjecture at half-filling: An improved proof},\ }\href
  {https://doi.org/10.1007/BF02199361} {\bibfield  {journal} {\bibinfo
  {journal} {Journal of Statistical Physics}\ }\textbf {\bibinfo {volume}
  {85}},\ \bibinfo {pages} {745} (\bibinfo {year} {1996})}\BibitemShut
  {NoStop}%
\bibitem [{\citenamefont {Nasu}\ \emph {et~al.}(2014)\citenamefont {Nasu},
  \citenamefont {Udagawa},\ and\ \citenamefont
  {Motome}}]{PhysRevLett.113.197205}%
  \BibitemOpen
  \bibfield  {author} {\bibinfo {author} {\bibfnamefont {J.}~\bibnamefont
  {Nasu}}, \bibinfo {author} {\bibfnamefont {M.}~\bibnamefont {Udagawa}},\ and\
  \bibinfo {author} {\bibfnamefont {Y.}~\bibnamefont {Motome}},\ }\bibfield
  {title} {\bibinfo {title} {Vaporization of {K}itaev spin liquids},\ }\href
  {https://doi.org/10.1103/PhysRevLett.113.197205} {\bibfield  {journal}
  {\bibinfo  {journal} {Phys. Rev. Lett.}\ }\textbf {\bibinfo {volume} {113}},\
  \bibinfo {pages} {197205} (\bibinfo {year} {2014})}\BibitemShut {NoStop}%
\bibitem [{\citenamefont {Nasu}\ and\ \citenamefont
  {Motome}(2015)}]{PhysRevLett.115.087203}%
  \BibitemOpen
  \bibfield  {author} {\bibinfo {author} {\bibfnamefont {J.}~\bibnamefont
  {Nasu}}\ and\ \bibinfo {author} {\bibfnamefont {Y.}~\bibnamefont {Motome}},\
  }\bibfield  {title} {\bibinfo {title} {Thermodynamics of chiral spin liquids
  with {A}belian and non-{A}belian anyons},\ }\href
  {https://doi.org/10.1103/PhysRevLett.115.087203} {\bibfield  {journal}
  {\bibinfo  {journal} {Phys. Rev. Lett.}\ }\textbf {\bibinfo {volume} {115}},\
  \bibinfo {pages} {087203} (\bibinfo {year} {2015})}\BibitemShut {NoStop}%
\bibitem [{\citenamefont {Nasu}\ \emph {et~al.}(2015)\citenamefont {Nasu},
  \citenamefont {Udagawa},\ and\ \citenamefont {Motome}}]{PhysRevB.92.115122}%
  \BibitemOpen
  \bibfield  {author} {\bibinfo {author} {\bibfnamefont {J.}~\bibnamefont
  {Nasu}}, \bibinfo {author} {\bibfnamefont {M.}~\bibnamefont {Udagawa}},\ and\
  \bibinfo {author} {\bibfnamefont {Y.}~\bibnamefont {Motome}},\ }\bibfield
  {title} {\bibinfo {title} {Thermal fractionalization of quantum spins in a
  {K}itaev model: Temperature-linear specific heat and coherent transport of
  {M}ajorana fermions},\ }\href {https://doi.org/10.1103/PhysRevB.92.115122}
  {\bibfield  {journal} {\bibinfo  {journal} {Phys. Rev. B}\ }\textbf {\bibinfo
  {volume} {92}},\ \bibinfo {pages} {115122} (\bibinfo {year}
  {2015})}\BibitemShut {NoStop}%
\bibitem [{\citenamefont {Nasu}\ \emph {et~al.}(2016)\citenamefont {Nasu},
  \citenamefont {Knolle}, \citenamefont {Kovrizhin}, \citenamefont {Motome},\
  and\ \citenamefont {Moessner}}]{nphys3809}%
  \BibitemOpen
  \bibfield  {author} {\bibinfo {author} {\bibfnamefont {J.}~\bibnamefont
  {Nasu}}, \bibinfo {author} {\bibfnamefont {J.}~\bibnamefont {Knolle}},
  \bibinfo {author} {\bibfnamefont {D.~L.}\ \bibnamefont {Kovrizhin}}, \bibinfo
  {author} {\bibfnamefont {Y.}~\bibnamefont {Motome}},\ and\ \bibinfo {author}
  {\bibfnamefont {R.}~\bibnamefont {Moessner}},\ }\bibfield  {title} {\bibinfo
  {title} {Fermionic response from fractionalization in an insulating
  two-dimensional magnet},\ }\href {https://doi.org/10.1038/nphys3809}
  {\bibfield  {journal} {\bibinfo  {journal} {Nature Physics}\ }\textbf
  {\bibinfo {volume} {12}},\ \bibinfo {pages} {912} (\bibinfo {year}
  {2016})}\BibitemShut {NoStop}%
\bibitem [{\citenamefont {Mishchenko}\ \emph {et~al.}(2017)\citenamefont
  {Mishchenko}, \citenamefont {Kato},\ and\ \citenamefont
  {Motome}}]{PhysRevB.96.125124}%
  \BibitemOpen
  \bibfield  {author} {\bibinfo {author} {\bibfnamefont {P.~A.}\ \bibnamefont
  {Mishchenko}}, \bibinfo {author} {\bibfnamefont {Y.}~\bibnamefont {Kato}},\
  and\ \bibinfo {author} {\bibfnamefont {Y.}~\bibnamefont {Motome}},\
  }\bibfield  {title} {\bibinfo {title} {Finite-temperature phase transition to
  a {K}itaev spin liquid phase on a hyperoctagon lattice: A large-scale quantum
  {M}onte {C}arlo study},\ }\href {https://doi.org/10.1103/PhysRevB.96.125124}
  {\bibfield  {journal} {\bibinfo  {journal} {Phys. Rev. B}\ }\textbf {\bibinfo
  {volume} {96}},\ \bibinfo {pages} {125124} (\bibinfo {year}
  {2017})}\BibitemShut {NoStop}%
\bibitem [{\citenamefont {Dwivedi}\ \emph {et~al.}(2018)\citenamefont
  {Dwivedi}, \citenamefont {Hickey}, \citenamefont {Eschmann},\ and\
  \citenamefont {Trebst}}]{PhysRevB.98.054432}%
  \BibitemOpen
  \bibfield  {author} {\bibinfo {author} {\bibfnamefont {V.}~\bibnamefont
  {Dwivedi}}, \bibinfo {author} {\bibfnamefont {C.}~\bibnamefont {Hickey}},
  \bibinfo {author} {\bibfnamefont {T.}~\bibnamefont {Eschmann}},\ and\
  \bibinfo {author} {\bibfnamefont {S.}~\bibnamefont {Trebst}},\ }\bibfield
  {title} {\bibinfo {title} {Majorana corner modes in a second-order {K}itaev
  spin liquid},\ }\href {https://doi.org/10.1103/PhysRevB.98.054432} {\bibfield
   {journal} {\bibinfo  {journal} {Phys. Rev. B}\ }\textbf {\bibinfo {volume}
  {98}},\ \bibinfo {pages} {054432} (\bibinfo {year} {2018})}\BibitemShut
  {NoStop}%
\bibitem [{\citenamefont {Eschmann}\ \emph {et~al.}(2019)\citenamefont
  {Eschmann}, \citenamefont {Mishchenko}, \citenamefont {Bojesen},
  \citenamefont {Kato}, \citenamefont {Hermanns}, \citenamefont {Motome},\ and\
  \citenamefont {Trebst}}]{PhysRevResearch.1.032011}%
  \BibitemOpen
  \bibfield  {author} {\bibinfo {author} {\bibfnamefont {T.}~\bibnamefont
  {Eschmann}}, \bibinfo {author} {\bibfnamefont {P.~A.}\ \bibnamefont
  {Mishchenko}}, \bibinfo {author} {\bibfnamefont {T.~A.}\ \bibnamefont
  {Bojesen}}, \bibinfo {author} {\bibfnamefont {Y.}~\bibnamefont {Kato}},
  \bibinfo {author} {\bibfnamefont {M.}~\bibnamefont {Hermanns}}, \bibinfo
  {author} {\bibfnamefont {Y.}~\bibnamefont {Motome}},\ and\ \bibinfo {author}
  {\bibfnamefont {S.}~\bibnamefont {Trebst}},\ }\bibfield  {title} {\bibinfo
  {title} {Thermodynamics of a gauge-frustrated {K}itaev spin liquid},\ }\href
  {https://doi.org/10.1103/PhysRevResearch.1.032011} {\bibfield  {journal}
  {\bibinfo  {journal} {Phys. Rev. Res.}\ }\textbf {\bibinfo {volume} {1}},\
  \bibinfo {pages} {032011} (\bibinfo {year} {2019})}\BibitemShut {NoStop}%
\bibitem [{\citenamefont {Mishchenko}\ \emph {et~al.}(2020)\citenamefont
  {Mishchenko}, \citenamefont {Kato}, \citenamefont {O'Brien}, \citenamefont
  {Bojesen}, \citenamefont {Eschmann}, \citenamefont {Hermanns}, \citenamefont
  {Trebst},\ and\ \citenamefont {Motome}}]{PhysRevB.101.045118}%
  \BibitemOpen
  \bibfield  {author} {\bibinfo {author} {\bibfnamefont {P.~A.}\ \bibnamefont
  {Mishchenko}}, \bibinfo {author} {\bibfnamefont {Y.}~\bibnamefont {Kato}},
  \bibinfo {author} {\bibfnamefont {K.}~\bibnamefont {O'Brien}}, \bibinfo
  {author} {\bibfnamefont {T.~A.}\ \bibnamefont {Bojesen}}, \bibinfo {author}
  {\bibfnamefont {T.}~\bibnamefont {Eschmann}}, \bibinfo {author}
  {\bibfnamefont {M.}~\bibnamefont {Hermanns}}, \bibinfo {author}
  {\bibfnamefont {S.}~\bibnamefont {Trebst}},\ and\ \bibinfo {author}
  {\bibfnamefont {Y.}~\bibnamefont {Motome}},\ }\bibfield  {title} {\bibinfo
  {title} {Chiral spin liquids with crystalline {$\mathbb{Z}_{2}$} gauge order
  in a three-dimensional {K}itaev model},\ }\href
  {https://doi.org/10.1103/PhysRevB.101.045118} {\bibfield  {journal} {\bibinfo
   {journal} {Phys. Rev. B}\ }\textbf {\bibinfo {volume} {101}},\ \bibinfo
  {pages} {045118} (\bibinfo {year} {2020})}\BibitemShut {NoStop}%
\bibitem [{\citenamefont {Eschmann}(2020)}]{eschmannthesis}%
  \BibitemOpen
  \bibfield  {author} {\bibinfo {author} {\bibfnamefont {T.}~\bibnamefont
  {Eschmann}},\ }\emph {\bibinfo {title} {Thermodynamics of Kitaev Spin
  Liquids}},\ \href@noop {} {Ph.D. thesis},\ \bibinfo  {school} {University of
  Cologne} (\bibinfo {year} {2020})\BibitemShut {NoStop}%
\bibitem [{\citenamefont {Eschmann}\ \emph
  {et~al.}(2020{\natexlab{a}})\citenamefont {Eschmann}, \citenamefont
  {Mishchenko}, \citenamefont {O'Brien}, \citenamefont {Bojesen}, \citenamefont
  {Kato}, \citenamefont {Hermanns}, \citenamefont {Motome},\ and\ \citenamefont
  {Trebst}}]{PhysRevB.102.075125}%
  \BibitemOpen
  \bibfield  {author} {\bibinfo {author} {\bibfnamefont {T.}~\bibnamefont
  {Eschmann}}, \bibinfo {author} {\bibfnamefont {P.~A.}\ \bibnamefont
  {Mishchenko}}, \bibinfo {author} {\bibfnamefont {K.}~\bibnamefont {O'Brien}},
  \bibinfo {author} {\bibfnamefont {T.~A.}\ \bibnamefont {Bojesen}}, \bibinfo
  {author} {\bibfnamefont {Y.}~\bibnamefont {Kato}}, \bibinfo {author}
  {\bibfnamefont {M.}~\bibnamefont {Hermanns}}, \bibinfo {author}
  {\bibfnamefont {Y.}~\bibnamefont {Motome}},\ and\ \bibinfo {author}
  {\bibfnamefont {S.}~\bibnamefont {Trebst}},\ }\bibfield  {title} {\bibinfo
  {title} {Thermodynamic classification of three-dimensional {K}itaev spin
  liquids},\ }\href {https://doi.org/10.1103/PhysRevB.102.075125} {\bibfield
  {journal} {\bibinfo  {journal} {Phys. Rev. B}\ }\textbf {\bibinfo {volume}
  {102}},\ \bibinfo {pages} {075125} (\bibinfo {year}
  {2020}{\natexlab{a}})}\BibitemShut {NoStop}%
\bibitem [{\citenamefont {Eschmann}\ \emph
  {et~al.}(2020{\natexlab{b}})\citenamefont {Eschmann}, \citenamefont
  {Dwivedi}, \citenamefont {Legg}, \citenamefont {Hickey},\ and\ \citenamefont
  {Trebst}}]{PhysRevResearch.2.043159}%
  \BibitemOpen
  \bibfield  {author} {\bibinfo {author} {\bibfnamefont {T.}~\bibnamefont
  {Eschmann}}, \bibinfo {author} {\bibfnamefont {V.}~\bibnamefont {Dwivedi}},
  \bibinfo {author} {\bibfnamefont {H.~F.}\ \bibnamefont {Legg}}, \bibinfo
  {author} {\bibfnamefont {C.}~\bibnamefont {Hickey}},\ and\ \bibinfo {author}
  {\bibfnamefont {S.}~\bibnamefont {Trebst}},\ }\bibfield  {title} {\bibinfo
  {title} {Partial flux ordering and thermal {M}ajorana metals in higher-order
  spin liquids},\ }\href {https://doi.org/10.1103/PhysRevResearch.2.043159}
  {\bibfield  {journal} {\bibinfo  {journal} {Phys. Rev. Res.}\ }\textbf
  {\bibinfo {volume} {2}},\ \bibinfo {pages} {043159} (\bibinfo {year}
  {2020}{\natexlab{b}})}\BibitemShut {NoStop}%
\bibitem [{\citenamefont {Wei\ss{}e}\ \emph {et~al.}(2006)\citenamefont
  {Wei\ss{}e}, \citenamefont {Wellein}, \citenamefont {Alvermann},\ and\
  \citenamefont {Fehske}}]{RevModPhys.78.275}%
  \BibitemOpen
  \bibfield  {author} {\bibinfo {author} {\bibfnamefont {A.}~\bibnamefont
  {Wei\ss{}e}}, \bibinfo {author} {\bibfnamefont {G.}~\bibnamefont {Wellein}},
  \bibinfo {author} {\bibfnamefont {A.}~\bibnamefont {Alvermann}},\ and\
  \bibinfo {author} {\bibfnamefont {H.}~\bibnamefont {Fehske}},\ }\bibfield
  {title} {\bibinfo {title} {The kernel polynomial method},\ }\href
  {https://doi.org/10.1103/RevModPhys.78.275} {\bibfield  {journal} {\bibinfo
  {journal} {Rev. Mod. Phys.}\ }\textbf {\bibinfo {volume} {78}},\ \bibinfo
  {pages} {275} (\bibinfo {year} {2006})}\BibitemShut {NoStop}%
\bibitem [{\citenamefont {Wei\ss{}e}(2009)}]{PhysRevLett.102.150604}%
  \BibitemOpen
  \bibfield  {author} {\bibinfo {author} {\bibfnamefont {A.}~\bibnamefont
  {Wei\ss{}e}},\ }\bibfield  {title} {\bibinfo {title} {Green-function-based
  {M}onte {C}arlo method for classical fields coupled to fermions},\ }\href
  {https://doi.org/10.1103/PhysRevLett.102.150604} {\bibfield  {journal}
  {\bibinfo  {journal} {Phys. Rev. Lett.}\ }\textbf {\bibinfo {volume} {102}},\
  \bibinfo {pages} {150604} (\bibinfo {year} {2009})}\BibitemShut {NoStop}%
\bibitem [{\citenamefont {Cassella}\ \emph {et~al.}(2023)\citenamefont
  {Cassella}, \citenamefont {d'Ornellas}, \citenamefont {Hodson}, \citenamefont
  {Natori},\ and\ \citenamefont {Knolle}}]{s41467-023-42105-9}%
  \BibitemOpen
  \bibfield  {author} {\bibinfo {author} {\bibfnamefont {G.}~\bibnamefont
  {Cassella}}, \bibinfo {author} {\bibfnamefont {P.}~\bibnamefont
  {d'Ornellas}}, \bibinfo {author} {\bibfnamefont {T.}~\bibnamefont {Hodson}},
  \bibinfo {author} {\bibfnamefont {W.~M.~H.}\ \bibnamefont {Natori}},\ and\
  \bibinfo {author} {\bibfnamefont {J.}~\bibnamefont {Knolle}},\ }\bibfield
  {title} {\bibinfo {title} {An exact chiral amorphous spin liquid},\ }\href
  {https://doi.org/10.1038/s41467-023-42105-9} {\bibfield  {journal} {\bibinfo
  {journal} {Nature Communications}\ }\textbf {\bibinfo {volume} {14}},\
  \bibinfo {pages} {6663} (\bibinfo {year} {2023})}\BibitemShut {NoStop}%
\bibitem [{\citenamefont {Bernevig}\ and\ \citenamefont
  {Hughes}(2013)}]{bernevigtextbook}%
  \BibitemOpen
  \bibfield  {author} {\bibinfo {author} {\bibfnamefont {B.~A.}\ \bibnamefont
  {Bernevig}}\ and\ \bibinfo {author} {\bibfnamefont {T.~L.}\ \bibnamefont
  {Hughes}},\ }\href@noop {} {\emph {\bibinfo {title} {Topological Insulators
  and Topological Superconductors}}}\ (\bibinfo  {publisher} {Princeton
  University Press},\ \bibinfo {address} {Princeton, New Jersey},\ \bibinfo
  {year} {2013})\BibitemShut {NoStop}%
\bibitem [{\citenamefont {Chern}\ \emph {et~al.}()\citenamefont {Chern},
  \citenamefont {Moessner},\ and\ \citenamefont {Castelnovo}}]{2607.12027}%
  \BibitemOpen
  \bibfield  {author} {\bibinfo {author} {\bibfnamefont {L.~E.}\ \bibnamefont
  {Chern}}, \bibinfo {author} {\bibfnamefont {R.}~\bibnamefont {Moessner}},\
  and\ \bibinfo {author} {\bibfnamefont {C.}~\bibnamefont {Castelnovo}},\
  }\bibfield  {title} {\bibinfo {title} {Mapping vortices to anyons in toric
  code phases of generalized {K}itaev models},\ }\href
  {https://arxiv.org/abs/2607.12027} {\ }\Eprint
  {https://arxiv.org/abs/2607.12027} {arXiv:2607.12027} \BibitemShut {NoStop}%
\bibitem [{\citenamefont {Wootton}(2015)}]{Wootton_2015}%
  \BibitemOpen
  \bibfield  {author} {\bibinfo {author} {\bibfnamefont {J.~R.}\ \bibnamefont
  {Wootton}},\ }\bibfield  {title} {\bibinfo {title} {A family of stabilizer
  codes for {$D(\mathbb{Z}_2)$} anyons and {M}ajorana modes},\ }\href
  {https://doi.org/10.1088/1751-8113/48/21/215302} {\bibfield  {journal}
  {\bibinfo  {journal} {Journal of Physics A: Mathematical and Theoretical}\
  }\textbf {\bibinfo {volume} {48}},\ \bibinfo {pages} {215302} (\bibinfo
  {year} {2015})}\BibitemShut {NoStop}%
\bibitem [{\citenamefont {Kitaev}(2003)}]{KITAEV20032}%
  \BibitemOpen
  \bibfield  {author} {\bibinfo {author} {\bibfnamefont {A.}~\bibnamefont
  {Kitaev}},\ }\bibfield  {title} {\bibinfo {title} {Fault-tolerant quantum
  computation by anyons},\ }\href
  {https://doi.org/10.1016/S0003-4916(02)00018-0} {\bibfield  {journal}
  {\bibinfo  {journal} {Annals of Physics}\ }\textbf {\bibinfo {volume}
  {303}},\ \bibinfo {pages} {2} (\bibinfo {year} {2003})}\BibitemShut {NoStop}%
\bibitem [{sup()}]{supply}%
  \BibitemOpen
  \href@noop {} {}\bibinfo {note} {See the Supplemental Material at [URL will
  be inserted by publisher] for details of the quantum Monte Carlo simulations
  and the analysis of band topology, as well as additional results on the toric
  code phase of the Kitaev maple-leaf model.}\BibitemShut {Stop}%
\bibitem [{Note2()}]{Note2}%
  \BibitemOpen
  \bibinfo {note} {For generic $n \geq 2$, the $2n-1$ local $\Gamma $ matrices
  can be represented in terms of $2n$ species of Majorana fermions \cite
  {PhysRevB.79.134427}.}\BibitemShut {Stop}%
\bibitem [{Note3()}]{Note3}%
  \BibitemOpen
  \bibinfo {note} {A more accurate name for $u_{ij}^\lambda $ would be the
  \protect \textit {gauge vector potential}. Here, we follow the more common
  convention in the literature and call $u_{ij}^\lambda $ the \protect \textit
  {gauge field}.}\BibitemShut {Stop}%
\bibitem [{Note4()}]{Note4}%
  \BibitemOpen
  \bibinfo {note} {It follows that the product of $\Gamma $ matrices involved
  in the pairwise interactions along any closed path, known as the Wilson loop,
  is also gauge invariant.}\BibitemShut {Stop}%
\bibitem [{\citenamefont {Motome}\ and\ \citenamefont
  {Furukawa}(1999)}]{JPSJ.68.3853}%
  \BibitemOpen
  \bibfield  {author} {\bibinfo {author} {\bibfnamefont {Y.}~\bibnamefont
  {Motome}}\ and\ \bibinfo {author} {\bibfnamefont {N.}~\bibnamefont
  {Furukawa}},\ }\bibfield  {title} {\bibinfo {title} {A {M}onte {C}arlo method
  for fermion systems coupled with classical degrees of freedom},\ }\href
  {https://doi.org/10.1143/JPSJ.68.3853} {\bibfield  {journal} {\bibinfo
  {journal} {Journal of the Physical Society of Japan}\ }\textbf {\bibinfo
  {volume} {68}},\ \bibinfo {pages} {3853} (\bibinfo {year}
  {1999})}\BibitemShut {NoStop}%
\bibitem [{Note5()}]{Note5}%
  \BibitemOpen
  \bibinfo {note} {The issue of domain walls is discussed in the Supplemental
  Material. The reader should also bear in mind that our conclusions are of
  course drawn from simulations performed only for a finite number of
  parameters.}\BibitemShut {Stop}%
\bibitem [{\citenamefont {Motome}\ and\ \citenamefont
  {Nasu}(2020)}]{JPSJ.89.012002}%
  \BibitemOpen
  \bibfield  {author} {\bibinfo {author} {\bibfnamefont {Y.}~\bibnamefont
  {Motome}}\ and\ \bibinfo {author} {\bibfnamefont {J.}~\bibnamefont {Nasu}},\
  }\bibfield  {title} {\bibinfo {title} {Hunting {M}ajorana fermions in
  {K}itaev magnets},\ }\href {https://doi.org/10.7566/JPSJ.89.012002}
  {\bibfield  {journal} {\bibinfo  {journal} {Journal of the Physical Society
  of Japan}\ }\textbf {\bibinfo {volume} {89}},\ \bibinfo {pages} {012002}
  (\bibinfo {year} {2020})}\BibitemShut {NoStop}%
\bibitem [{Note6()}]{Note6}%
  \BibitemOpen
  \bibinfo {note} {The gauge fields themselves are unphysical, but different
  gauge-field configurations yielding the same gauge-flux configuration must
  give the same set of physical quantities, such as the energy spectrum of the
  itinerant fermions and the Chern number indicating the net chiral edge
  mode(s).}\BibitemShut {Stop}%
\bibitem [{Note7()}]{Note7}%
  \BibitemOpen
  \bibinfo {note} {We also verified by explicit calculation that the fermion
  gap remains invariant, while the Chern number changes sign, upon reversing
  $W_p$ from $+i$ to $-i$ for all unit triangles.}\BibitemShut {Stop}%
\bibitem [{\citenamefont {Fukui}\ \emph {et~al.}(2005)\citenamefont {Fukui},
  \citenamefont {Hatsugai},\ and\ \citenamefont {Suzuki}}]{JPSJ.74.1674}%
  \BibitemOpen
  \bibfield  {author} {\bibinfo {author} {\bibfnamefont {T.}~\bibnamefont
  {Fukui}}, \bibinfo {author} {\bibfnamefont {Y.}~\bibnamefont {Hatsugai}},\
  and\ \bibinfo {author} {\bibfnamefont {H.}~\bibnamefont {Suzuki}},\
  }\bibfield  {title} {\bibinfo {title} {{C}hern numbers in discretized
  {B}rillouin zone: Efficient method of computing (spin) {H}all conductances},\
  }\href {https://doi.org/10.1143/JPSJ.74.1674} {\bibfield  {journal} {\bibinfo
   {journal} {Journal of the Physical Society of Japan}\ }\textbf {\bibinfo
  {volume} {74}},\ \bibinfo {pages} {1674} (\bibinfo {year}
  {2005})}\BibitemShut {NoStop}%
\bibitem [{Note8()}]{Note8}%
  \BibitemOpen
  \bibinfo {note} {Fig.~\ref {figure:trellisgap} suggests that the system is
  also gapless on segments of the boundary of the triangular parameter space
  where $J_\lambda = 0$ for some $\lambda $, which are excluded from our
  considerations as the lattice geometry would be altered by missing
  bonds.}\BibitemShut {Stop}%
\bibitem [{Note9()}]{Note9}%
  \BibitemOpen
  \bibinfo {note} {They are originally known as the electric charge and the
  magnetic vortex. To avoid confusion, we reserve the terminology ``vortex''
  for a generic flux excitation, which can be either $e$ or $m$, rather than
  specifically referring to $m$ with it.}\BibitemShut {Stop}%
\bibitem [{\citenamefont {Petrova}\ \emph {et~al.}(2014)\citenamefont
  {Petrova}, \citenamefont {Mellado},\ and\ \citenamefont
  {Tchernyshyov}}]{PhysRevB.90.134404}%
  \BibitemOpen
  \bibfield  {author} {\bibinfo {author} {\bibfnamefont {O.}~\bibnamefont
  {Petrova}}, \bibinfo {author} {\bibfnamefont {P.}~\bibnamefont {Mellado}},\
  and\ \bibinfo {author} {\bibfnamefont {O.}~\bibnamefont {Tchernyshyov}},\
  }\bibfield  {title} {\bibinfo {title} {Unpaired {M}ajorana modes on
  dislocations and string defects in {K}itaev's honeycomb model},\ }\href
  {https://doi.org/10.1103/PhysRevB.90.134404} {\bibfield  {journal} {\bibinfo
  {journal} {Phys. Rev. B}\ }\textbf {\bibinfo {volume} {90}},\ \bibinfo
  {pages} {134404} (\bibinfo {year} {2014})}\BibitemShut {NoStop}%
\bibitem [{\citenamefont {Simon}(2023)}]{simontextbook}%
  \BibitemOpen
  \bibfield  {author} {\bibinfo {author} {\bibfnamefont {S.~H.}\ \bibnamefont
  {Simon}},\ }\href@noop {} {\emph {\bibinfo {title} {Topological Quantum}}}\
  (\bibinfo  {publisher} {Oxford University Press},\ \bibinfo {year}
  {2023})\BibitemShut {NoStop}%
\bibitem [{\citenamefont {Zhang}\ \emph {et~al.}(2020)\citenamefont {Zhang},
  \citenamefont {Batista},\ and\ \citenamefont
  {Hal\'asz}}]{PhysRevResearch.2.023334}%
  \BibitemOpen
  \bibfield  {author} {\bibinfo {author} {\bibfnamefont {S.-S.}\ \bibnamefont
  {Zhang}}, \bibinfo {author} {\bibfnamefont {C.~D.}\ \bibnamefont {Batista}},\
  and\ \bibinfo {author} {\bibfnamefont {G.~B.}\ \bibnamefont {Hal\'asz}},\
  }\bibfield  {title} {\bibinfo {title} {Toward {K}itaev's sixteenfold way in a
  honeycomb lattice model},\ }\href
  {https://doi.org/10.1103/PhysRevResearch.2.023334} {\bibfield  {journal}
  {\bibinfo  {journal} {Phys. Rev. Res.}\ }\textbf {\bibinfo {volume} {2}},\
  \bibinfo {pages} {023334} (\bibinfo {year} {2020})}\BibitemShut {NoStop}%
\bibitem [{\citenamefont {Pedrocchi}\ \emph {et~al.}(2011)\citenamefont
  {Pedrocchi}, \citenamefont {Chesi},\ and\ \citenamefont
  {Loss}}]{PhysRevB.84.165414}%
  \BibitemOpen
  \bibfield  {author} {\bibinfo {author} {\bibfnamefont {F.~L.}\ \bibnamefont
  {Pedrocchi}}, \bibinfo {author} {\bibfnamefont {S.}~\bibnamefont {Chesi}},\
  and\ \bibinfo {author} {\bibfnamefont {D.}~\bibnamefont {Loss}},\ }\bibfield
  {title} {\bibinfo {title} {Physical solutions of the {K}itaev honeycomb
  model},\ }\href {https://doi.org/10.1103/PhysRevB.84.165414} {\bibfield
  {journal} {\bibinfo  {journal} {Phys. Rev. B}\ }\textbf {\bibinfo {volume}
  {84}},\ \bibinfo {pages} {165414} (\bibinfo {year} {2011})}\BibitemShut
  {NoStop}%
\bibitem [{\citenamefont {Zschocke}\ and\ \citenamefont
  {Vojta}(2015)}]{PhysRevB.92.014403}%
  \BibitemOpen
  \bibfield  {author} {\bibinfo {author} {\bibfnamefont {F.}~\bibnamefont
  {Zschocke}}\ and\ \bibinfo {author} {\bibfnamefont {M.}~\bibnamefont
  {Vojta}},\ }\bibfield  {title} {\bibinfo {title} {Physical states and
  finite-size effects in {K}itaev's honeycomb model: Bond disorder, spin
  excitations, and {NMR} line shape},\ }\href
  {https://doi.org/10.1103/PhysRevB.92.014403} {\bibfield  {journal} {\bibinfo
  {journal} {Phys. Rev. B}\ }\textbf {\bibinfo {volume} {92}},\ \bibinfo
  {pages} {014403} (\bibinfo {year} {2015})}\BibitemShut {NoStop}%
\bibitem [{Note10()}]{Note10}%
  \BibitemOpen
  \bibinfo {note} {Ideally, the distance between the vortices should be much
  greater than the correlation length, such that their braiding statistics is
  well-defined.}\BibitemShut {Stop}%
\bibitem [{Note11()}]{Note11}%
  \BibitemOpen
  \bibinfo {note} {We refer interested readers to Ref.~\protect \rev@citealp
  {2607.12027} for a detailed discussion of Proposition 2.}\BibitemShut {Stop}%
\bibitem [{\citenamefont {Kasahara}\ \emph {et~al.}(2018)\citenamefont
  {Kasahara}, \citenamefont {Ohnishi}, \citenamefont {Mizukami}, \citenamefont
  {Tanaka}, \citenamefont {Ma}, \citenamefont {Sugii}, \citenamefont {Kurita},
  \citenamefont {Tanaka}, \citenamefont {Nasu}, \citenamefont {Motome},
  \citenamefont {Shibauchi},\ and\ \citenamefont
  {Matsuda}}]{s41586-018-0274-0}%
  \BibitemOpen
  \bibfield  {author} {\bibinfo {author} {\bibfnamefont {Y.}~\bibnamefont
  {Kasahara}}, \bibinfo {author} {\bibfnamefont {T.}~\bibnamefont {Ohnishi}},
  \bibinfo {author} {\bibfnamefont {Y.}~\bibnamefont {Mizukami}}, \bibinfo
  {author} {\bibfnamefont {O.}~\bibnamefont {Tanaka}}, \bibinfo {author}
  {\bibfnamefont {S.}~\bibnamefont {Ma}}, \bibinfo {author} {\bibfnamefont
  {K.}~\bibnamefont {Sugii}}, \bibinfo {author} {\bibfnamefont
  {N.}~\bibnamefont {Kurita}}, \bibinfo {author} {\bibfnamefont
  {H.}~\bibnamefont {Tanaka}}, \bibinfo {author} {\bibfnamefont
  {J.}~\bibnamefont {Nasu}}, \bibinfo {author} {\bibfnamefont {Y.}~\bibnamefont
  {Motome}}, \bibinfo {author} {\bibfnamefont {T.}~\bibnamefont {Shibauchi}},\
  and\ \bibinfo {author} {\bibfnamefont {Y.}~\bibnamefont {Matsuda}},\
  }\bibfield  {title} {\bibinfo {title} {Majorana quantization and half-integer
  thermal quantum {H}all effect in a {K}itaev spin liquid},\ }\href
  {https://doi.org/10.1038/s41586-018-0274-0} {\bibfield  {journal} {\bibinfo
  {journal} {Nature}\ }\textbf {\bibinfo {volume} {559}},\ \bibinfo {pages}
  {227} (\bibinfo {year} {2018})}\BibitemShut {NoStop}%
\bibitem [{\citenamefont {Yokoi}\ \emph {et~al.}(2021)\citenamefont {Yokoi},
  \citenamefont {Ma}, \citenamefont {Kasahara}, \citenamefont {Kasahara},
  \citenamefont {Shibauchi}, \citenamefont {Kurita}, \citenamefont {Tanaka},
  \citenamefont {Nasu}, \citenamefont {Motome}, \citenamefont {Hickey},
  \citenamefont {Trebst},\ and\ \citenamefont {Matsuda}}]{science.aay5551}%
  \BibitemOpen
  \bibfield  {author} {\bibinfo {author} {\bibfnamefont {T.}~\bibnamefont
  {Yokoi}}, \bibinfo {author} {\bibfnamefont {S.}~\bibnamefont {Ma}}, \bibinfo
  {author} {\bibfnamefont {Y.}~\bibnamefont {Kasahara}}, \bibinfo {author}
  {\bibfnamefont {S.}~\bibnamefont {Kasahara}}, \bibinfo {author}
  {\bibfnamefont {T.}~\bibnamefont {Shibauchi}}, \bibinfo {author}
  {\bibfnamefont {N.}~\bibnamefont {Kurita}}, \bibinfo {author} {\bibfnamefont
  {H.}~\bibnamefont {Tanaka}}, \bibinfo {author} {\bibfnamefont
  {J.}~\bibnamefont {Nasu}}, \bibinfo {author} {\bibfnamefont {Y.}~\bibnamefont
  {Motome}}, \bibinfo {author} {\bibfnamefont {C.}~\bibnamefont {Hickey}},
  \bibinfo {author} {\bibfnamefont {S.}~\bibnamefont {Trebst}},\ and\ \bibinfo
  {author} {\bibfnamefont {Y.}~\bibnamefont {Matsuda}},\ }\bibfield  {title}
  {\bibinfo {title} {Half-integer quantized anomalous thermal {H}all effect in
  the {K}itaev material candidate {$\alpha$-RuCl$_3$}},\ }\href
  {https://doi.org/10.1126/science.aay5551} {\bibfield  {journal} {\bibinfo
  {journal} {Science}\ }\textbf {\bibinfo {volume} {373}},\ \bibinfo {pages}
  {568} (\bibinfo {year} {2021})}\BibitemShut {NoStop}%
\bibitem [{\citenamefont {Bruin}\ \emph {et~al.}(2022)\citenamefont {Bruin},
  \citenamefont {Claus}, \citenamefont {Matsumoto}, \citenamefont {Kurita},
  \citenamefont {Tanaka},\ and\ \citenamefont {Takagi}}]{s41567-021-01501-y}%
  \BibitemOpen
  \bibfield  {author} {\bibinfo {author} {\bibfnamefont {J.~A.~N.}\
  \bibnamefont {Bruin}}, \bibinfo {author} {\bibfnamefont {R.~R.}\ \bibnamefont
  {Claus}}, \bibinfo {author} {\bibfnamefont {Y.}~\bibnamefont {Matsumoto}},
  \bibinfo {author} {\bibfnamefont {N.}~\bibnamefont {Kurita}}, \bibinfo
  {author} {\bibfnamefont {H.}~\bibnamefont {Tanaka}},\ and\ \bibinfo {author}
  {\bibfnamefont {H.}~\bibnamefont {Takagi}},\ }\bibfield  {title} {\bibinfo
  {title} {Robustness of the thermal {H}all effect close to half-quantization
  in $\alpha$-{RuCl}$_3$},\ }\href {https://doi.org/10.1038/s41567-021-01501-y}
  {\bibfield  {journal} {\bibinfo  {journal} {Nature Physics}\ }\textbf
  {\bibinfo {volume} {18}},\ \bibinfo {pages} {401} (\bibinfo {year}
  {2022})}\BibitemShut {NoStop}%
\bibitem [{\citenamefont {Tanaka}\ \emph {et~al.}(2022)\citenamefont {Tanaka},
  \citenamefont {Mizukami}, \citenamefont {Harasawa}, \citenamefont
  {Hashimoto}, \citenamefont {Hwang}, \citenamefont {Kurita}, \citenamefont
  {Tanaka}, \citenamefont {Fujimoto}, \citenamefont {Matsuda}, \citenamefont
  {Moon},\ and\ \citenamefont {Shibauchi}}]{s41567-021-01488-6}%
  \BibitemOpen
  \bibfield  {author} {\bibinfo {author} {\bibfnamefont {O.}~\bibnamefont
  {Tanaka}}, \bibinfo {author} {\bibfnamefont {Y.}~\bibnamefont {Mizukami}},
  \bibinfo {author} {\bibfnamefont {R.}~\bibnamefont {Harasawa}}, \bibinfo
  {author} {\bibfnamefont {K.}~\bibnamefont {Hashimoto}}, \bibinfo {author}
  {\bibfnamefont {K.}~\bibnamefont {Hwang}}, \bibinfo {author} {\bibfnamefont
  {N.}~\bibnamefont {Kurita}}, \bibinfo {author} {\bibfnamefont
  {H.}~\bibnamefont {Tanaka}}, \bibinfo {author} {\bibfnamefont
  {S.}~\bibnamefont {Fujimoto}}, \bibinfo {author} {\bibfnamefont
  {Y.}~\bibnamefont {Matsuda}}, \bibinfo {author} {\bibfnamefont {E.-G.}\
  \bibnamefont {Moon}},\ and\ \bibinfo {author} {\bibfnamefont
  {T.}~\bibnamefont {Shibauchi}},\ }\bibfield  {title} {\bibinfo {title}
  {Thermodynamic evidence for a field-angle-dependent {M}ajorana gap in a
  {K}itaev spin liquid},\ }\href {https://doi.org/10.1038/s41567-021-01488-6}
  {\bibfield  {journal} {\bibinfo  {journal} {Nature Physics}\ }\textbf
  {\bibinfo {volume} {18}},\ \bibinfo {pages} {429} (\bibinfo {year}
  {2022})}\BibitemShut {NoStop}%
\bibitem [{\citenamefont {Czajka}\ \emph {et~al.}(2023)\citenamefont {Czajka},
  \citenamefont {Gao}, \citenamefont {Hirschberger}, \citenamefont
  {Lampen-Kelley}, \citenamefont {Banerjee}, \citenamefont {Quirk},
  \citenamefont {Mandrus}, \citenamefont {Nagler},\ and\ \citenamefont
  {Ong}}]{s41563-022-01397-w}%
  \BibitemOpen
  \bibfield  {author} {\bibinfo {author} {\bibfnamefont {P.}~\bibnamefont
  {Czajka}}, \bibinfo {author} {\bibfnamefont {T.}~\bibnamefont {Gao}},
  \bibinfo {author} {\bibfnamefont {M.}~\bibnamefont {Hirschberger}}, \bibinfo
  {author} {\bibfnamefont {P.}~\bibnamefont {Lampen-Kelley}}, \bibinfo {author}
  {\bibfnamefont {A.}~\bibnamefont {Banerjee}}, \bibinfo {author}
  {\bibfnamefont {N.}~\bibnamefont {Quirk}}, \bibinfo {author} {\bibfnamefont
  {D.~G.}\ \bibnamefont {Mandrus}}, \bibinfo {author} {\bibfnamefont {S.~E.}\
  \bibnamefont {Nagler}},\ and\ \bibinfo {author} {\bibfnamefont {N.~P.}\
  \bibnamefont {Ong}},\ }\bibfield  {title} {\bibinfo {title} {Planar thermal
  {H}all effect of topological bosons in the {K}itaev magnet
  {$\alpha$-RuCl$_3$}},\ }\href {https://doi.org/10.1038/s41563-022-01397-w}
  {\bibfield  {journal} {\bibinfo  {journal} {Nature Materials}\ }\textbf
  {\bibinfo {volume} {22}},\ \bibinfo {pages} {36} (\bibinfo {year}
  {2023})}\BibitemShut {NoStop}%
\bibitem [{\citenamefont {Matsuda}\ \emph {et~al.}(2025)\citenamefont
  {Matsuda}, \citenamefont {Shibauchi},\ and\ \citenamefont {Kee}}]{3m4m-3v59}%
  \BibitemOpen
  \bibfield  {author} {\bibinfo {author} {\bibfnamefont {Y.}~\bibnamefont
  {Matsuda}}, \bibinfo {author} {\bibfnamefont {T.}~\bibnamefont {Shibauchi}},\
  and\ \bibinfo {author} {\bibfnamefont {H.-Y.}\ \bibnamefont {Kee}},\
  }\bibfield  {title} {\bibinfo {title} {Kitaev quantum spin liquids},\ }\href
  {https://doi.org/10.1103/3m4m-3v59} {\bibfield  {journal} {\bibinfo
  {journal} {Rev. Mod. Phys.}\ }\textbf {\bibinfo {volume} {97}},\ \bibinfo
  {pages} {045003} (\bibinfo {year} {2025})}\BibitemShut {NoStop}%
\bibitem [{\citenamefont {Inui}\ and\ \citenamefont
  {Motome}(2024)}]{PhysRevResearch.6.033080}%
  \BibitemOpen
  \bibfield  {author} {\bibinfo {author} {\bibfnamefont {K.}~\bibnamefont
  {Inui}}\ and\ \bibinfo {author} {\bibfnamefont {Y.}~\bibnamefont {Motome}},\
  }\bibfield  {title} {\bibinfo {title} {Inverse {H}amiltonian design of highly
  entangled quantum systems},\ }\href
  {https://doi.org/10.1103/PhysRevResearch.6.033080} {\bibfield  {journal}
  {\bibinfo  {journal} {Phys. Rev. Res.}\ }\textbf {\bibinfo {volume} {6}},\
  \bibinfo {pages} {033080} (\bibinfo {year} {2024})}\BibitemShut {NoStop}%
\bibitem [{\citenamefont {Ramchandani}\ \emph {et~al.}(2025)\citenamefont
  {Ramchandani}, \citenamefont {Trebst},\ and\ \citenamefont
  {Hickey}}]{7psd-1zvr}%
  \BibitemOpen
  \bibfield  {author} {\bibinfo {author} {\bibfnamefont {S.}~\bibnamefont
  {Ramchandani}}, \bibinfo {author} {\bibfnamefont {S.}~\bibnamefont
  {Trebst}},\ and\ \bibinfo {author} {\bibfnamefont {C.}~\bibnamefont
  {Hickey}},\ }\bibfield  {title} {\bibinfo {title} {Constructing emergent
  {U(1)} symmetries in the ${\mathrm{\ensuremath{\Gamma}}}^{\ensuremath{'}}$
  model},\ }\href {https://doi.org/10.1103/7psd-1zvr} {\bibfield  {journal}
  {\bibinfo  {journal} {Phys. Rev. B}\ }\textbf {\bibinfo {volume} {112}},\
  \bibinfo {pages} {054444} (\bibinfo {year} {2025})}\BibitemShut {NoStop}%
\bibitem [{\citenamefont {Kugel'}\ and\ \citenamefont
  {Khomski\u{i}}(1982)}]{PU1982v025n04ABEH004537}%
  \BibitemOpen
  \bibfield  {author} {\bibinfo {author} {\bibfnamefont {K.~I.}\ \bibnamefont
  {Kugel'}}\ and\ \bibinfo {author} {\bibfnamefont {D.~I.}\ \bibnamefont
  {Khomski\u{i}}},\ }\bibfield  {title} {\bibinfo {title} {The {J}ahn-{T}eller
  effect and magnetism: transition metal compounds},\ }\href
  {https://doi.org/10.1070/PU1982v025n04ABEH004537} {\bibfield  {journal}
  {\bibinfo  {journal} {Soviet Physics Uspekhi}\ }\textbf {\bibinfo {volume}
  {25}},\ \bibinfo {pages} {231} (\bibinfo {year} {1982})}\BibitemShut
  {NoStop}%
\bibitem [{\citenamefont {Churchill}\ \emph {et~al.}(2025)\citenamefont
  {Churchill}, \citenamefont {Zhang},\ and\ \citenamefont
  {Kee}}]{s41535-025-00744-9}%
  \BibitemOpen
  \bibfield  {author} {\bibinfo {author} {\bibfnamefont {D.}~\bibnamefont
  {Churchill}}, \bibinfo {author} {\bibfnamefont {E.~Z.}\ \bibnamefont
  {Zhang}},\ and\ \bibinfo {author} {\bibfnamefont {H.-Y.}\ \bibnamefont
  {Kee}},\ }\bibfield  {title} {\bibinfo {title} {Microscopic roadmap to a
  {K}itaev-{Y}ao-{L}ee spin-orbital liquid},\ }\href
  {https://doi.org/10.1038/s41535-025-00744-9} {\bibfield  {journal} {\bibinfo
  {journal} {npj Quantum Materials}\ }\textbf {\bibinfo {volume} {10}},\
  \bibinfo {pages} {26} (\bibinfo {year} {2025})}\BibitemShut {NoStop}%
\bibitem [{\citenamefont {Sakurai}(1994)}]{sakuraitextbook}%
  \BibitemOpen
  \bibfield  {author} {\bibinfo {author} {\bibfnamefont {J.~J.}\ \bibnamefont
  {Sakurai}},\ }\href@noop {} {\emph {\bibinfo {title} {Modern Quantum
  Mechanics}}}\ (\bibinfo  {publisher} {Addison-Wesley},\ \bibinfo {year}
  {1994})\BibitemShut {NoStop}%
\bibitem [{\citenamefont {Jackeli}\ and\ \citenamefont
  {Khaliullin}(2009)}]{PhysRevLett.102.017205}%
  \BibitemOpen
  \bibfield  {author} {\bibinfo {author} {\bibfnamefont {G.}~\bibnamefont
  {Jackeli}}\ and\ \bibinfo {author} {\bibfnamefont {G.}~\bibnamefont
  {Khaliullin}},\ }\bibfield  {title} {\bibinfo {title} {Mott insulators in the
  strong spin-orbit coupling limit: From {H}eisenberg to a quantum compass and
  {K}itaev models},\ }\href {https://doi.org/10.1103/PhysRevLett.102.017205}
  {\bibfield  {journal} {\bibinfo  {journal} {Phys. Rev. Lett.}\ }\textbf
  {\bibinfo {volume} {102}},\ \bibinfo {pages} {017205} (\bibinfo {year}
  {2009})}\BibitemShut {NoStop}%
\bibitem [{\citenamefont {Huang}\ \emph {et~al.}(2014)\citenamefont {Huang},
  \citenamefont {Chen},\ and\ \citenamefont
  {Hermele}}]{PhysRevLett.112.167203}%
  \BibitemOpen
  \bibfield  {author} {\bibinfo {author} {\bibfnamefont {Y.-P.}\ \bibnamefont
  {Huang}}, \bibinfo {author} {\bibfnamefont {G.}~\bibnamefont {Chen}},\ and\
  \bibinfo {author} {\bibfnamefont {M.}~\bibnamefont {Hermele}},\ }\bibfield
  {title} {\bibinfo {title} {Quantum spin ices and topological phases from
  dipolar-octupolar doublets on the pyrochlore lattice},\ }\href
  {https://doi.org/10.1103/PhysRevLett.112.167203} {\bibfield  {journal}
  {\bibinfo  {journal} {Phys. Rev. Lett.}\ }\textbf {\bibinfo {volume} {112}},\
  \bibinfo {pages} {167203} (\bibinfo {year} {2014})}\BibitemShut {NoStop}%
\bibitem [{\citenamefont {Hukushima}\ and\ \citenamefont
  {Nemoto}(1996)}]{JPSJ.65.1604}%
  \BibitemOpen
  \bibfield  {author} {\bibinfo {author} {\bibfnamefont {K.}~\bibnamefont
  {Hukushima}}\ and\ \bibinfo {author} {\bibfnamefont {K.}~\bibnamefont
  {Nemoto}},\ }\bibfield  {title} {\bibinfo {title} {Exchange {M}onte {C}arlo
  method and application to spin glass simulations},\ }\href
  {https://doi.org/10.1143/JPSJ.65.1604} {\bibfield  {journal} {\bibinfo
  {journal} {Journal of the Physical Society of Japan}\ }\textbf {\bibinfo
  {volume} {65}},\ \bibinfo {pages} {1604} (\bibinfo {year}
  {1996})}\BibitemShut {NoStop}%
\bibitem [{Note12()}]{Note12}%
  \BibitemOpen
  \bibinfo {note} {We define $s$ to be the maximum plus a $5 \%$ allowance. As
  a side remark, the eigenvalues of $i A (\lbrace u_{ij}^\lambda \rbrace )$ can
  indeed be covered by a very large $s$, but due to finite energy resolution,
  taking $s$ larger than necessary leads to poorer numerical
  approximation.}\BibitemShut {Stop}%
\bibitem [{Note13()}]{Note13}%
  \BibitemOpen
  \bibinfo {note} {Similarly, for $G_{k+il,k+il}$, the expectation value of
  $T_m (H / s)$ is taken with respect to $\lvert k \rangle + i \lvert l \rangle
  $}\BibitemShut {NoStop}%
\bibitem [{Note14()}]{Note14}%
  \BibitemOpen
  \bibinfo {note} {The maple-leaf (trellis) lattice has $N = L \times L \times
  6$ ($L \times L \times 4$) sites in total.}\BibitemShut {Stop}%
\bibitem [{Note15()}]{Note15}%
  \BibitemOpen
  \bibinfo {note} {This stems from the identity $\Gamma ^{x0} \Gamma ^{y0}
  \Gamma ^{zx} \Gamma ^{zy} \Gamma ^{zz} = -1$.}\BibitemShut {Stop}%
\bibitem [{\citenamefont {Self}\ \emph {et~al.}(2019)\citenamefont {Self},
  \citenamefont {Knolle}, \citenamefont {Iblisdir},\ and\ \citenamefont
  {Pachos}}]{PhysRevB.99.045142}%
  \BibitemOpen
  \bibfield  {author} {\bibinfo {author} {\bibfnamefont {C.~N.}\ \bibnamefont
  {Self}}, \bibinfo {author} {\bibfnamefont {J.}~\bibnamefont {Knolle}},
  \bibinfo {author} {\bibfnamefont {S.}~\bibnamefont {Iblisdir}},\ and\
  \bibinfo {author} {\bibfnamefont {J.~K.}\ \bibnamefont {Pachos}},\ }\bibfield
   {title} {\bibinfo {title} {Thermally induced metallic phase in a gapped
  quantum spin liquid: {M}onte {C}arlo study of the {K}itaev model with parity
  projection},\ }\href {https://doi.org/10.1103/PhysRevB.99.045142} {\bibfield
  {journal} {\bibinfo  {journal} {Phys. Rev. B}\ }\textbf {\bibinfo {volume}
  {99}},\ \bibinfo {pages} {045142} (\bibinfo {year} {2019})}\BibitemShut
  {NoStop}%
\bibitem [{Note16()}]{Note16}%
  \BibitemOpen
  \bibinfo {note} {If we had to exactly diagonalize the Hamiltonian for every
  bond-flip proposal, the kernel polynomial method would be rendered completely
  useless!}\BibitemShut {Stop}%
\bibitem [{\citenamefont {Zhang}\ \emph {et~al.}(2019)\citenamefont {Zhang},
  \citenamefont {Wang}, \citenamefont {Hal\'asz},\ and\ \citenamefont
  {Batista}}]{PhysRevLett.123.057201}%
  \BibitemOpen
  \bibfield  {author} {\bibinfo {author} {\bibfnamefont {S.-S.}\ \bibnamefont
  {Zhang}}, \bibinfo {author} {\bibfnamefont {Z.}~\bibnamefont {Wang}},
  \bibinfo {author} {\bibfnamefont {G.~B.}\ \bibnamefont {Hal\'asz}},\ and\
  \bibinfo {author} {\bibfnamefont {C.~D.}\ \bibnamefont {Batista}},\
  }\bibfield  {title} {\bibinfo {title} {Vison crystals in an extended {K}itaev
  model on the honeycomb lattice},\ }\href
  {https://doi.org/10.1103/PhysRevLett.123.057201} {\bibfield  {journal}
  {\bibinfo  {journal} {Phys. Rev. Lett.}\ }\textbf {\bibinfo {volume} {123}},\
  \bibinfo {pages} {057201} (\bibinfo {year} {2019})}\BibitemShut {NoStop}%
\bibitem [{Note17()}]{Note17}%
  \BibitemOpen
  \bibinfo {note} {An equivalent way to think about this is that
  $u_{ij}^\lambda = i b_i^\lambda b_j^\lambda $ can freely assume the values
  $+1$ or $-1$.}\BibitemShut {Stop}%
\bibitem [{\citenamefont {Gubernatis}\ \emph {et~al.}(2016)\citenamefont
  {Gubernatis}, \citenamefont {Kawashima},\ and\ \citenamefont
  {Werner}}]{gubernatistextbook}%
  \BibitemOpen
  \bibfield  {author} {\bibinfo {author} {\bibfnamefont {J.~E.}\ \bibnamefont
  {Gubernatis}}, \bibinfo {author} {\bibfnamefont {N.}~\bibnamefont
  {Kawashima}},\ and\ \bibinfo {author} {\bibfnamefont {P.}~\bibnamefont
  {Werner}},\ }\href@noop {} {\emph {\bibinfo {title} {Quantum Monte Carlo
  Methods: Algorithms for Lattice Models}}}\ (\bibinfo  {publisher} {Cambridge
  University Press},\ \bibinfo {year} {2016})\BibitemShut {NoStop}%
\bibitem [{\citenamefont {Ambegaokar}\ and\ \citenamefont
  {Troyer}(2010)}]{1.3247985}%
  \BibitemOpen
  \bibfield  {author} {\bibinfo {author} {\bibfnamefont {V.}~\bibnamefont
  {Ambegaokar}}\ and\ \bibinfo {author} {\bibfnamefont {M.}~\bibnamefont
  {Troyer}},\ }\bibfield  {title} {\bibinfo {title} {Estimating errors reliably
  in {M}onte {C}arlo simulations of the {E}hrenfest model},\ }\href
  {https://doi.org/10.1119/1.3247985} {\bibfield  {journal} {\bibinfo
  {journal} {American Journal of Physics}\ }\textbf {\bibinfo {volume} {78}},\
  \bibinfo {pages} {150} (\bibinfo {year} {2010})}\BibitemShut {NoStop}%
\bibitem [{\citenamefont {Becca}\ and\ \citenamefont
  {Sorella}(2017)}]{beccatextbook}%
  \BibitemOpen
  \bibfield  {author} {\bibinfo {author} {\bibfnamefont {F.}~\bibnamefont
  {Becca}}\ and\ \bibinfo {author} {\bibfnamefont {S.}~\bibnamefont
  {Sorella}},\ }\href@noop {} {\emph {\bibinfo {title} {Quantum Monte Carlo
  Approaches for Correlated Systems}}}\ (\bibinfo  {publisher} {Cambridge
  University Press},\ \bibinfo {year} {2017})\BibitemShut {NoStop}%
\bibitem [{Note18()}]{Note18}%
  \BibitemOpen
  \bibinfo {note} {Each bootstrap data set yields a sample of $C$.}\BibitemShut
  {Stop}%
\bibitem [{Note19()}]{Note19}%
  \BibitemOpen
  \bibinfo {note} {Note that here we adopt the notation $\sigma _{1,2,3}$ for
  the Pauli matrices, instead of $\sigma ^{x,y,z}$ as in the main
  text.}\BibitemShut {Stop}%
\bibitem [{Note20()}]{Note20}%
  \BibitemOpen
  \bibinfo {note} {Here, the Berry connection is defined in accordance with the
  convention for the Chern number in Ref.~\protect \rev@citealp {KITAEV20062}.
  We caution readers that the definition in Ref.~\protect \rev@citealp
  {bernevigtextbook} differs from ours by a minus sign.}\BibitemShut {Stop}%
\bibitem [{Note21()}]{Note21}%
  \BibitemOpen
  \bibinfo {note} {A useful trick to evaluate these integrals is the
  substitution $x=\protect \qopname \relax o{tan}\theta $.}\BibitemShut {Stop}%
\bibitem [{Note22()}]{Note22}%
  \BibitemOpen
  \bibinfo {note} {In theory the increments $\lvert \protect \hat {1} \rvert $
  and $\lvert \protect \hat {2} \rvert $ are infinitesimal, but in practice we
  let them be the spacing between neighboring points on the momentum grid, $2
  \pi / L$.}\BibitemShut {Stop}%
\bibitem [{Note23()}]{Note23}%
  \BibitemOpen
  \bibinfo {note} {This can be shown using the fact that $H_0$ is antisymmetric
  and $H_1$ is symmetric, which we omit here for brevity. Furthermore, we have
  dropped a possible contribution to $H^\protect \mathrm {eff} ( \protect
  \mathbf {q} )$ that is proportional to the identity $\protect \varmathbb
  {1}$}\BibitemShut {NoStop}%
\bibitem [{Note24()}]{Note24}%
  \BibitemOpen
  \bibinfo {note} {The direct computation of the Chern number using the
  formula~\protect \textup {\hbox {\mathsurround \z@ \protect \normalfont
  (\ignorespaces \ref {chernnonabelcompute}\unskip \@@italiccorr )}} can be
  done with oblique coordinates $(q_1 , q_2)$ in the momentum space. Here,
  since we are applying the result~\protect \textup {\hbox {\mathsurround \z@
  \protect \normalfont (\ignorespaces \ref {chernnumberhalf}\unskip
  \@@italiccorr )}}, whose derivation involves an integral in polar
  coordinates, we express the effective Hamiltonian in terms of the orthogonal
  coordinates $(q_x , q_y)$, which are related to $(q_1 , q_2)$ via a linear
  transformation as shown in~\protect \textup {\hbox {\mathsurround \z@
  \protect \normalfont (\ignorespaces \ref {genericmomentum}\unskip
  \@@italiccorr )}}.}\BibitemShut {Stop}%
\end{thebibliography}%

\clearpage

\onecolumngrid

\begin{center}
\textbf{\large Supplemental Material: \\ Exact quantum spin liquids with topological order on maple-leaf and trellis lattices}
\end{center}
\begin{center}
Li Ern Chern$^{1}$, Roderich Moessner$^{1}$, and Claudio Castelnovo$^{2}$
\end{center}
\begin{center}
{\small
\textit{$^1$Max Planck Institute for the Physics of Complex Systems, 01187 Dresden, Germany} \\
\textit{$^2$T.C.M.~Group, Cavendish Laboratory, University of Cambridge, Cambridge CB3 0HE, United Kingdom}}
\end{center}

\setcounter{equation}{0}
\setcounter{figure}{0}
\setcounter{table}{0}
\setcounter{page}{1}
\setcounter{section}{0}

\renewcommand{\thesection}{S\arabic{section}}
\renewcommand{\theequation}{S\arabic{equation}}
\renewcommand{\thefigure}{S\arabic{figure}}
\renewcommand{\thetable}{S\arabic{table}}

\section{\label{section:symmetry}Symmetry Analysis}

In this section, we provide details of the point-group symmetries that we implement in our models~\eqref{kitaevmodel}. First, we treat each local degree of freedom as a composite of two $S=1/2$ moments, which live on the same site but are otherwise independent. The corresponding operators are $\bm{\tau}_i$ and $\bm{\sigma}_i$, which enter the definitions of the local $\Gamma$ matrices via~\eqref{gammadefine}. Next, we impose threefold and twofold rotational symmetries about the out-of-plane axis for the Kitaev maple-leaf and trellis models, respectively. We assume spin-orbit couplings for both $\bm{\tau}_i$ and $\bm{\sigma}_i$, so that any rotation in real space is accompanied by a corresponding rotation in spin space. Recall that an $SU(2)$ rotation of an ordinary $S=1/2$ object is equal to an $SO(3)$ transformation on the vector of the three spin components~\cite{sakuraitextbook},
\begin{equation}
e^{i \bm{\sigma} \cdot \hat{\mathbf{n}} \theta / 2} \bm{\sigma} e^{-i \bm{\sigma} \cdot \hat{\mathbf{n}} \theta / 2} = R (\hat{\mathbf{n}} , \theta) \bm{\sigma}, \quad R (\hat{\mathbf{n}} \, , \theta) \in SO (3) \, ,
\end{equation}
where $\theta$ is the angle of rotation and $\hat{\mathbf{n}}$ is the (three-dimensional) unit vector along the axis of rotation. In general, a symmetry operator $X$ acts on a local moment $\bm{\sigma}_i$ as $X: \bm{\sigma}_i \longrightarrow R_X \bm{\sigma}_{X^{-1}(i)}$, with $R_X \in SO (3)$ being the spin rotation associated with $X$. 

For the Kitaev maple-leaf model, we choose $\bm{\sigma}_i$ to be an effective $S=1/2$ moment with its $[111]$ axis aligned with the out-of-plane direction of the lattice~\cite{PhysRevLett.102.017205}, and $\bm{\tau}_i$ a dipolar-octupolar moment as defined in Ref.~\onlinecite{PhysRevLett.112.167203} with its $z$-axis aligned with the out-of-plane direction of the lattice. The $x$, $y$, and $z$ components of $\bm{\sigma}$ are permuted by $R(\hat{\mathbf{n}} \parallel [111],\theta=2 \pi /3)$, which we denote by $R_{C_3}$ for simplicity. Explicitly,
\begin{equation}
R_{C_3} = \begin{pmatrix} 0 & 0 & 1 \\ 1 & 0 & 0 \\ 0 & 1 & 0 \end{pmatrix}
\end{equation}
and $\sigma^x_i \longrightarrow \sigma^y_{C_3 (i)} \longrightarrow \sigma^z_{C_3^2 (i)}$ under successive applications of $C_3^{-1}$. On the other hand, the components of $\bm{\tau}$ are invariant under $C_3$, owing to the distinct symmetry transformation properties of dipolar-octupolar moments~\cite{PhysRevLett.112.167203}. The requirement $C_3 H C_3^{-1} = H$ enforces $J_{zx}=J_{zy}=J_{zz}$, reducing the number of independent couplings by two. 

For the Kitaev trellis model, we choose both $\bm{\sigma}_i$ and $\bm{\tau}_i$ to be effective $S=1/2$ moments with their $[\overline{1}10]$ axes aligned with the out-of-plane direction of the lattice. The $x$ and $y$ components are exchanged, while every component acquires a minus sign, under $R (\hat{\mathbf{n}} \parallel [\overline{1}10] , \theta=\pi)$, which we denote by $R_{C_2}$ for simplicity. Explicitly,
\begin{equation}
R_{C_2} = \begin{pmatrix} 0 & -1 & 0 \\ -1 & 0 & 0 \\ 0 & 0 & -1 \end{pmatrix}
\end{equation}
and $(\sigma^x , \sigma^y , \sigma^z)_i \longrightarrow - (\sigma^y , \sigma^x , \sigma^z)_{C_2 (i)}$ under the action of $C_2^{-1}$, with $\bm{\tau}_i$ transforming in the same manner. The requirement $C_2 H C_2^{-1} = H$ enforces $J_{x0}=J_{y0}$ and $J_{zx}=J_{zy}$, reducing the number of independent couplings by two.

\section{\label{section:montecarlo}Quantum Monte Carlo}

In this section, we provide details of the replica exchange quantum Monte Carlo method~\cite{PhysRevLett.113.197205,PhysRevLett.115.087203,PhysRevB.92.115122,nphys3809,PhysRevB.96.125124,PhysRevB.98.054432,PhysRevResearch.1.032011,PhysRevB.101.045118,eschmannthesis,PhysRevB.102.075125,PhysRevResearch.2.043159} as well as its numerical implementation in our study of the Kitaev maple-leaf and trellis models.

\subsection{\label{section:preliminary}Preliminaries}

Let $H$ be the quadratic Hamiltonian of Majorana fermions~\eqref{kitaevmodelquadratic} for a given set of couplings and $N$ be the total number of sites in the system. At each temperature $T$, the partition function is given by~\cite{PhysRevLett.113.197205}
\begin{equation}
Z = \mathrm{Tr}_{\lbrace u_{ij}^\lambda \rbrace} \mathrm{Tr}_{\lbrace c_i \rbrace} e^{- \beta H} \, , \qquad \beta \equiv \frac{1}{T} \, .
\end{equation}
For a fixed gauge-field configuration $\lbrace u_{ij}^\lambda \rbrace$, the Hamiltonian can be brought into canonical form by a transformation $Q \in O (N)$ and further diagonalized~\cite{KITAEV20062},
\begin{equation}
H (\lbrace u_{ij}^\lambda \rbrace) = \frac{i}{2} \sum_{k=1}^{N/2} \varepsilon_k \gamma_{2k-1} \gamma_{2k} = \sum_{k=1}^{N/2} \varepsilon_k \left( \psi_k^\dagger \psi_k - \frac{1}{2} \right) \, , \quad \varepsilon_k \geq 0 \, ,
\end{equation}
where $\gamma_i = \sum_j c_j Q_{ji}$ is a Majorana fermion, $\psi_k = (\gamma_{2k-1} + i \gamma_{2k}) / 2$ is a complex fermion, and the dependence of the energies $\varepsilon_k$ on $\lbrace u_{ij}^\lambda \rbrace$ is suppressed for brevity. The trace over the itinerant Majorana fermions $\lbrace c_i \rbrace$ is thus equivalent to the trace over the independent complex fermion modes $\lbrace \psi_k \rbrace$, each of which can be either occupied or unoccupied, leading to
\begin{equation}
\mathrm{Tr}_{\lbrace c_i \rbrace} e^{- \beta H (\lbrace u_{ij}^\lambda \rbrace)} = \prod_{k=1}^{N/2} \sum_{n_k = 0}^1 e^{- \beta \varepsilon_k (n_k - 1/2)} = \prod_{k=1}^{N/2} \left[ 2 \cosh \left( \frac{\beta \varepsilon_k}{2} \right) \right] \equiv e^{- \beta F (\lbrace u_{ij}^\lambda \rbrace , \beta)} \, ,
\end{equation}
where we have defined the free energy for a given gauge-field configuration,
\begin{equation} \label{freeenergy}
F (\lbrace u_{ij}^\lambda \rbrace , \beta) = - T \sum_{k=1}^{N/2} \ln \left[ 2 \cosh \left( \frac{\beta \varepsilon_k}{2} \right) \right] \, ,
\end{equation}
which also depends on the temperature. The issue of unphysical states arising from the Majorana fermion representation will be addressed later in Sec.~\ref{section:implement}.

\subsection{\label{section:replica}Replica Exchange}

We simulate $R$ replicas of the system at $R$ logarithmically spaced temperatures $T_1 < T_2 < \ldots < T_{R-1} < T_R$~\cite{JPSJ.65.1604}. At each temperature $T_n$, we begin with a random gauge-field configuration $\lbrace u_{ij}^\lambda \rbrace_n$. Then, a bond $\langle ij \rangle_\lambda$ is randomly selected, and the corresponding gauge field $u_{ij}^\lambda$ is multiplied by $-1$. We introduce the terminology ``bond flip'' for such a process to simplify the discussions. The change in free energy $\Delta F$ is calculated, and the bond-flip proposal is accepted with the probability $\min \lbrace 1 , \exp (-\Delta F/T) \rbrace$, which is the Metropolis criterion. Let $N_b$ be the total number of bonds. After each Monte Carlo sweep, which consists of $N_b$ bond-flip proposals, we attempt a replica exchange for each pair of adjacent temperatures $T_n$ and $T_{n+1}$, starting from $n=1$~\cite{eschmannthesis}. We calculate
\begin{equation}
p = \frac{\exp [ - \beta_n F (\lbrace u_{ij}^\lambda \rbrace_{n+1} , \beta_n) - \beta_{n+1} F (\lbrace u_{ij}^\lambda \rbrace_n , \beta_{n+1}) ]}{{\exp [ - \beta_n F (\lbrace u_{ij}^\lambda \rbrace_n , \beta_n) - \beta_{n+1} F (\lbrace u_{ij}^\lambda \rbrace_{n+1} , \beta_{n+1}) ]}} \, ,
\end{equation}
and swap the gauge-field configurations $\lbrace u_{ij}^\lambda \rbrace_n$ and $\lbrace u_{ij}^\lambda \rbrace_{n+1}$ at $T_n$ and $T_{n+1}$ with the probability $\min \lbrace 1 , p \rbrace$~\cite{PhysRevLett.113.197205}. Compared to the traditional Monte Carlo method where one only simulates a single system and gradually lowers the temperature, the replica exchange method enables the system to avoid local minima and reach equilibrium more efficiently~\cite{JPSJ.65.1604}. 

Although the procedure outlined above is straightforward to implement in theory, numerical simulations can be extremely time consuming for a system of few hundred sites. The bottleneck is that, every time a bond flip is proposed, the Hamiltonian has to be diagonalized to obtain the full spectrum $\lbrace \varepsilon_k \rbrace$ for an exact evaluation of the free energy~\eqref{freeenergy}. One can contrast this with Monte Carlo simulations of classical systems, where changes of a local degree of freedom only affect the local energy. The numerical diagonalization of an $N \times N$ matrix has a time complexity that scales as $N^3$. An approximation scheme that enables a significant speedup of the computation would be desirable. The Green-function-based kernel polynomial method~\cite{RevModPhys.78.275,PhysRevLett.102.150604,PhysRevB.96.125124,PhysRevB.102.075125,eschmannthesis} is precisely designed for this purpose.

\subsection{\label{section:kernel}Green-Function-Based Kernel Polynomial Method}

The content of this subsection is mainly based on Refs.~\onlinecite{PhysRevLett.102.150604,eschmannthesis}. Let the $N$-dimensional matrix $G(E)$ be the Green function of the Hamiltonian $H \equiv iA$ at the energy $E$, i.e., $G(E) (H-E) = \vmathbb{1}$. Suppose that $\langle ij \rangle_\lambda$ is selected for the bond-flip proposal, i.e., $u_{ij}^\lambda \longrightarrow - u_{ij}^\lambda$. We express the change in the Hamiltonian due to this bond flip as $H \longrightarrow H + \Delta$, where only the $(i,j)$ and $(j,i)$ entries of $\Delta$ are nonzero. Explicitly, $\Delta_{ij}=-4i J_\lambda u_{ij}^\lambda$ and $\Delta_{ji}=-\Delta_{ij}$. We introduce the function
\begin{equation} \label{dfunction}
d(E) \equiv \det [ \vmathbb{1} + G(E) \Delta ] = \det \begin{pmatrix} 1 + G_{ij} \Delta_{ji} & G_{ii} \Delta_{ij} \\ G_{jj} \Delta_{ji} & 1 + G_{ji} \Delta_{ij} \end{pmatrix} = 1 + G_{ij} \Delta_{ji} + G_{ji} \Delta_{ij} - \Delta_{ji} \Delta_{ij} ( G_{ii} G_{jj} - G_{ij} G_{ji} ) \, ,
\end{equation}
where in the second equality we have exploited the sparsity of $\Delta$, and we have suppressed the $E$ dependence of the various entries of $G (E)$ for brevity. Since $\vmathbb{1}+G(E)\Delta = G(E) (H + \Delta - E)$, we have
\begin{equation} \label{dfunctionalter}
d(E) = \det G(E) \det (H + \Delta - E) = \frac{\prod_{l=1}^{N} ( \epsilon_l' - E )}{\prod_{l=1}^{N} ( \epsilon_l - E )} \, ,
\end{equation}
where $\lbrace \epsilon_l \rbrace$ and $\lbrace \epsilon_l' \rbrace$ are the complete sets of eigenvalues (i.e., including both positive and negative ones) of $H$ and $H' \equiv H + \Delta$, respectively. Replacing $E \in \mathbb{R}$ by $z \equiv E + i 0^+ \in \mathbb{C}$, we evaluate
\begin{equation} \label{densityofstate}
\frac{1}{\pi} \mathrm{Im} \frac{\mathrm{d} \ln d (z)}{\mathrm{d} z} = \sum_{l=1}^{N} \frac{1}{\pi} \mathrm{Im} \left[ \frac{1}{\epsilon_l - E - i 0^+} - \frac{1}{\epsilon_l' - E - i 0^+} \right] = \sum_{l=1}^N \delta (E - \epsilon_l) - \sum_{l=1}^N \delta (E - \epsilon_l') \equiv \rho (E) - \rho' (E) \, .
\end{equation}
From~\eqref{freeenergy}, the change in free energy due to the bond flip can be expressed as
\begin{equation} \label{freeenergychange}
\Delta F = \frac{T}{2} \int_{- \infty}^{\infty} \mathrm{d} E \, \ln \left[ 2 \cosh \left( \frac{\beta E}{2} \right) \right] [ \rho (E) - \rho' (E) ] = - \frac{1}{2 \pi} \int_0^{\infty} \mathrm{d} E \, \tanh \left( \frac{\beta E}{2} \right) \mathrm{Im} \ln d (E + i 0^+) \, ,
\end{equation}
where, in the second equality, we have used integration by parts, the validity of which can be demonstrated by the decay of $\ln [ 2 \cosh ( \beta E / 2 ) ] \mathrm{Im} \ln d (E + i 0^+)$ being not slower than $1/E$ as $E \longrightarrow \pm \infty$. Also, from~\eqref{dfunctionalter} and the fact that the eigenvalues of either $H$ or $H'$ come in positive-negative pairs, one can show $d (- E + i 0^+) = [d (E + i 0^+)]^*$, which implies that the integrand on the RHS is even in $E$. 

We have thus related $\Delta F$ to $G (z)$ via~\eqref{freeenergychange} and~\eqref{dfunction}. The present task is to efficiently compute the four entries of $G (z)$ involved in the definition~\eqref{dfunction} of $d (z)$, which is accomplished by the Chebyshev approximation. For the diagonal components of the Green function, we have
\begin{equation} \label{chebyshevexpand}
G_{ii} (z) \approx \frac{i}{\sqrt{s^2 - z^2}} \left[ \mu_0 + 2 \sum_{m=1}^{M-1} \mu_m e^{- i m \cos^{-1} (z / s)} \right] \, ,
\end{equation}
where $s > 0$ is the bandwidth defined such that all eigenvalues of $H$ fall within the range $(-s , s)$. $s$ is calculated in the beginning by exactly diagonalizing $H ( \lbrace u_{ij}^\lambda \rbrace )$ for a large number ($\sim 10^4$) of randomly generated gauge-field configurations $\lbrace u_{ij}^\lambda \rbrace$, and taking the maximum of the largest eigenvalues~\footnote{We define $s$ to be the maximum plus a $5 \%$ allowance. As a side remark, the eigenvalues of $i A (\lbrace u_{ij}^\lambda \rbrace)$ can indeed be covered by a very large $s$, but due to finite energy resolution, taking $s$ larger than necessary leads to poorer numerical approximation.}. On the other hand, the Chebyshev moments are given by
\begin{equation} \label{chebyshevmoment}
\mu_m = g_m \langle i \lvert T_m (H / s) \rvert i \rangle , \quad g_m = \frac{1}{M+1} \left[ (M - m + 1) \cos \left( \frac{m \pi}{M+1} \right) + \sin \left( \frac{m \pi}{M+1} \right) \cot \left( \frac{\pi}{M+1} \right) \right] \, ,
\end{equation}
where $T_m$ is the $m$th order Chebyshev polynomial of the first kind, which satisfies the recursion relation
\begin{equation} \label{recursion}
T_m (x) = 2 x T_{m-1} (x) - T_{m-2} (x) \, ,
\end{equation}
and $\lvert i \rangle$ is the $N$-dimensional unit vector with the $i$th component equal to 1. Given $T_0 (x)=1$ and $T_1 (x)=x$, we can generate higher order Chebyshev polynomials iteratively via~\eqref{recursion}. $g_m$ as defined in~\eqref{chebyshevmoment} is known as the Jackson kernel~\cite{RevModPhys.78.275}, which is introduced to damp the Gibbs oscillations due to the truncation of the series expansion at some finite order $M$. With $x = H / s$ in~\eqref{recursion}, $\lvert \alpha_0 \rangle \equiv \lvert i \rangle$, and $\lvert \alpha_1 \rangle \equiv ( H / s ) \lvert i \rangle$, we carry out successive matrix-vector multiplications $\lvert \alpha_m \rangle = 2 ( H / s ) \lvert \alpha_{m-1} \rangle - \lvert \alpha_{m-2} \rangle$ and compute $\langle i \lvert T_m ( H / s ) \rvert i \rangle = \langle i \vert \alpha_m \rangle$ for $m \geq 2$. The number of multiplications can be reduced by half through the relations
\begin{subequations}
\begin{align}
\langle i \lvert T_{2m} (H / s) \rvert i \rangle &= 2 \langle \alpha_m \vert \alpha_m \rangle - \langle i \vert \alpha_0 \rangle \, , 
\\
\langle i \lvert T_{2m+1} (H / s) \rvert i \rangle &= 2 \langle \alpha_{m+1} \vert \alpha_{m} \rangle - \langle i \lvert \alpha_1 \rangle \, ,
\end{align}
\end{subequations}
which arise from the identity $2 T_m (x) T_n (x) = T_{m + n} (x) + T_{m - n} (x)$~\cite{RevModPhys.78.275}. Finally, the off-diagonal entries of the Green function are obtained by
\begin{subequations}
\begin{align}
G_{k,l} &= \frac{1}{2} \left[ G_{k+l,k+l} - i G_{k+il,k+il} - (1-i) (G_{k,k} + G_{l,l}) \right] \, , 
\\
G_{l,k} &= \frac{1}{2} \left[ G_{k+l,k+l} + i G_{k+il,k+il} - (1+i) (G_{k,k} + G_{l,l}) \right] \, ,
\end{align}
\end{subequations}
where, by abuse of notation, $G_{k+l,k+l}$ means that the expectation value of $T_m ( H / s )$ is taken with respect to $\lvert k \rangle + \lvert l \rangle$ in the calculation of $\mu_m$~\eqref{chebyshevmoment}, but not $\lvert k + l \rangle$ which would give the $(k+l,k+l)$ entry of $G$~\footnote{Similarly, for $G_{k+il,k+il}$, the expectation value of $T_m (H / s)$ is taken with respect to $\lvert k \rangle + i \lvert l \rangle$}. 

The time complexity of a sparse matrix-vector multiplication in $N$ dimensions scales as $N$, but we have to perform $O (M)$ times of it, so the total time complexity is $O (MN)$. If $M \lesssim N$, this is a substantial improvement over the exact diagonalization approach, which has a time complexity of $O (N^3)$. We remark that the Green-function-based kernel polynomial method is only used to calculate the change in free energy due to a bond flip. After each Monte Carlo sweep, we still have to carry out exact diagonalization of the Hamiltonian for the replica exchange process.

\subsection{\label{section:implement}Numerical Implementation}

\begin{figure}
\includegraphics[scale=0.23]{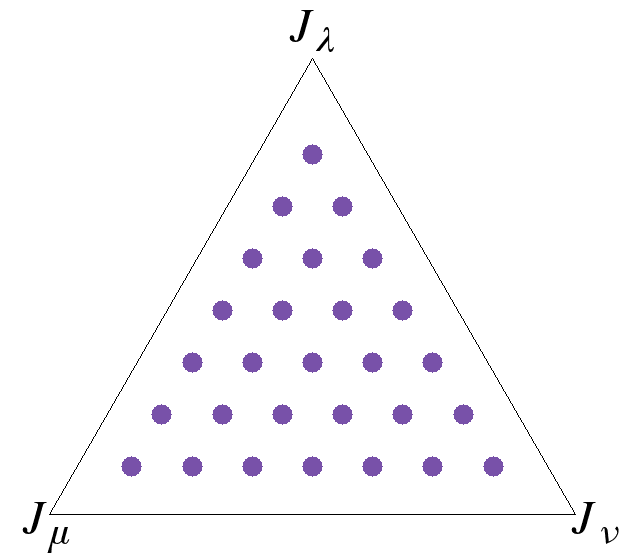}
\caption{\label{figure:parameter}The set of discrete points at which we perform numerical simulations to determine the ground-state flux sector of the Kitaev maple-leaf or trellis model, in the triangular parameter space $J_\mu + J_\nu + J_\lambda = 1$.}
\end{figure}

We perform the replica exchange quantum Monte Carlo simulations on a regular grid of points $(J_\mu , J_\nu , J_\lambda)$ in the triangular parameter space $J_\mu + J_\nu + J_\lambda = 1$, where $(J_\mu , J_\nu , J_\lambda) = (J_{x0} , J_{y0} , J_{zx} = J_{zy} = J_{zz})$ and $( J_{x0} = J_{y0} , J_{zx} = J_{zy} , J_{zz})$ for the Kitaev maple-leaf and trellis models, respectively. These points, which are shown in Fig.~\ref{figure:parameter}, are parametrized by two integers $t_1 \in [0 , 7)$ and $t_2 \in [0 , 7 - t_1)$ as
\begin{equation} \label{selectpoint}
J_\mu = \frac{4}{35} t_1 + \frac{11}{105} , \quad J_\nu = \frac{4}{35} t_2 + \frac{11}{105} , \quad J_\lambda = 1 - \frac{4}{35} (t_1 + t_2) - \frac{22}{105} \, .
\end{equation}
We will refer to the selected points simply by $(t_1 , t_2)$ instead of $( J_\mu , J_\nu , J_\lambda )$ in later discussions. For the Kitaev maple-leaf model, we observe that the spatial distributions of the coupling strengths at $(J_{x0} , J_{y0} , J_{zx})$ and $(J_{y0} , J_{x0} , J_{zx})$ are related by a $C_2$ symmetry, so it is sufficient to study the regime $J_{x0} \geq J_{y0}$ or $t_1 \geq t_2$, which corresponds to the left half (including the vertical bisector) of the triangular parameter space. 

We run the simulations on systems of $L \times L$ unit cells~\footnote{The maple-leaf (trellis) lattice has $N = L \times L \times 6$ ($L \times L \times 4$) sites in total.} with periodic boundary conditions in both the $\mathbf{a}_1$ and $\mathbf{a}_2$ directions. In Secs.~\ref{section:replica} and~\ref{section:kernel}, we have seen that the change in free energy due to a bond flip can be either computed exactly by diagonalizing the full Hamiltonian, or approximated by the Green-function-based kernel polynomial method. We apply the exact diagonalization method to $L=4$ ($L=5$) and the kernel polynomial method to $L=6,8$ ($L=7,9$) for the Kitaev maple-leaf (trellis) model, with $M=256$ or $M=512$ for the order of Chebyshev expansion~\eqref{chebyshevexpand}. In either approach, we run $16,000$ Monte Carlo sweeps for each replica. We allow the system to thermalize during the first half of the Monte Carlo sweeps, and take measurements in the second half. For points nearer to the center of the triangular parameter space, which have weaker anisotropy, we simulate $R = 64$ replicas between the temperatures $T_1 = 10^{-3}$ and $T_R = 10$. For points nearer to the boundary of the triangular parameter space, which have stronger anisotropy, we simulate $R = 72$ replicas between the temperatures $T_1 = 10^{-5}$ and $T_R = 10$, as the transition to the ground state typically occurs at lower temperatures. 

One ought to treat the Majorana fermion representation of the $\Gamma$ matrices carefully, as it gives rise to unphysical states which enlarge the Hilbert space. The projection to the physical subspace for the Kitaev honeycomb model is analyzed in Ref.~\onlinecite{PhysRevB.84.165414}, and a straightforward extension for generalized Kitaev models on graphs with odd coordination numbers can be found in Ref.~\onlinecite{2607.12027}. Here, we specialize to the coordination number $z=5$, which is relevant to the Kitaev maple-leaf and trellis models. From the constraint~\footnote{This stems from the identity $\Gamma^{x0} \Gamma^{y0} \Gamma^{zx} \Gamma^{zy} \Gamma^{zz} = -1$.}
\begin{equation}
D_i \lvert \psi \rangle = \lvert \psi \rangle, \quad D_i \equiv - i b_i^{x0} b_i^{y0} b_i^{zx} b_i^{zy} b_i^{zz} c_i
\end{equation}
on a physical wavefunction $\lvert \psi \rangle$, one finds that an eigenstate of the Hamiltonian $H (\lbrace u_{ij}^\lambda \rbrace)$ constitutes a basis for $\lvert \psi \rangle$ if and only if it has eigenvalue $+1$ under the operator
\begin{equation} \label{tripod}
\prod_i D_i = (-1)^\xi \left( \prod_{\langle ij \rangle_\lambda} u_{ij}^\lambda \right) \det Q \prod_{k=1}^{N/2} \left( 2 \psi_k^\dagger \psi_k - 1 \right) \, ,
\end{equation}
where $\xi \in \mathbb{Z}$ depends on the geometry of the lattice. Eq.~\eqref{tripod} implies that, once the geometry, the couplings, and the gauge fields are specified, physical and unphysical wavefunctions differ in the parity of matter fermions: One has odd fermion parity while the other has even. The projection can indeed be implemented directly in the simulations~\cite{PhysRevB.99.045142}, but this would require the calculation of the determinant of $Q$, in addition to the exact diagonalization of the Hamiltonian, for every bond-flip proposal~\footnote{If we had to exactly diagonalize the Hamiltonian for every bond-flip proposal, the kernel polynomial method would be rendered completely useless!}. Instead, we apply a simple trick from Ref.~\onlinecite{PhysRevB.92.014403} (see also Ref.~\onlinecite{PhysRevLett.123.057201}), which is described as follows. We remove a bond $\langle ij \rangle_\lambda$ from the lattice by setting $J_\lambda = 0$ only for that bond, thus decoupling two Majorana fermions $b_i^\lambda$ and $b_j^\lambda$ from the Hamiltonian. They form a zero energy complex fermion mode $b_i^\lambda + i b_j^\lambda$, which can be freely occupied or unoccupied to adjust the total fermion parity~\footnote{An equivalent way to think about this is that $u_{ij}^\lambda = i b_i^\lambda b_j^\lambda$ can freely assume the values $+1$ or $-1$.} according to the physical constraint. By such single-bond removal, the numerics can proceed as described in Secs.~\ref{section:replica} and~\ref{section:kernel} without any other modification. In our simulations of either the Kitaev maple-leaf or trellis model, we remove a bond between two unit triangles, see Fig.~\ref{figure:wall} for example. 

In the kernel polynomial method, Eq.~\eqref{freeenergychange} is evaluated by a numerical integration. From~\eqref{densityofstate}, $\mathrm{d} \, \mathrm{Im} \ln d (z=E + i 0^+) / \mathrm{d} E$ yields the delta functions $\delta (E - \epsilon_l)$ and $\delta (E - \epsilon_l')$, so $\mathrm{d} \,  \mathrm{Im} \ln d (z) / \mathrm{d} E$ must be zero for $E > s$, as the bandwidth $s > 0$ is defined such that $\epsilon_l , \epsilon_l' \in ( -s , s )$. In other words, $\mathrm{Im} \ln d (z)$ is a constant throughout the domain $E \in [s , \infty)$. We evaluate
\begin{equation} \label{showzero}
\begin{aligned}[b]
\mathrm{Im} \ln d (z) &= \mathrm{Im} \left[ \sum_{l=1}^N \ln ( \epsilon _l' - E - i 0^+ ) - ( \epsilon_l' \longleftrightarrow \epsilon_l ) \right] = \sum_{l=1}^N \left[ \tan^{-1} \left( \frac{0^+}{E - \epsilon_l'} \right) - ( \epsilon_l' \longleftrightarrow \epsilon_l ) \right] \\
&= \sum_{l=1}^N \left[ \frac{0^+}{E - \epsilon_l'} - \frac{1}{3} \left( \frac{0^+}{E - \epsilon_l'} \right)^3 + \ldots - ( \epsilon_l' \longleftrightarrow \epsilon_l ) \right] \, ,
\end{aligned}
\end{equation}
which shows that $\mathrm{Im} \ln d (z) \longrightarrow 0$ as $E \longrightarrow \infty$. Therefore, we deduce that $\mathrm{Im} \ln d (z) = 0$ throughout the domain $E \in [s , \infty)$, and the upper limit $\infty$ of the integral on the RHS of~\eqref{freeenergychange} can be replaced by $s$. We choose $M$ equally spaced points $E_m$ within the semi-open interval $E \in [0 , s)$, and carry out the numerical integration with the trapezoidal scheme~\cite{eschmannthesis}
\begin{equation}
\int_{E_0=0}^{E_M=s} \mathrm{d} E \, f (E) \approx \left[ \frac{1}{2} f (E_0) + f (E_1) + \ldots + f (E_{M-2}) + \frac{3}{2} f (E_{M-1}) \right] \Delta E \, .
\end{equation}
We also find, phenomenologically, $0^+ = 4 s / M$ to be a good choice in our numerical calculations~\cite{PhysRevLett.102.150604}. 

\begin{figure}
\includegraphics[scale=0.24]{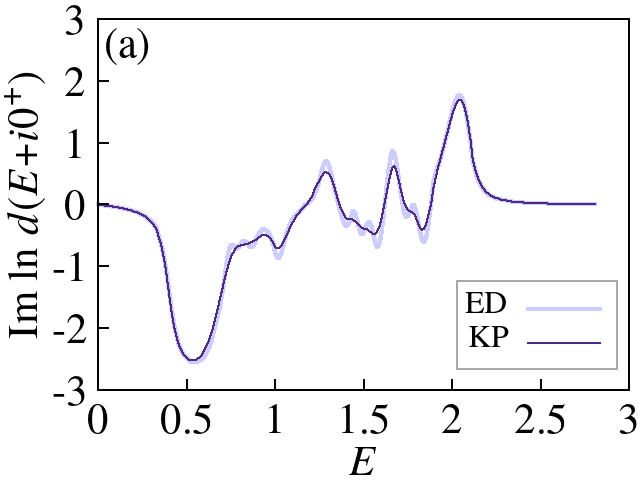} \qquad
\includegraphics[scale=0.24]{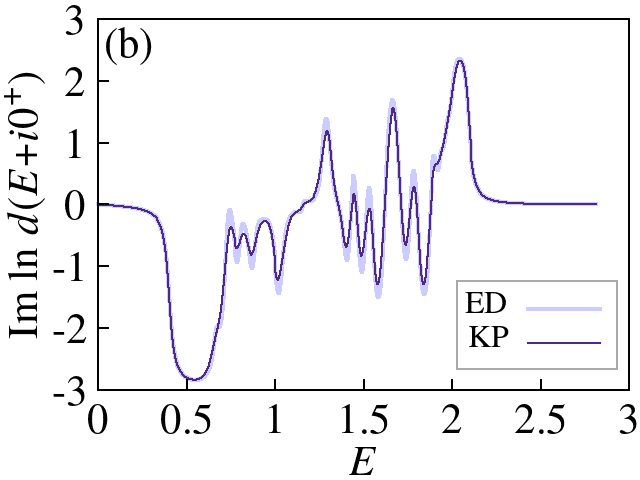}
\caption{\label{figure:bench}Checking the accuracies of Chebyshev expansions to the orders (a) $M=256$ and (b) $M=512$ against exact results. The function $\mathrm{Im} \ln d (E + i0^+)$, which is defined here via a bond flip on a hexagon in the ground-state flux sector of the isotropic Kitaev maple-leaf model, is computed using the kernel polynomial (KP) and exact diagonalization (ED) methods, the results of which are plotted with thin, dark-purple and thick, light-purple lines, respectively.}
\end{figure}

We provide an instance of comparing the kernel polynomial and exact diagonalization methods. Starting from the Kitaev maple-leaf model at the isotropic point, where $J_\lambda=1/3$ for all $\lambda$, in the ground-state flux sector, we flip a bond of a hexagon. On the one hand, we compute $d (E + i 0^+)$ using the rightmost expression of~\eqref{dfunction}, where the Green function is approximated by the Chebyshev expansion~\eqref{chebyshevexpand} to order $M=256$ or $512$. On the other hand, we compute the same quantity using the rightmost expression of~\eqref{dfunctionalter}, where the eigenvalues $\epsilon_l$ ($\epsilon_l'$) are obtained by exact diagonalization of the Hamiltonian before (after) the bond flip. The results are plotted in Figs.~\ref{figure:bench}a and~\ref{figure:bench}b. We see that the kernel polynomial method is able to capture the major features of $\mathrm{Im} \ln d (E + i0^+)$ though it may miss some minor details, and its accuracy can be improved by increasing $M$. Note that both the discretization of the $E$ axis and the infinitesimal number $0^+$ depend on $M$ in our calculations, such that a larger $M$ resolves a richer behavior of $\mathrm{Im} \ln d (E + i0^+)$. Note also that $\mathrm{Im} \ln d (E + i0^+)$ becomes zero at sufficiently large $E$, see~\eqref{showzero}.

\subsection{\label{section:measure}Measurements and Error Analyses}

Apart from the spatially averaged fluxes of the elementary plaquettes, we also measure the energy of, and the heat capacity due to, the itinerant fermions,
\begin{subequations}
\begin{align}
E &= \frac{\partial}{\partial \beta} \beta F (\lbrace u_{ij}^\lambda \rbrace, \beta) = - \sum_{k=1}^{N/2} \frac{\varepsilon_k}{2} \tanh \left( \frac{\beta \varepsilon_k}{2} \right) \, , \label{fermionenergy} \\
C_\psi &= \frac{\partial E}{\partial T} = \frac{1}{T^2} \sum_{k=1}^{N/2} \left( \frac{\varepsilon_k}{2} \right)^2 \left[ 1 - \tanh^2 \left( \frac{\beta \varepsilon_k}{2} \right) \right] 
\, ,
\label{fermionheat} 
\end{align}
\end{subequations}
where we have suppressed the dependence of these quantities on $\lbrace u_{ij}^\lambda \rbrace$ and $T$. The total heat capacity, including contributions from both itinerant fermions and gauge fluxes, is given by~\cite{PhysRevLett.113.197205,PhysRevB.102.075125}
\begin{equation} \label{heatcapacity}
C = \langle C_\psi \rangle + \frac{\langle E ^2 \rangle - \langle E \rangle^2}{T^2} \, ,
\end{equation}
where $\langle \ldots \rangle$ denotes the thermal average. 

For each observable of interest, one measurement is taken per Monte Carlo sweep after allowing the system to thermalize. At high temperatures, bond flips occur frequently, so we expect negligible correlations between subsequent measurements. However, near a phase transition or at low temperatures, correlations may persist over a large number of measurements, which reduce the number of statistically independent samples. To treat correlation effects, we use the binning analysis outlined in Ref.~\onlinecite{gubernatistextbook} (see also Ref.~\onlinecite{1.3247985}), with an arithmetically increasing bin size up to $160$, so that we have at least $50$ binned samples. 

The binning analysis alone is sufficient for estimating the error bars of the spatially averaged fluxes. On the other hand, to avoid cumbersome error propagation, we carry out an additional treatment to the total heat capacity~\eqref{heatcapacity} as follows. At each temperature, we first apply the binning analysis on the measured energy~\eqref{fermionenergy} and determine the appropriate bin size. Using the same bin size, we obtain binned data of $E$, $E^2$, and $C_\psi$, which are required for the evaluation of $C$. We then employ the bootstrap method outlined in Ref.~\onlinecite{beccatextbook} to resample the binned data~\cite{gubernatistextbook}. We generate $1,000$ bootstrap data sets~\footnote{Each bootstrap data set yields a sample of $C$.}, from which we estimate the mean and error bar of $C$, regardless of the bin size.

\subsection{\label{section:wall}Domain Walls}

\begin{figure}
\subfloat[]{\label{figure:wall}
\includegraphics[scale=0.4]{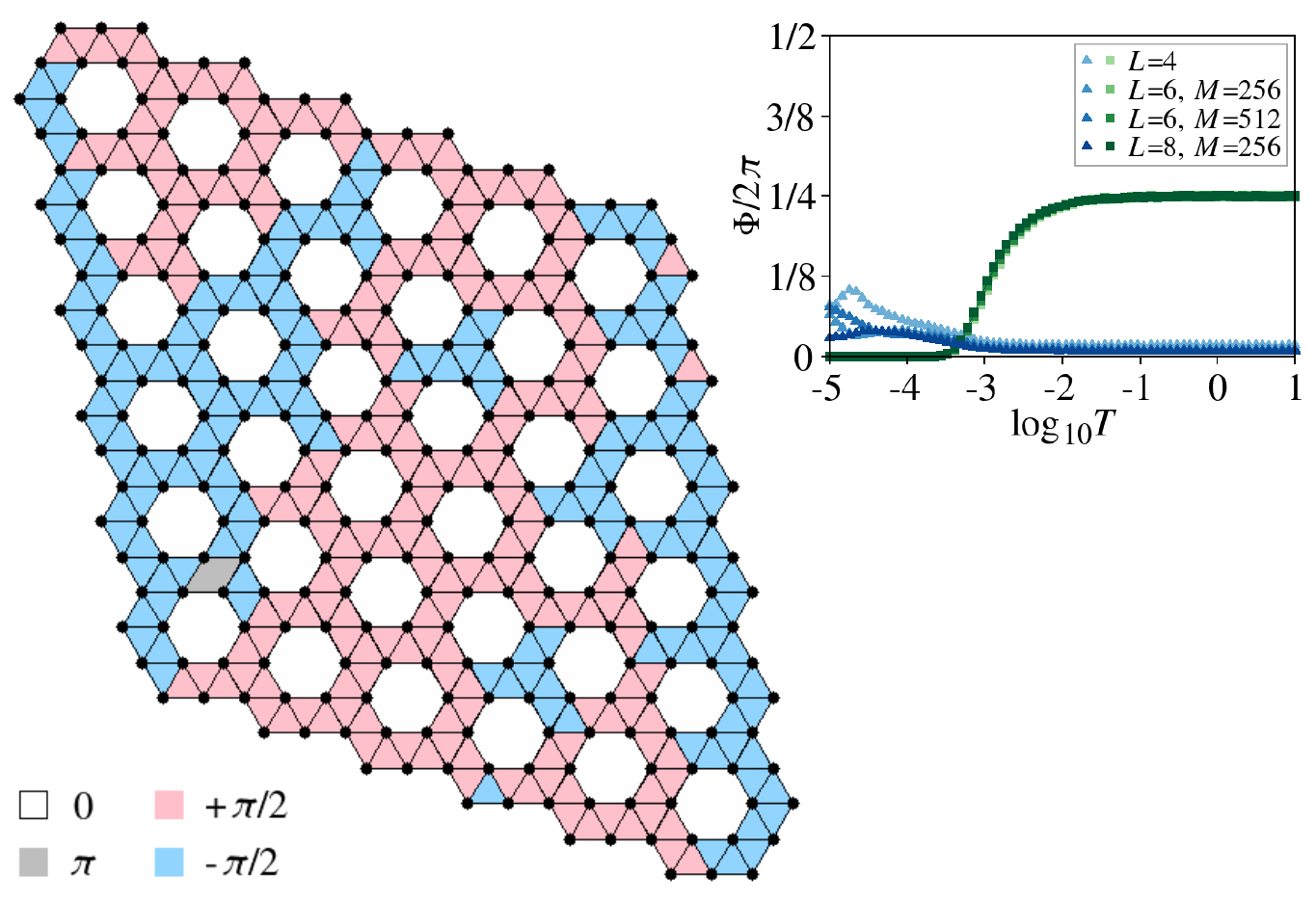}} \qquad
\subfloat[]{\label{figure:domain}
\includegraphics[scale=0.4]{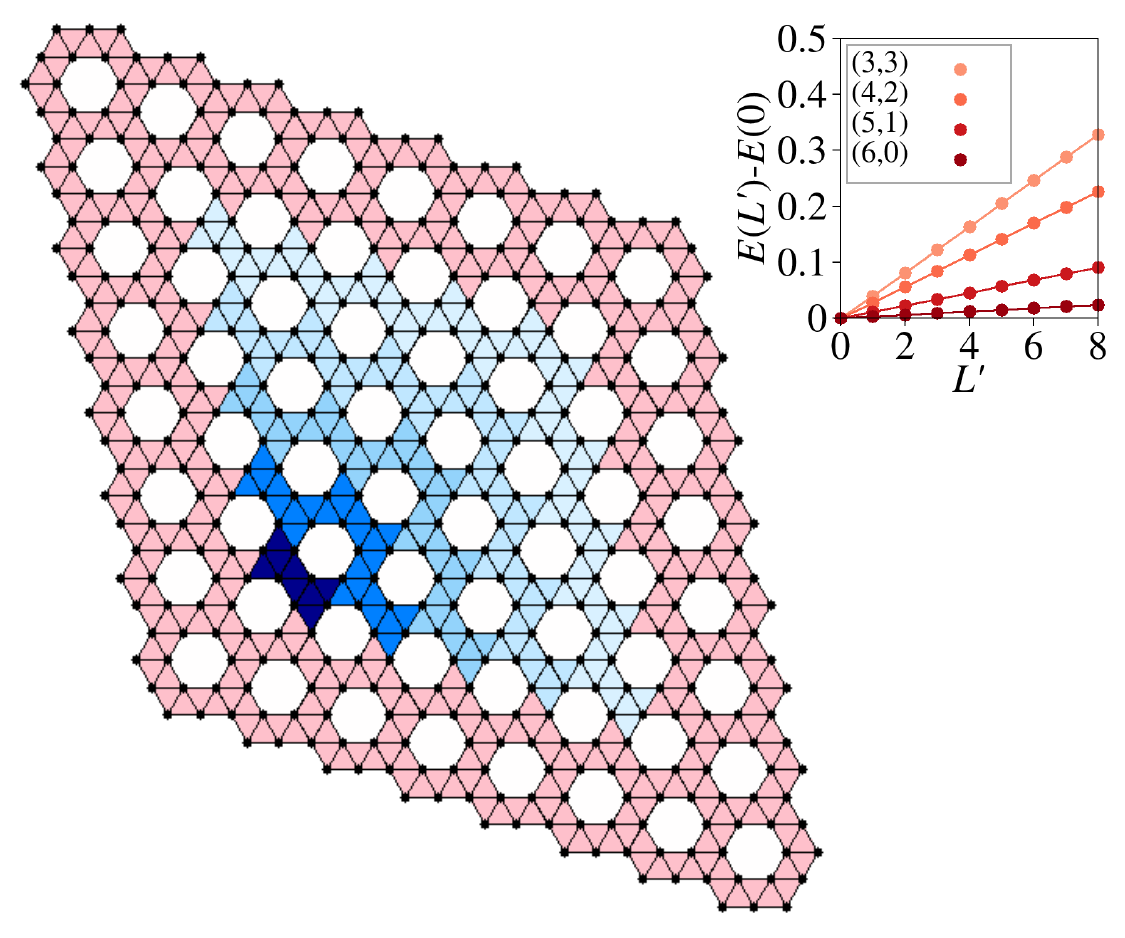}}
\caption{(a) A snapshot of the flux configuration of the Kitaev maple-leaf model at the parameter $(t_1 , t_2) = (6 , 0)$ and the temperature $T=10^{-5}$. The colors of the elementary plaquettes indicate their fluxes, with white, gray, red, and blue for $\Phi_p=0$, $\pi$, $+ \pi / 2$, and $- \pi / 2$, respectively. Note that the rhombus filled with gray color in the lower left part results from the removal of a bond between two triangles, in order to enforce the physical constraint, see Sec.~\ref{section:implement}. The inset shows the spatially averaged fluxes of the unit hexagons and triangles as functions of the temperature, for the Kitaev maple-leaf model at $(t_1 , t_2) = (6 , 0)$. The green squares and blue triangles represent $\Phi_{\protect\hexago}$ and $\lvert \Phi_{\protect\triangle} \rvert$, respectively. $L$ and $M$ are the linear size of the system and the order of Chebyshev expansion, respectively. Data with (without) $M$ are obtained via the kernel polynomial (exact diagonalization) method. (b) Domains of $\Phi_p = - \pi / 2$ with varying sizes, indicated by different shades of blue, in a background of $\Phi_p = + \pi / 2$ for the unit triangles. All the unit hexagons have $\Phi_p = 0$. The darkest blue domain has a linear size of $L'=1$. The inset shows the ground-state energy $E (L')$ as a function of the linear size $L' \leq 8$ of the domain, relative to $E (0)$, of the Kitaev maple-leaf model at various parameters $(t_1 , t_2)$ on a torus of linear size $L = 10$.}
\end{figure}

From our simulations of the Kitaev maple-leaf model, we find that the hexagon and triangle fluxes are ordered in the low temperature regime at all selected parameters except those near the bottom edge of the triangular parameter space, where $J_{x0} + J_{y0}$ is much larger than $J_{zx}=J_{zy}=J_{zz}$, see Fig.~\ref{figure:parameter}. At these parameters, the hexagon fluxes are ordered at low temperatures, but the triangle fluxes are not fully ordered down to $T=10^{-5}$. For instance, $\lvert \Phi_{\triang} \rvert$ only reaches a fraction of the saturated value $\pi / 2$ at the parameter $(t_1 , t_2) = (6 , 0)$, as shown in the inset of Fig.~\ref{figure:wall}. Snapshots of the flux configuration at $T=10^{-5}$ reveal connected regions, or $\textit{domains}$, of unit triangles with $\Phi_p = + \pi / 2$ or $\Phi_p = - \pi / 2$, see Fig.~\ref{figure:wall} for example. We interpret this situation as arising from the small cost of creating domain walls between regions of opposite fluxes, due to the small energy scale of the bonds between unit triangles, thus obstructing the formation of a single flux domain. To support our interpretation, we compute the free energy for the nominal ground-state flux sector, where $\lvert \Phi_{\triang} \rvert = \pi / 2$ and $\Phi_{\hexago} = 0$, at $T=10^{-5}$. We confirm that it provides a lower bound for the free energies obtained in our simulations, albeit the difference is small. We further investigate how the ground-state energy $E = - \sum_k \varepsilon_k / 2$ changes as we reverse the triangle fluxes in a domain of linear size $0 \leq L' \leq 8$ within a system of linear size $L=10$, see Fig.~\ref{figure:domain}. The inset shows that $E (L') - E (0)$ is non-negative and scales linearly with $L'$, i.e., the perimeter of the domain wall. These calculations strongly suggest that the ground-state flux sector is dictated by the flux phase conjecture within the parameter space of interest.

From our simulations of the Kitaev trellis model, we find that the square and triangle fluxes are ordered in the low temperature regime at all selected parameters except those near the left edge of the triangular parameter space, where $J_{zx}$ is much smaller than $J_{zz} + J_{x0}$, see Fig.~\ref{figure:parameter}. At these parameters, the square fluxes are usually saturated at the lowest temperatures, but the triangle fluxes may not be saturated, i.e., $\Phi_{\squa}=\pi$ and $\lvert \Phi_{\triang} \rvert < \pi/2$, especially when the system size is large. Snapshots of the flux configuration at $T=10^{-5}$ reveal rows of triangular plaquettes with fluxes opposite to the rest, which cost little energy as different rows are weakly coupled by $J_{zx}=J_{zy}$. A direct comparison between the ground-state energies of such flux sectors and the nominal ground-state flux sector, where $\Phi_{\squa}=\pi$ and $\lvert \Phi_{\triang} \rvert = \pi/2$, shows that the latter is indeed more energetically favorable.

\section{\label{section:band}Band Topology}

In this section, we provide details of the analysis of band topology and the computation of the Chern number, for the Kitaev maple-leaf and trellis models in their respective ground-state flux sectors.

\subsection{\label{section:two}Two-Band Model}

We review the calculation of the Chern number for a two band model based on Ref.~\onlinecite{bernevigtextbook}. Consider a Hamiltonian of free fermions $H = \sum_{\mathbf{q}} c_{\mathbf{q} \alpha}^\dagger [ H (\mathbf{q}) ]_{\alpha \beta} c_{\mathbf{q} \beta}$ expressed as a sum over momenta $\mathbf{q}$, where $H ( \mathbf{q} )$ is a $2 \times 2$ matrix, and $c_{\mathbf{q} \alpha}$ can be either complex or real (i.e., Majorana) fermions. By hermicity, we can generically express $H (\mathbf{q}) = d_0 (\mathbf{q}) \vmathbb{1} + \mathbf{d} (\mathbf{q}) \cdot \bm{\sigma}$, where $\mathbf{d} (\mathbf{q}) = ( d_1 (\mathbf{q}) , d_2 (\mathbf{q}) , d_3(\mathbf{q}) )$, $d_{0,1,2,3}(\mathbf{q})$ are real functions of $\mathbf{q}$, and $\bm{\sigma}=( \sigma_1 , \sigma_2 , \sigma_3 )$ is the vector of Pauli matrices~\footnote{Note that here we adopt the notation $\sigma_{1,2,3}$ for the Pauli matrices, instead of $\sigma^{x,y,z}$ as in the main text.}. $d_0 (\mathbf{q})$ is of no importance to the band topology, so we can simply neglect it. For brevity, we will often suppress the $\mathbf{q}$ dependence of $d_{1,2,3}$ and various other quantities in subsequent discussions. The eigenvalues of $H (\mathbf{q})$ are $\pm d$, where $d = (d_1^2 + d_2^2 + d_3^2)^{1/2}$ is the norm of the vector $\mathbf{d}$, and the system is gapless (i.e., $d ( \mathbf{q})=0$) if and only if all components of $\mathbf{d}( \mathbf{q})$ vanish (i.e., $d_i (\mathbf{q})=0$ for all $i$) at some $\mathbf{q}$. Assuming that the system is gapped, the normalized eigenvector corresponding to the eigenvalue $-d$ is given by 
\begin{equation}
\lvert \psi_- (\mathbf{q}) \rangle = \frac{1}{\sqrt{2d (d - d_3)}} \begin{pmatrix} d_3 - d \\ d_1 + i d_2 \end{pmatrix} \, .
\end{equation}
Let us define the Berry connection $\mathcal{A}_\lambda (\mathbf{q}) = - i \langle \psi_- \lvert \partial_{q_\lambda} \rvert \psi_- \rangle$~\footnote{Here, the Berry connection is defined in accordance with the convention for the Chern number in Ref.~\onlinecite{KITAEV20062}. We caution readers that the definition in Ref.~\onlinecite{bernevigtextbook} differs from ours by a minus sign.} for the state $\lvert \psi_- (\mathbf{q}) \rangle$. After some algebra, we arrive at
\begin{equation} \label{berryconnection}
\mathcal{A}_\lambda = \frac{1}{2d (d - d_3)} ( d_1 \partial_{q_\lambda} d_2 - d_2 \partial_{q_\lambda} d_1 ) \, .
\end{equation}
We then proceed to calculate the Berry curvature $\mathcal{F}_{\mu \nu} (\mathbf{q}) \equiv \partial_{q_\mu} \mathcal{A}_\nu - \partial_{q_\nu} \mathcal{A}_\mu$. After some algebra, we arrive at
\begin{equation} \label{berrycurvature}
\mathcal{F}_{\mu \nu} = - \frac{1}{2d^3} \epsilon_{ijk} d_i \partial_{q_\mu} d_j \partial_{q_\nu} d_k \, ,
\end{equation}
where $\epsilon_{ijk}$ is the totally antisymmetric tensor and repeated indices are summed over. Finally, the Chern number of the lower band is given by
\begin{equation} \label{chernnumber}
\nu = \frac{1}{2 \pi} \int \mathrm{d}^2 \mathbf{q} \, \mathcal{F}_{\mu \nu} (\mathbf{q}) \, .
\end{equation}

Let us specialize to the case $d_1 = a_{11} q_x + a_{21} q_y$, $d_2 = a_{12} q_x + a_{22} q_y$, $d_3=\mathsf{m}$, where $a_{ij}$ are entries of a matrix $\mathsf{A} \in \mathrm{GL} (2 , \mathbb{R})$ that has a nonzero determinant. The resulting Hamiltonian describes a gapless dispersion with a conic singularity at $\mathbf{q}=0$ if $\mathsf{m}=0$, and a gapped dispersion otherwise. With~\eqref{berrycurvature}, we obtain $\mathcal{F}_{xy} = - \mathsf{m} \det \mathsf{A} / 2 d^3$. To calculate the Chern number, we use polar coordinates $(q , \phi)$ in the integral~\eqref{chernnumber} with the ranges $q \in [ 0 , \infty )$ and $\phi \in [ 0 , 2 \pi )$, and apply the formula
\begin{equation}
\int_0^{2 \pi} \frac{\mathrm{d} \phi}{a \cos^2 \phi + b \sin^2 \phi + 2c \cos \phi \sin \phi} = \frac{2 \pi}{\sqrt{ab - c^2}} \, .
\end{equation}
Eventually, we are led to~\footnote{A useful trick to evaluate these integrals is the substitution $x=\tan \theta$.}
\begin{equation} \label{chernnumberhalf}
\nu = - \frac{1}{2} \mathrm{sgn} (\mathsf{m}) \mathrm{sgn} (\det \mathsf{A}) \, .
\end{equation}
The Chern number not being an integer but taking values of $\pm 1/2$ might seem like a spurious result at first sight. The reason behind this is our treatment of the problem in the continuum, where we have integrated $\mathbf{q}$ over the noncompact manifold $\mathbb{R}^2$ in~\eqref{chernnumber}. On a lattice, in contrast, the bandwidth is finite, and the so-called spectator fermions at higher energies (in absolute values) contribute the missing half of the Chern number. The lesson is that one cannot fully determine the Chern number by just analyzing a small neighborhood in the Brillouin zone where the gap is the smallest. Nonetheless, the \textit{change} in Chern number due to the closing and reopening of a gap in this small neighborhood is a meaningful quantity. For instance, suppose that $\det \mathsf{A} = -1$ is fixed. Then, by tuning the mass term $\mathsf{m}$ from positive to negative, the gap disappears and reappears, and the Chern number changes by $-1$ according to~\eqref{chernnumberhalf}. If the Chern number before the transition is known to be $\nu$, then the Chern number after the transition will be $\nu-1$.

\subsection{\label{section:fourier}Fourier Transform}

From the definition~\eqref{fourier}, one can check that $\lbrace c_{\mathbf{q} \alpha} , c_{\mathbf{q}' \alpha'}^\dagger \rbrace = \delta_{\mathbf{q}\mathbf{q}'} \delta_{\alpha \alpha'}$. It follows from $c_{\mathbf{R} \alpha}^\dagger = c_{\mathbf{R} \alpha}$ that $c_{\mathbf{q} \alpha}^\dagger = c_{-\mathbf{q} \alpha}$. Besides, the inverse Fourier transform is given by
\begin{equation}
c_{\mathbf{R} \alpha} = \sqrt{\frac{2}{\mathcal{N}}} \sum_\mathbf{q} c_{\mathbf{q} \alpha} e^{i \mathbf{q} \cdot \mathbf{R}} \, .
\end{equation}

The fermion spectrum $\varepsilon (\mathbf{q})$ is given by the eigenvalues of the matrix $i A (\mathbf{q})$ that appears in~\eqref{kitaevmodelfourier}. One finds $\varepsilon (- \mathbf{q}) = - \varepsilon ( \mathbf{q} )$ due to the antisymmetry of the matrix $A_{ij}$, which reflects the built-in particle-hole redundancy of the Majorana fermion representation. Therefore, the complete set of eigenvalues can be called the ``double'' spectrum, while the ``single'' or excitation spectrum contains only the non-negative half of the eigenvalues. 

In the next two subsections, we provide details of mapping out the topological phase diagrams of the Kitaev maple-leaf and trellis models (Figs.~\ref{figure:maplemass} and~\ref{figure:trellismass}) in their respective triangular parameter spaces $J_\mu + J_\nu + J_\lambda = 1$. For numerical calculations of the fermion gap $\Delta_\psi$ and the Chern number $\nu$, we employ a uniform grid containing $L \times L$ discrete momenta in the first Brillouin zone, with increasing $L = 100, 200, 400$ to ensure convergence.

\subsection{\label{section:mapletopology}Kitaev Maple-Leaf Model}

The maple-leaf lattice has six sites per unit cell, so $i A (\mathbf{q})$ is a $6 \times 6$ matrix, the diagonalization of which is difficult by hand but easy by computer. We thus numerically diagonalize
\begin{equation} \label{genericmomentum}
\begin{aligned}[b]
& i A ( \mathbf{q} ) = 2 i \begin{pmatrix}
0 & J_{y0} & J_{zy} e^{i q_1} & J_{zz} e^{i q_1} & - J_{zx} e^{i (q_1 + q_2)} & J_{x0} \\
- J_{y0} & 0 & - J_{x0} & J_{zx} e^{i (q_1 + q_2)} & - J_{zy} e^{i (q_1 + q_2)} & - J_{zz} e^{i q_2} \\
- J_{zy} e^{- i q_1} & J_{x0} & 0 & J_{y0} & J_{zz} e^{i q_2} & J_{zx} e^{i q_2} \\
- J_{zz} e^{- i q_1} & - J_{zx} e^{- i (q_1 + q_2)} & - J_{y0} & 0 & - J_{x0} & J_{zy} e^{- i q_1} \\
J_{zx} e^{- i (q_1 + q_2)} & J_{zy} e^{- i (q_1 + q_2)} & - J_{zz} e^{- i q_2} & J_{x0} & 0 & J_{y0} \\
- J_{x0} & J_{zz} e^{- i q_2} & - J_{zx} e^{- i q_2} & - J_{zy} e^{i q_1} & - J_{y0} & 0
\end{pmatrix} , \\
& q_i \in [0 , 2 \pi) , \quad \begin{pmatrix} q_1 \\ q_2 \end{pmatrix} = \mathsf{B} \begin{pmatrix} q_x \\ q_y \end{pmatrix} , \quad \mathsf{B} \equiv \begin{pmatrix} \sqrt{3}/2 & - 1/2 \\ 0 & 1 \end{pmatrix} , \quad J_{zx} = J_{zy} = J_{zz} \, ,
\end{aligned}
\end{equation}
which is constructed according to the $C_3$ symmetric gauge in Fig.~\ref{figure:maplegauge}. Allowing these bands to cross each other as long as $\Delta_\psi$ remains finite, we consider the non-Abelian Berry connection and curvature,
\begin{equation}
\mathcal{A} (\mathbf{q}) = - i \psi^\dagger (\mathbf{q}) \mathrm{d} \psi (\mathbf{q}) , \quad \mathcal{F} (\mathbf{q}) = \mathrm{d} \mathcal{A} (\mathbf{q}) + i \mathcal{A} (\mathbf{q}) \wedge \mathcal{A} (\mathbf{q})
\, ,
\end{equation}
where $\psi ( \mathbf{q} )$ is a $6 \times 3$ matrix whose columns are the three orthonormal eigenvectors of $i A (\mathbf{q})$ corresponding to the negative eigenvalues, so $\mathcal{A} ( \mathbf{q} )$ and $\mathcal{F} ( \mathbf{q} )$ are $3 \times 3$ matrices. We have borrowed notations from differential forms, where $\mathrm{d}$ and $\wedge$ represent the exterior derivative and the exterior product, respectively. The Chern number is given by
\begin{equation} \label{chernnonabel}
\nu = \frac{1}{2 \pi} \int_\mathrm{FBZ} \mathrm{Tr} [\mathcal{F (\mathbf{q})}] = \frac{1}{2 \pi} \int_\mathrm{FBZ} \mathrm{Tr} [\mathrm{d} \mathcal{A (\mathbf{q})}] \, ,
\end{equation}
where the integration is over the first Brillouin zone (FBZ), and the second equality follows from the tracelessness of $i \mathcal{A} (\mathbf{q}) \wedge \mathcal{A} (\mathbf{q})$. Note that $\nu$ is invariant under the transformation $\psi ( \mathbf{q} ) \longrightarrow \psi ( \mathbf{q} ) U (\mathbf{q})$ for $U(\mathbf{q}) \in U(3)$, owing to $\mathcal{F} ( \mathbf{q} ) \longrightarrow U^\dagger ( \mathbf{q} ) \mathcal{F} ( \mathbf{q} ) U ( \mathbf{q} )$ and the cyclic property of trace. We evaluate~\eqref{chernnonabel} numerically using the manifestly gauge invariant expression~\cite{JPSJ.74.1674},
\begin{equation} \label{chernnonabelcompute}
\nu \approx \frac{1}{2 \pi} \sum_{\mathbf{q} \in \mathrm{FBZ}} \mathrm{Im} \ln \det \left[ \psi^\dagger (\mathbf{q}) \psi (\mathbf{q} + \hat{1}) \psi^\dagger (\mathbf{q} + \hat{1}) \psi (\mathbf{q} + \hat{1} + \hat{2}) \psi^\dagger (\mathbf{q} + \hat{1} + \hat{2}) \psi (\mathbf{q} + \hat{2}) \psi^\dagger (\mathbf{q} + \hat{2}) \psi (\mathbf{q}) \right] \, ,
\end{equation}
where $\hat{1}$ and $\hat{2}$ represent small increments of the momentum along the $\mathbf{b}_1$ and $\mathbf{b}_2$ directions, respectively, on the reciprocal lattice~\footnote{In theory the increments $\lvert \hat{1} \rvert$ and $\lvert \hat{2} \rvert$ are infinitesimal, but in practice we let them be the spacing between neighboring points on the momentum grid, $2 \pi / L$.}. 

\begin{figure}
\includegraphics[scale=0.24]{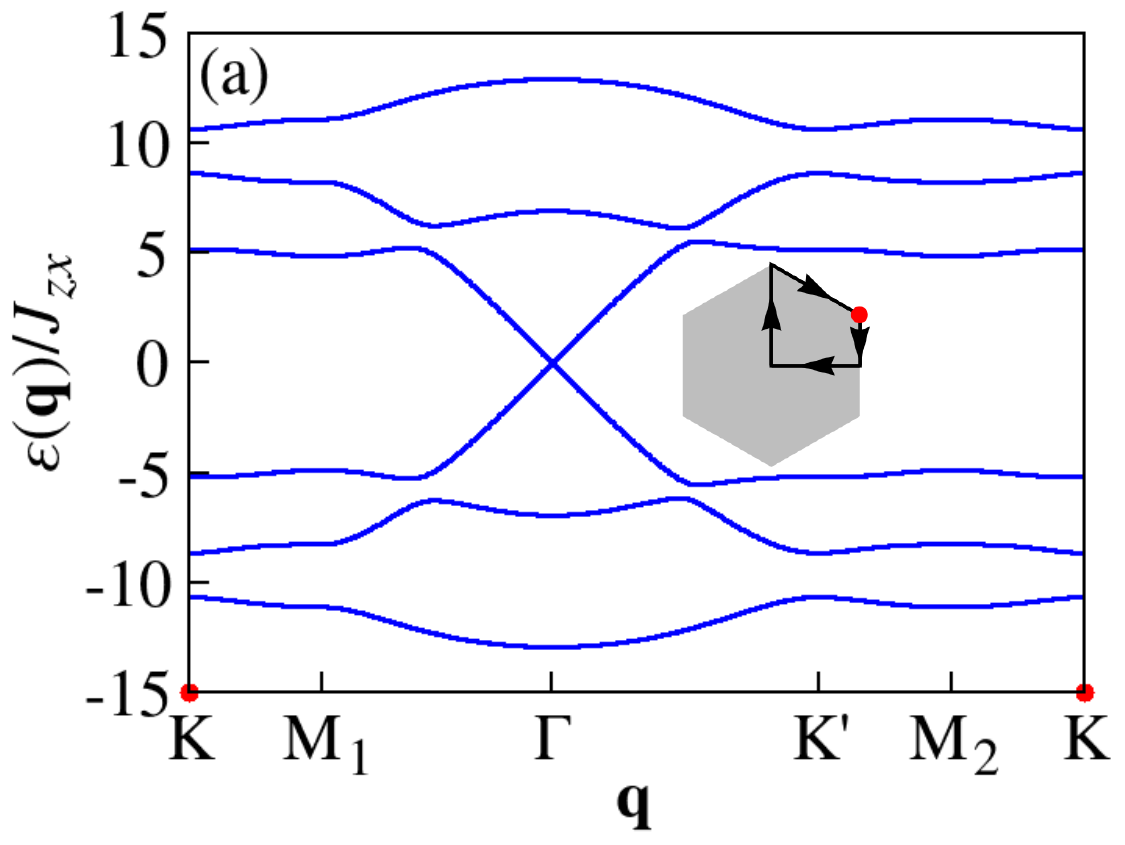} \qquad
\includegraphics[scale=0.24]{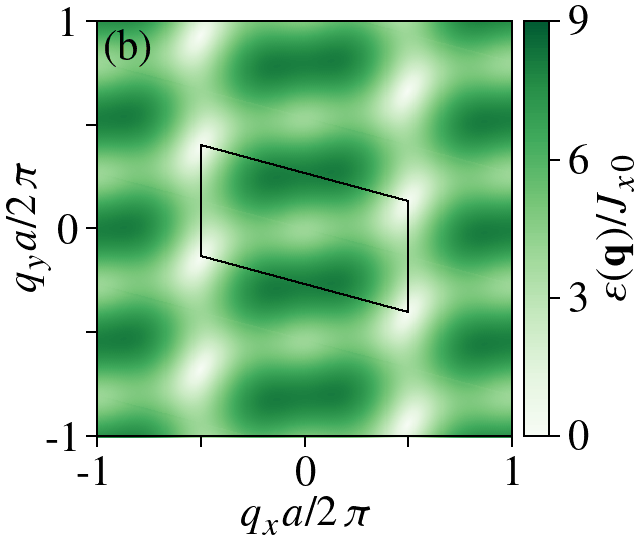}
\caption{\label{figure:gapless}(a) The (doubled) fermion spectrum $\varepsilon (\mathbf{q})$, including both positive and negative energies, of the Kitaev maple-leaf model with the couplings $J_{x0} = J_{y0} = (1 + \sqrt{3}) J_{zx}$, where the gap closes at the $\Gamma$ point. The inset shows the high-symmetry path (black lines with arrows) in the first Brillouin zone (gray hexagon) along which $\varepsilon (\mathbf{q})$ is plotted, starting and ending at the $\mathrm{K}$ point (red dot). (b) The fermion spectrum $\varepsilon (\mathbf{q})$, including only the non-negative (i.e., excitation) energies, of the Kitaev trellis model with the couplings $J_{x0} = J_{zx} = J_{zz}/2$, where the gap closes at the $\mathrm{M}$ point. The parallelogram bounded by the black lines represents the first Brillouin zone.}
\end{figure}

We find that, on the boundary between the $\nu=-1$ and $\nu=0$ regimes, the fermion gap closes at the $\Gamma$ point $\mathbf{q}=0$, see Fig.~\ref{figure:gapless}a for instance. This further allows for an analytical understanding of the topological phase transition in terms of the sign change of a mass term, as follows. We first diagonalize the matrix
\begin{equation} \label{zeromomentum}
i A (\mathbf{q}=0) = 2 i \begin{pmatrix}
0 & J_{y0} & J_{zx} & J_{zx} & - J_{zx} & J_{x0} \\
- J_{y0} & 0 & - J_{x0} & J_{zx} & - J_{zx} & - J_{zx} \\
- J_{zx} & J_{x0} & 0 & J_{y0} & J_{zx} & J_{zx} \\
- J_{zx} & - J_{zx} & - J_{y0} & 0 & - J_{x0} & J_{zx} \\
J_{zx} & J_{zx} & - J_{zx} & J_{x0} & 0 & J_{y0} \\
- J_{x0} & J_{zx} & - J_{zx} & - J_{zx} & - J_{y0} & 0
\end{pmatrix}
\end{equation}
and determine its eigenvalues, which come in three positive-negative pairs
\begin{subequations}
\begin{align}
\pm \lambda_1 &= \pm \sqrt{2} \left \lvert \left[ (J_{zx} - J_{x0})^2 + (J_{x0} - J_{y0})^2 + (J_{y0} - J_{zx})^2 \right]^{1/2} - \sqrt{6} J_{zx} \right \rvert \, , \\
\pm \lambda_2 &= \pm \sqrt{2} \left \lbrace \left[ (J_{zx} - J_{x0})^2 + (J_{x0} - J_{y0})^2 + (J_{y0} - J_{zx})^2 \right]^{1/2} + \sqrt{6} J_{zx} \right \rbrace \, , \\
\pm \lambda_3 &= \pm 2 ( J_{x0} + J_{y0} + J_{zx} ) = \pm 2 \, .
\end{align}
\end{subequations}
A quick inspection reveals that it is possible for $\lambda_1$ to vanish upon varying the couplings, while $\lambda_{2,3}$ are always nonzero, within the parameter space of interest. Let the normalized eigenvectors corresponding to $\pm \lambda_1$ be $\lvert \pm \rangle$. To derive a low-energy effective Hamiltonian describing the gap-closing transition, we perform a small $\mathbf{q}$ expansion up to the linear order such that $i A (\mathbf{q} \approx 0) \approx H_0 + H_1 (\mathbf{q})$, where $H_0 \equiv i A (\mathbf{q}=0)$ is given by \eqref{zeromomentum} and
\begin{equation}
H_1 ( \mathbf{q} ) = 2 J_{zx} \begin{pmatrix}
0 & 0 & - q_1 & - q_1 & q_1 + q_2 & 0 \\
0 & 0 & 0 & - q_1 - q_2 & q_1 + q_2 & q_2 \\
- q_1 & 0 & 0 & 0 & - q_2 & - q_2 \\
- q_1 & - q_1 - q_2 & 0 & 0 & 0 & q_1 \\
q_1 + q_2 & q_1 + q_2 & - q_2 & 0 & 0 & 0 \\
0 & q_2 & - q_2 & q_1 & 0 & 0
\end{pmatrix} \, .
\end{equation}
The matrix elements of the $2 \times 2$ effective Hamiltonian is then given by $H^\mathrm{eff}_{\alpha \beta} ( \mathbf{q} ) = \langle \alpha \lvert H_0 +H_1 ( \mathbf{q} ) \rvert \beta \rangle$ where $\alpha, \beta \in \lbrace + , - \rbrace$, which has the form of a Dirac Hamiltonian $H^\mathrm{eff} ( \mathbf{q} ) =  ( q_x, q_y )^\mathrm{T} \mathsf{A} ( \sigma_x , \sigma_y ) + \mathsf{m} \sigma_z$~\footnote{This can be shown using the fact that $H_0$ is antisymmetric and $H_1$ is symmetric, which we omit here for brevity. Furthermore, we have dropped a possible contribution to $H^\mathrm{eff} ( \mathbf{q} )$ that is proportional to the identity $\vmathbb{1}$}. Since $\pm \lambda_1$ are given by the positive and negative of an absolute value, we are free to choose the sign of the mass term $\mathsf{m}$. We thus adopt the definition
\begin{equation}
\mathsf{m} = \sqrt{2} \left \lbrace \left[ (J_{zx} - J_{x0})^2 + (J_{x0} - J_{y0})^2 + (J_{y0} - J_{zx})^2 \right]^{1/2} - \sqrt{6} J_{zx} \right \rbrace \, ,
\end{equation}
such that $\mathsf{m}$ changing from positive to negative corresponds to $\nu$ changing by $-1$, see Fig.~\ref{figure:maplemass}. For consistency, let us numerically check that $\det \mathsf A < 0 $ for two parameters on different sides of the topological phase transition, as required by~\eqref{chernnumberhalf}. When $J_{x0} = J_{y0}$, we find that $\mathsf{m}=0$ if and only if $J_{x0}=(1 + \sqrt{3}) J_{zx}$, so we consider the cases $J_{x0} = (1 + \sqrt{3} + \delta) J_{zx}$ with $\delta = \pm 0.05$. For $\delta > 0$, we have $\mathsf{m}>0$, so $+$ ($-$) is identified with the index 1 (2) in the construction of $H^\mathrm{eff} ( \mathbf{q} )$. For $\delta < 0$, we have $\mathsf{m}<0$, so $-$ ($+$) is identified with the index 1 (2) in the construction of $H^\mathrm{eff} ( \mathbf{q} )$. In both cases, we find
\begin{equation}
\mathsf{A} = \mathsf{B}^\mathrm{T} \begin{pmatrix} 3.15 & 1.15 \\ 2.58 & -2.15 \end{pmatrix} J_{zx} \, ,
\end{equation}
which shows that $\det \mathsf{A}$ is indeed negative~\footnote{The direct computation of the Chern number using the formula~\eqref{chernnonabelcompute} can be done with oblique coordinates $(q_1 , q_2)$ in the momentum space. Here, since we are applying the result~\eqref{chernnumberhalf}, whose derivation involves an integral in polar coordinates, we express the effective Hamiltonian in terms of the orthogonal coordinates $(q_x , q_y)$, which are related to $(q_1 , q_2)$ via a linear transformation as shown in~\eqref{genericmomentum}.}.

\subsection{\label{section:trellistopology}Kitaev Trellis Model}

We solve for the gap-closing condition of the Hamiltonian~\eqref{kitaevmodelfourier} constructed according to the gauge choice in Fig.~\ref{figure:trellisgauge}. One obtains $i A (\mathbf{q}) = \sum_i d_i ( \mathbf{q} ) \sigma_i$ with
\begin{subequations}
\begin{align}
d_1 ( \mathbf{q} ) &= - 2 J_{zx} \sin q_2 + 2 J_{zz} \sin (q_1 - q_2) \, , \\
d_2 ( \mathbf{q} ) &= - 2 J_{zx} ( 1 - \cos q_2 ) + 2 J_{zz} \cos (q_1 - q_2) \, , \\
d_3 ( \mathbf{q} ) &= - 4 J_{x0} \sin q_1 \, .
\end{align}
\end{subequations}

First of all, note that $\Delta_\psi = 0$ iff $d_i ( \mathbf{q} ) = 0$ for all $i$ at some $\mathbf{q}$. Since $J_{x0} > 0$ is nonzero, we have $d_3 = 0$ iff $q_1=0 , \pi$.

\vspace{0.2cm}
~\hspace{1.0cm}
\begin{minipage}{15.0cm}
\begin{itemize}
\item[\textbf{Case 1.}] $q_1 = 0$. Then, $d_1 = - 2 ( J_{zx} + J_{zz} ) \sin q_2$. Since $J_{zx} + J_{zz} > 0$ is nonzero, $d_1 = 0$ iff $q_2 = 0 , \pi$.

\item[\textbf{Case 1.1.}] $q_2 = 0$. Then, $d_2 = 2 J_{zz} > 0$ is nonzero. We discard this case.

\item[\textbf{Case 1.2.}] $q_2 = \pi$. Then, $d_2 = - 4 J_{zx} - 2 J_{zz} < 0$ is nonzero. We discard this case.

\item[\textbf{Case 2.}] $q_1 = \pi$. Then, $d_1 = 2 (- J_{zx} + J_{zz}) \sin q_2$, which is zero iff $J_{zx}=J_{zz}$ or $q_2 = 0 , \pi$.

\item[\textbf{Case 2.1.}] $J_{zx} = J_{zz}$. Then, $d_2 = - 2 J_{zx} < 0$ is nonzero. We discard this case.

\item[\textbf{Case 2.2.}] $q_2 = 0$. Then, $d_2 = - 2 J_{zz} < 0$ is nonzero. We discard this case.

\item[\textbf{Case 2.3.}] $q_2 = \pi$. Then, $d_2 = - 4 J_{zx} + 2 J_{zz}$. This is the only valid case. 
\end{itemize}
\end{minipage}

\vspace{0.2 cm}
We conclude that the fermion gap closes at $(q_1 , q_2) = (\pi , \pi)$ iff the couplings satisfy $- 4 J_{zx} + 2 J_{zz} = 0$, see Fig.~\ref{figure:gapless}b for example. 

We verify our analysis in Sec.~\ref{section:topology} by computing the Chern number of the lower band for each gapped regime. We use the manifestly gauge invariant expression~\cite{JPSJ.74.1674}
\begin{equation}
\nu \approx \frac{1}{2 \pi} \sum_\mathbf{q \in \mathrm{FBZ}} \mathrm{Im} \ln \left[ \psi^\dagger ( \mathbf{q} ) \psi ( \mathbf{q} + \hat{1} ) \psi^\dagger ( \mathbf{q} + \hat{1} ) \psi ( \mathbf{q} + \hat{1} + \hat{2} ) \psi^\dagger ( \mathbf{q} + \hat{1} + \hat{2} ) \psi ( \mathbf{q} + \hat{2} ) \psi^\dagger ( \mathbf{q} + \hat{2} ) \psi ( \mathbf{q} ) \right] \, ,
\end{equation}
where $\psi ( \mathbf{q} )$ is the eigenvector of $i A (\mathbf{q})$ corresponding to the eigenvalue $- d ( \mathbf{q} )$, while $\hat{1}$ and $\hat{2}$ represent small increments of the momentum along the $\mathbf{b}_1$ and $\mathbf{b}_2$ directions, respectively, on the reciprocal lattice.

\section{\label{section:toriccode}Toric Code Phase}

In this section, we provide further analysis of the toric code (i.e., $\nu \, \mathrm{mod} \, 16=0$) phase of the Kitaev maple-leaf model.

\begin{figure}
\subfloat[]{\label{figure:mapleperturb}
\includegraphics[scale=0.2]{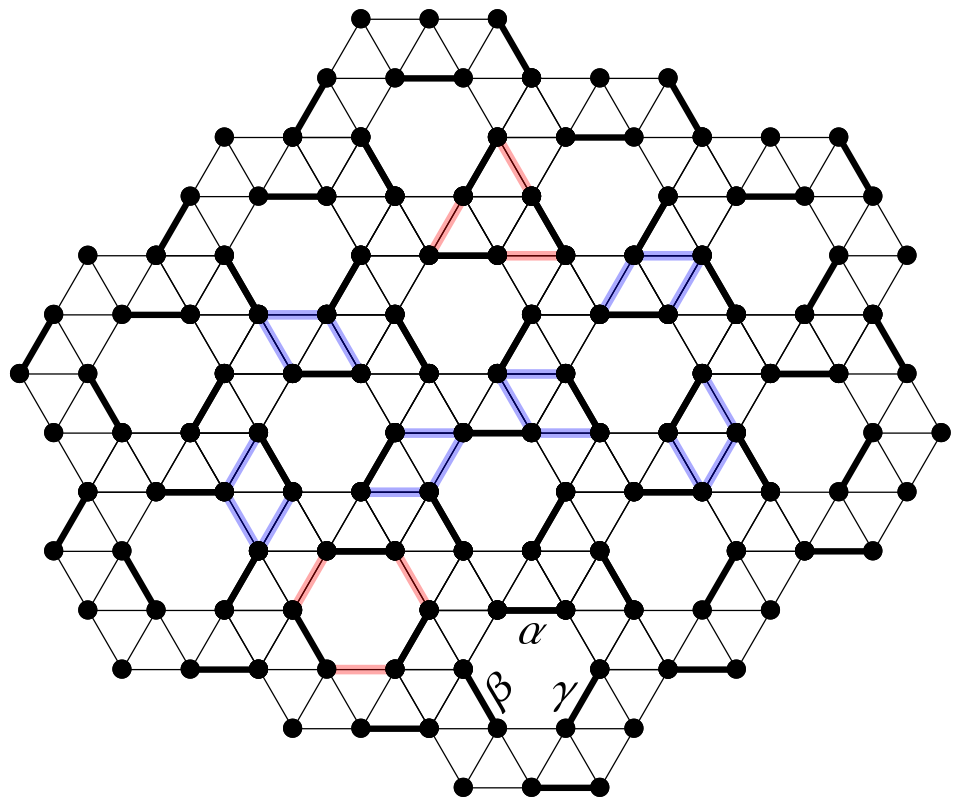}} \qquad
\subfloat[]{\label{figure:path}
\includegraphics[scale=0.23]{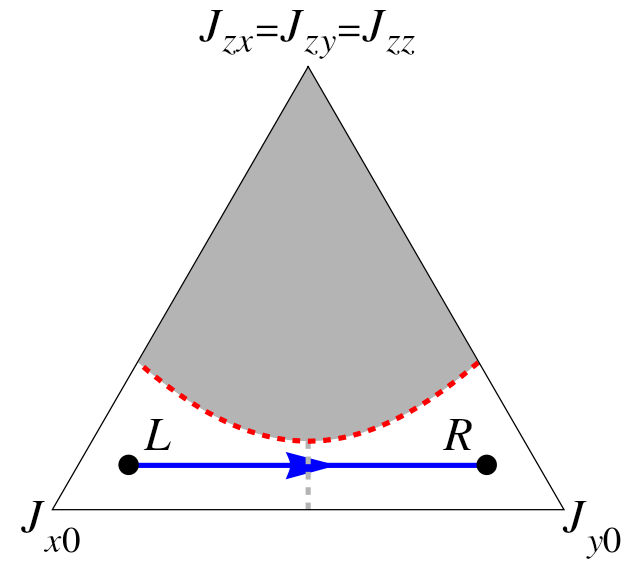}} \qquad
\subfloat[]{\label{figure:cut}
\includegraphics[scale=0.24]{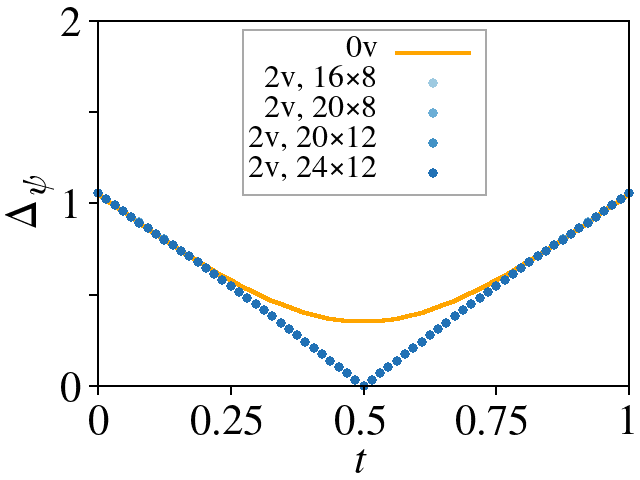}}
\caption{(a) The Kitaev maple-leaf model in the strong $J_{x0}$ limit. The strong (i.e., $x0$) and weak (i.e., $\lambda \neq x0$) bonds are indicated by thick and thin black lines, respectively. The three dimers in a unit cell are labelled by $\alpha$, $\beta$ and $\gamma$. The weak bonds highlighted by red (blue) indicate the perturbations that yield the length-$6$($4$) plaquette operators appearing in the effective Hamiltonian~\eqref{thirdorder}. (b) The path $LR$, which is indicated by the blue solid line with an arrow, in the triangular parameter space of the Kitaev maple-leaf model. The red dashed line indicates the vanishing of the fermion gap in the vortex-free sector and all two-vortex sectors. The gray filled area (gray dashed line) indicates the vanishing of the fermion gap in all (some) two-vortex sectors, while the vortex-free sector has a finite gap. See also Figs.~\ref{figure:maplegap}, \ref{figure:maplegapht}, and \ref{figure:maplegaptt}. (c) The averaged fermion gap of two-vortex ($2\mathrm{v}$) sectors with at least one vortex hosted by a unit hexagon along the path $LR$, which is linearly parametrized by $t \in [0,1]$. The product of two integers associated with each set of $2\mathrm{v}$ data indicates the size of the system ($L_1 \times L_2$ unit cells) on which the computations are performed. The data of various system sizes overlap, and the error bars are smaller than the symbols. The fermion gap of the vortex-free (0v) sector, which is calculated in the thermodynamic limit, is also plotted for comparison.}
\end{figure}

In any dimer limit of any generalized Kitaev model, it is proved that the effective Hamiltonian derived from the standard degenerate perturbation theory cannot contain an operator that acts nontrivially only on an odd-length elementary plaquette~\cite{2607.12027}. The coefficient of such an operator is exactly $0$ at all orders. It is thus not possible for the anyon species of each unit triangle, which has a length $3$, in the Kitaev maple-leaf or trellis model to be determined via degenerate perturbation theory. For instance, up to the third order, the effective Hamiltonian in the strong $J_{x0}$ limit of the Kitaev maple-leaf model reads
\begin{equation} \label{thirdorder}
\begin{aligned}[b]
H_\mathrm{eff}^{(3)} = \sum_\mathbf{R} \Bigg[ &- \frac{3}{8} \frac{J_{y0}^3}{J_{x0}^2} (\sigma^y \otimes \vmathbb{1} \otimes \vmathbb{1})_{\mathbf{R} ; \alpha} (\sigma^y \otimes \vmathbb{1} \otimes \vmathbb{1})_{\mathbf{R} ; \beta} (\sigma^y \otimes \vmathbb{1} \otimes \vmathbb{1})_{\mathbf{R} ; \gamma} \\
&+ \frac{3}{8} \frac{J_{zx} J_{zy} J_{zz}}{J_{x0}^2} (\sigma^y \otimes \sigma^x \otimes \sigma^y)_{\mathbf{R} ; \alpha}  (\sigma^y \otimes \sigma^y \otimes \sigma^z)_{\mathbf{R} + \mathbf{a}_1 + \mathbf{a}_2 ; \beta} (\sigma^y \otimes \sigma^z \otimes \sigma^x)_{\mathbf{R} + \mathbf{a}_2 ; \gamma} \\
&- \frac{1}{8} \frac{J_{zx}^2 J_{zy}}{J_{x0}^2} (\sigma^y \otimes \sigma^x \otimes \sigma^x)_{\mathbf{R} ; \alpha} (\vmathbb{1} \otimes \vmathbb{1} \otimes \sigma^z)_{\mathbf{R} + \mathbf{a}_1 + \mathbf{a}_2; \beta} (\vmathbb{1} \otimes \vmathbb{1} \otimes \sigma^z)_{\mathbf{R} + \mathbf{a}_2 ; \gamma} \\
&- \frac{1}{8} \frac{J_{zy}^2 J_{zz}}{J_{x0}^2} (\vmathbb{1} \otimes \vmathbb{1} \otimes \sigma^x)_{\mathbf{R} ; \alpha} (\sigma^y \otimes \sigma^y \otimes \sigma^y)_{\mathbf{R} + \mathbf{a}_1 + \mathbf{a}_2 ; \beta} (\vmathbb{1} \otimes \vmathbb{1} \otimes \sigma^x)_{\mathbf{R} + \mathbf{a}_2 ; \gamma} \\
&- \frac{1}{8} \frac{J_{zz}^2 J_{zx}}{J_{x0}^2} (\vmathbb{1} \otimes \vmathbb{1} \otimes \sigma^y)_{\mathbf{R} ; \alpha} (\vmathbb{1} \otimes \vmathbb{1} \otimes \sigma^y)_{\mathbf{R} + \mathbf{a}_1 + \mathbf{a}_2 ; \beta} (\sigma^y \otimes \sigma^z \otimes \sigma^z)_{\mathbf{R} + \mathbf{a}_2 ; \gamma} \\
&- \frac{1}{8} \frac{J_{zy}^2 J_{y0}}{J_{x0}^2} (\sigma^x \otimes \sigma^y \otimes \vmathbb{1})_{\mathbf{R} ; \alpha} (\sigma^x \otimes \vmathbb{1} \otimes \sigma^y)_{\mathbf{R} ; \beta} (\sigma^y \otimes \sigma^y \otimes \sigma^y)_{\mathbf{R} - \mathbf{a}_1 ; \gamma} \\
&- \frac{1}{8} \frac{J_{zz}^2 J_{y0}}{J_{x0}^2} (\sigma^y \otimes \sigma^z \otimes \sigma^z)_{\mathbf{R} - \mathbf{a}_2 ; \alpha} (\sigma^x \otimes \sigma^z \otimes \vmathbb{1})_{\mathbf{R} ; \beta} (\sigma^x \otimes \vmathbb{1} \otimes \sigma^z)_{\mathbf{R} ; \gamma} \\
&- \frac{1}{8} \frac{J_{zx}^2 J_{y0}}{J_{x0}^2} (\sigma^x \otimes \vmathbb{1} \otimes \sigma^x)_{\mathbf{R} ; \alpha} (\sigma^y \otimes \sigma^x \otimes \sigma^x)_{\mathbf{R} + \mathbf{a}_1 + \mathbf{a}_2 ; \beta} (\sigma^x \otimes \sigma^x \otimes \vmathbb{1})_{\mathbf{R} ; \gamma} \Bigg] \, ,
\end{aligned}
\end{equation}
where $\vmathbb{1}$ denotes the $2 \times 2$ identity matrix, $\mathbf{R}$ represents the coordinates of the unit cell, and $\alpha , \beta , \gamma$ label the three dimers in a unit cell, see Fig.~\ref{figure:mapleperturb}. The dimension of the Hilbert space per dimer is $(4 \times 4) / 2 = 8$ in the ground-state subspace of the unperturbed Hamiltonian, so $H_\mathrm{eff}^{(3)}$ is given in terms of local operators of dimension $8$ acting on the dimers. We include only non-constant operators in~\eqref{thirdorder}, which are illustrated in Fig.~\ref{figure:mapleperturb}. Note that each operator acts on a plaquette of even length ($4$ or $6$), which contains an even number ($0$, $2$, or $4$) of triangular plaquettes. Analyzing such an effective Hamiltonian is not only tedious, but also fails to achieve our goal to determine the anyon species. We thus employ the mapping from Ref.~\onlinecite{2607.12027}, which is based on the fusion rule of Abelian anyons and the fermion parity of physical states, in Sec.~\ref{section:anyon}. 

For the Kitaev maple-leaf model, we have shown that the strong $J_{x0}$ and $J_{y0}$ limits are separated by a fermion-gap-closing transition in the two-vortex sector with at least one vortex hosted by a unit hexagon. Here, we study the convergence of the averaged fermion gap of such two-vortex sectors with respect to the system size, along the path $LR$ with the endpoints $\mathbf{J}^{(L)} = (0.8,0.1,0.1)$ and $\mathbf{J}^{(R)}=(0.1,0.8,0.1)$, as shown in Fig.~\ref{figure:path}. The computational setup is similar to that in Ref.~\onlinecite{2607.12027}, which we briefly describe as follows. We create two well-separated vortices from the ground state in a torus of $L_1 \times L_2$ unit cells. We fix one vortex at a unit hexagon and move the other over a region of $L_1' \times L_2'$ unit cells, such that $L_1' \times L_2' \times 9$ two-vortex sectors are generated. We average the fermion gaps of these two-vortex sectors and take the standard deviation as the error. We also parametrize $LR$ as $\mathbf{J} (t) = (1-t) \mathbf{J}^{(L)} + t \mathbf{J}^{(R)}$ by $t \in [0,1]$, and choose $65$ equally spaced $t$ points for our computations.

To simplify the following discussion, we denote the fermion gap of the vortex-free sector by $\Delta_\psi^{(0\mathrm{v})}$ and the averaged fermion gap of the two-vortex sectors by $\Delta_\psi^{(2\mathrm{v})}$, which are functions of $t$. Fig.~\ref{figure:cut} shows $\Delta_\psi^{(2\mathrm{v})} (t)$ computed with (i) $L_1 \times L_2 = 16 \times 8$, $L_1' \times L_2' = 2 \times 4$, (ii) $L_1 \times L_2 = 20 \times 8$, $L_1' \times L_2' = 4 \times 4$, (iii) $L_1 \times L_2 = 20 \times 12$, $L_1' \times L_2' = 4 \times 6$, and (iv) $L_1 \times L_2 = 24 \times 12$, $L_1' \times L_2' = 6 \times 6$. The data of various system sizes overlap rather perfectly, and the smallness of the error bars (not visible) indicates that the fermion gaps being averaged over are approximately equal. As $t$ increases from $0$ to $1/2$, $\Delta_\psi^{(2\mathrm{v})} (t)$ decreases monotonically from $\sim 1$ to $0$. As $t$ further increases from $1/2$ to $1$, $\Delta_\psi^{(2\mathrm{v})} (t)$ increases monotonically from $0$ to $\sim 1$. In contrast, $\Delta_\psi^{(0\mathrm{v})} (t)$, which is calculated in the thermodynamic limit (see Sec.~\ref{section:topology}) and also plotted in Fig.~\ref{figure:cut}, remains finite throughout $LR$.

\end{document}